\documentclass[a4paper,11pt]{article}
\pdfoutput=1 
\usepackage{jheppub} 
\usepackage{bbm,bm,graphicx,bbold,mathtools,color,hyperref,slashed}
\usepackage[T1]{fontenc}

\DeclareMathOperator\ee{e}

\renewcommand\bar[1]{\overline{#1}}
\renewcommand\epsilon{\varepsilon}
\renewcommand\phi{\varphi}
\renewcommand\O{\text{O}}
\renewcommand\Im{\mathrm{Im}}
\renewcommand\Re{\mathrm{Re}}

\newcommand\1{\mathbb{1}}
\newcommand\SU{\text{SU}}

\newcommand\SO{\text{SO}}
\newcommand\U{\text{U}}

\newcommand\der{\partial}
\newcommand\calO{\mathcal{O}}
\newcommand{\RR}{\mathbb{R}}
\newcommand{\CC}{\mathbb{C}}
\newcommand{\ZZ}{\mathbb{Z}}
\newcommand\JJ{\mathcal{J}}
\newcommand\KK{\mathcal{K}}
\newcommand\JJJ{\mathfrak{J}}
\newcommand\kk{\hat{k}}
\newcommand\dd{\mathrm{d}}

\newcommand{\ckakko}[1]{\left\{{#1}\right\}}
\newcommand{\mkakko}[1]{\left({#1}\right)}
\newcommand{\kkakko}[1]{\left[{#1}\right]}

\preprint{RIKEN-QHP-174}

\title{Structure of Lefschetz thimbles in simple fermionic systems}

\author[a]{Takuya Kanazawa}
\author[b,c]{and Yuya Tanizaki}

\affiliation[a]{iTHES Research Group and Quantum Hadron Physics
  Laboratory, RIKEN, Wako, Saitama 351-0198, Japan}
\affiliation[b]{Department of Physics, The University of Tokyo, Tokyo 113-0033, Japan}
\affiliation[c]{Theoretical Research Division, Nishina Center, RIKEN, Wako, Saitama 351-0198, Japan}

\emailAdd{takuya.kanazawa@riken.jp}
\emailAdd{yuya.tanizaki@riken.jp}

\abstract{
The Picard-Lefschetz theory offers a promising tool to 
solve the sign problem in QCD and other field theories with complex 
path-integral weight. In this paper the Lefschetz-thimble approach is examined  
in simple fermionic models which share some features with QCD. 
In zero-dimensional versions of 
the Gross-Neveu model and the Nambu-Jona-Lasinio model,   
we study the structure of Lefschetz thimbles and its variation across the chiral 
phase transition.  We map out a phase diagram 
in the complex four-fermion coupling plane using 
a thimble decomposition of the path integral, and demonstrate 
an interesting link between anti-Stokes lines and Lee-Yang zeros. 
In the case of nonzero mass, it is shown that the 
approach to the chiral limit is singular because of intricate 
cancellation between competing thimbles, which implies the necessity 
to sum up multiple thimbles related by symmetry.  
We also consider a Chern-Simons theory 
with fermions in $0+1$-dimension and show how Lefschetz thimbles  
solve the complex phase problem caused by a topological term. 
These prototypical examples would aid future application of this framework 
to \emph{bona fide} QCD. 
}

\begin{document}
\maketitle
\section{Introduction}

Path integrals with complex weight appear in many branches of physics. 
Examples include the Minkowski path integral, QCD with chemical potential, 
Chern-Simons gauge theory, Yang-Mills theory in the theta vacuum  
and chiral gauge theories. 
Interest in quantum theories with complex actions was also stimulated 
by the advent of $\mathcal{PT}$ symmetry \cite{Bender:1998ke,Bender:2002vv}. 
Despite the overwhelming significance of these theories, 
only partial progress has been made towards their first-principle understanding 
partly due to the incapability 
of numerical simulations based on Monte Carlo sampling to 
deal with complex weights, which impede a probabilistic interpretation. 

A promising approach for handling complex actions is to complexify 
the field space. In a one-dimensional integral, the method of steepest descent 
(or stationary phase method) is well known, in which one deforms 
an integration contour into a more general path on the complex plane so that 
it passes through a critical point (saddle) of the integrand. This allows for an 
asymptotic evaluation of exponential integrals. The generalization of this method 
to higher dimensions is provided by the Morse theory (or Picard-Lefschetz theory) 
\cite{Pham1983,Nicolaescu2011}, 
where contours of steepest descent are generalized to higher-dimensional 
curved manifolds called \emph{Lefschetz thimbles}. 
Recently, a direct application of the Picard-Lefschetz theory to infinite-dimensional 
path integral in quantum field theory (QFT) 
was made by Witten \cite{Witten:2010cx, Witten:2010zr}.  
He showed that Chern-Simons path integral can be defined nonperturbatively 
for complex gauge fields by taking Lefschetz thimbles as integration cycles.  
This work has provoked subsequent developments of numerical approaches 
to complex path integrals on the basis of Picard-Lefschetz theory 
\cite{Cristoforetti:2012su,Cristoforetti:2013wha,Mukherjee:2013aga,
Aarts:2013fpa,Fujii:2013sra,Cristoforetti:2014gsa,Mukherjee:2014hsa,Aarts:2014nxa}. 
They are reminiscent of the complex Langevin method 
\cite{Parisi:1984cs,Klauder:1983zm,Klauder:1983sp}, 
which is also based on the idea of field complexification, but there seem to be  
fundamental differences \cite{Aarts:2013fpa,Aarts:2014nxa}. 

The Picard-Lefschetz theory also provides a useful framework for obtaining 
a visual understanding of a subtle interplay between perturbative and 
nonperturbative saddles in asymptotic series \cite{Berry1991}.  
See \cite{David:1992za,Felder:2004uy,Eynard:2008yb,Marino:2008ya,
Marino:2009dp,Pasquetti:2009jg,Chan:2010rw,Chan:2012ud,
Marino:2012zq,Schiappa:2013opa} for related works in matrix models  
with applications to quantum gravity, 
Dijkgraaf-Vafa theory, ABJM theory and non-critical string theory. 
More recently, the relevance of Lefschetz thimbles is discussed 
in the context of semiclassical expansion in 
asymptotically free QFTs \cite{Dunne:2012ae,Basar:2013eka,Cherman:2014ofa,
Cherman:2014xia,Dorigoni:2014hea}. 
For an inexhaustive list of references on complex path integral 
and Lefschetz thimbles, see \cite{Guralnik:2007rx,Alexanian:2008kd, 
Denbleyker:2010sv,Nagao:2011za,Harlow:2011ny,Ferrante:2013hg,Nishimura:2014rxa,
Tanizaki:2014xba,Cherman:2014sba,Gorsky:2014lia,Sexty:2014dxa,Tanizaki:2014tua}. 

So far many of the works on Lefschetz-thimble approach to QFT seem to have been 
centered around bosonic theories 
\cite{Witten:2010cx,Cristoforetti:2012su,Cristoforetti:2013wha,Mukherjee:2013aga,
Ferrante:2013hg,Aarts:2013fpa,Fujii:2013sra,Cristoforetti:2014gsa,Cherman:2014xia}. 
On the other hand, in view of possible future applications 
of the Picard-Lefschetz theory to QCD and QCD-like theories, 
it would be important to understand the behavior of Lefschetz thimbles 
for path integrals with fermionic degrees of freedom. 
In \cite{Mukherjee:2014hsa,Aarts:2014nxa} Lefschetz thimbles in presence of 
a fermion determinant were studied numerically.  These works were specifically focused 
on the Hubbard model \cite{Mukherjee:2014hsa} and on lattice fermions with chemical 
potential \cite{Aarts:2014nxa}, respectively. It would be certainly worthwhile to extend 
these analyses to a more general setting.  

In this paper, we investigate the structure of Lefschetz thimbles in a variety of  
fermionic systems in zero and one dimension. We obtain complex critical points,  
determine associated Lefschetz thimbles and discuss their Stokes jumps, in a fermionic 
model with discrete chiral symmetry (with or without small mass term), in a fermionic model 
with continuous chiral symmetry (with or without small mass term), and in 
a Chern-Simons-like theory with fermions where a sign problem is caused by  
a topological term in the action. We expect that examples worked out here and 
lessons learned therefrom will be of value in future attempts to study  
the complex phase problem and spontaneous chiral symmetry breaking 
in QCD and QCD-like theories on the basis of Picard-Lefschetz theory. 

This paper is organized as follows. 
In Section \ref{sc:generic} a brief review of the Lefschetz-thimble approach to 
path integrals is given. We use simple toy integrals to illustrate that zeros and poles 
in the functional determinant in field theories do not obstruct the application of 
the Picard-Lefschetz framework. 
In Section \ref{sc:gross} we study a zero-dimensional Gross-Neveu-like model 
with discrete chiral symmetry. 
We investigate chiral symmetry breaking, Stokes lines, monodromy and Lee-Yang 
zeros in the complex coupling space from a viewpoint of Lefschetz thimbles. 
In Section \ref{sc:NJLm} a zero-dimensional Nambu-Jona-Lasinio-like model 
with continuous chiral symmetry is analyzed. We add a small mass term that breaks 
chiral symmetry explicitly and find that the approach to chiral limit 
is singular in terms of thimbles.  In Section \ref{sc:CSm} a one-dimensional theory 
with a topological term is considered. Positions of infinitely many critical points are 
determined and the dependence of associated Lefschetz thimbles 
on the coefficient of the topological term is elucidated. We conclude in 
Section \ref{sc:conclusion}.

\section{General remarks}
\label{sc:generic}

In this section we give a short summary of the Picard-Lefschetz approach to complex integrals 
and discuss applicability of the method in presence of zeros and poles of the integrand, which 
is commonly caused by functional determinant in QFTs. 
We aim to provide a minimal background for later sections of this paper and 
to fix our notations and terminology. A more in-depth review on this topic may be found in 
\cite{Witten:2010cx}, to which we refer the interested reader for further details.  

In physics we frequently confront the need to evaluate integrals of the form 
$\displaystyle \int_\mathcal{D}\!\! \dd x~\ee^{-\frac{1}{g}f(x)}$ where 
$g$ is a parameter ($\sim\hbar$ in quantum physics, $\sim k_{\rm B}T$ in statistical mechanics, 
$\sim 1/N$ in $N\times N$ matrix models and $N$-vector models, and so on).  
The domain $\mathcal{D}$ is a subset of $\RR^n$ for simplicity. 
When $f(x)$ is real and $g$ is small, an asymptotic estimate of the integral is 
available by means of a saddle-point approximation. The saddle points of $f$ 
in $\mathcal{D}$ dominate the integral, while those that lie 
outside of $\mathcal{D}$ do not play a role. In addition, for real $f$ one can apply 
Monte Carlo sampling techniques by interpreting the measure $\ee^{-\frac{1}{g}f(x)}$ as a 
probabilistic weight, which is useful especially when $n$ is large.    

Things change substantially once $f$ becomes a complex function. Although  
the integrand at small $g>0$ is suppressed for large $\Re\,f$ and 
enhanced for small $\Re\,f$, a naive saddle-point approximation based on saddles of $\Re\,f$ 
fails because $\Im\,f$ causes a rapid oscillation of the integrand which 
lessens contributions from the vicinity of these saddles. A viable asymptotic expansion 
should thus be done around a point where both $\Re\,f$ and $\Im\,f$ are stationary, but such point 
might not exist on $\mathcal{D}$. Moreover, the complex phase of the integrand makes 
a numerical evaluation of the integral quite challenging.  

As is widely known, for $n=1$ the correct way of handling this analytically is to 
promote $f(x)$ to a holomorphic function on $\CC$, identify saddle points of $f$ on $\CC$ and 
deform the original integration contour so that it passes through the saddle in the direction 
of stationary phase: $\Im\;f$ is constant on the deformed path. 
However, in the presence of multiple saddles 
it is far from trivial to see which one contributes to the integral and which does not; 
this is even more so when considering multidimensional integrals, i.e., $n>1$. 

A lucid way to organize asymptotic expansions for complex integrals is provided by 
the Picard-Lefschetz theory. For this to work we require that $f(z)$ is a holomorphic 
function on $\CC^n$ and that all critical points of $f$ are 
non-degenerate.%
\footnote{The presence of degenerate saddles that stem from 
a symmetry of the action does not necessarily invalidate 
the Picard-Lefschetz theory, 
but requires a special treatment as we discuss in Section \ref{sc:NJLnomass}.} 
The \emph{downward flow equation} is then defined as
\begin{align}
  \frac{\dd z^i(\tau)}{\dd \tau}=-\bar{\mkakko{\frac{\der f(z)}{\der z^i}}}\,, 
  \qquad \tau\in\RR\,. 
  \label{eq:floweq}
\end{align}
The essential property of \eqref{eq:floweq} is that $\Im\,f$ is constant but 
$\Re\,f$ is monotonically \emph{decreasing} 
along the downward flow. (This is a Morse flow with $\Re\,f$ the Morse function.) 
Every critical point of $f$ is evidently a fixed point of the flow. 
For each critical point $z_\sigma$\, ($\sigma=1,2,\dots$)\,, we can define a 
\emph{Lefschetz thimble} $\JJ(z_\sigma)$ 
as the union of all flows that end at $z_\sigma$ in the limit $\tau\to +\infty$. 
This $\JJ(z_\sigma)$ is a manifold of real dimension $n$ due to the fact that 
there are precisely $n$ directions around $z_\sigma$ in which $\Re\,f$ is increasing. 
(For $f(z)=z^2$, for instance, $\Re\,f(z)=\Re\,(x+iy)^2=x^2-y^2$ 
has one increasing direction and one decreasing direction. This generalizes  
to higher dimensions thanks to the holomorphy of $f$.) 
Since $\Re\,f$ is strictly decreasing along the flow, it must be 
that $\Re\,f(z)$ tends to $+\infty$ in the limit $\tau\to -\infty$.%
\footnote{An exception is when the flow meets another critical point. 
This is called the \emph{Stokes phenomenon} \cite{Berry1989} 
but does not occur for generic case and we momentarily ignore this.} 
This implies that $\ee^{-\frac{1}{g}f(z)}$ goes to zero at the ends of 
$\JJ(z_\sigma)$. In other words, $\JJ(z_\sigma)$ gives an element of the relative homology 
$H_n(\CC^n,(\CC^n)^T;\ZZ)$ for very large $T$, where $(\CC^n)^T:=\{z\in\CC^n\,|\,\Re\,f(z)\geq T\}$ 
is the ``good'' regions in which the integrand decreases rapidly.   
It then follows, that for any linear combination of Lefschetz thimbles $\sum_\sigma n_\sigma \JJ(z_\sigma)$ 
with integer coefficients $n_\sigma\in\ZZ$, the integral 
\begin{align}
  \int_{\sum_\sigma n_\sigma \JJ(z_\sigma)}\dd z~\ee^{-\frac{1}{g}f(z)} 
  = \sum_{\sigma} n_\sigma \int_{\JJ(z_\sigma)} \dd z~\ee^{-\frac{1}{g}f(z)} 
\end{align} 
is convergent and well-defined. 
Importantly, the converse is also true: $\{\JJ(z_\sigma)\}_\sigma$ actually constitutes 
a basis of $H_n(\CC^n,(\CC^n)^T;\ZZ)$ so that \emph{any cycle} of real dimension $n$  
without a boundary which is suitable as an integration cycle for 
$\displaystyle\ee^{-\frac{1}{g}f(z)}$ can be decomposed into a sum of Lefschetz thimbles.  
Therefore, as long as the original integration cycle $\mathcal{D}\subset \CC^n$ belongs to $H_n(\CC^n,(\CC^n)^T;\ZZ)$, 
one can always express an integral over it in the form
\begin{align}
  \int_{\mathcal{D}} \dd x~\ee^{-\frac{1}{g}f(x)} & = 
  \int_{\sum_\sigma n_\sigma \JJ(z_\sigma)}\dd z~\ee^{-\frac{1}{g}f(z)} 
  = 
  \sum_{\sigma} n_\sigma \ee^{-\frac{i}{g}\Im\,f(z_\sigma)} 
  \int_{\JJ(z_\sigma)} \dd z~\ee^{-\frac{1}{g}\Re\,f(z)}\,.  
  \label{eq:23}
\end{align}
Now the integral on the RHS has a real positive weight (albeit on a curved 
manifold $\JJ(z_\sigma)$) and is easily amenable to usual asymptotic analysis 
around saddles.  At the same time the complex phase problem is resolved if we could perform 
efficient Monte Carlo sampling on the Lefschetz thimbles with $n_\sigma\ne 0$. 

Of course, this method would be useless if we do not know how to determine 
$\{n_\sigma\}_\sigma$. Fortunately this is known at least for a finite-dimensional integral. 
Let us consider an \emph{upward flow equation}
\begin{align}
  \frac{\dd z^i(\tau)}{\dd \tau}= \bar{\mkakko{\frac{\der f(z)}{\der z^i}}}\,, 
  \qquad \tau\in\RR 
  \label{eq:flowup}
\end{align}
in which the sign on the RHS is flipped as compared to \eqref{eq:floweq}. 
As a result of this, along the upward flow, $\Im\,f(z)$ is conserved as in \eqref{eq:floweq} 
while $\Re\,f(z)$ is monotonically \emph{increasing}.  For each critical point $z_\sigma$ of $f$, 
we define $\KK(z_\sigma)$ as the union of all upward flows which end in 
$z_\sigma$ in the limit $\tau\to+\infty$.  Again this is a manifold of real dimension $n$. 
The cycles $\{\KK(z_\sigma)\}_\sigma$ may be seen as a dual of the Lefschetz thimbles 
$\{\JJ(z_\sigma)\}_\sigma$ in a homological sense and 
there is a natural intersection pairing between $\JJ$ and $\KK$ \cite{Witten:2010cx}. Actually 
$\{\KK(z_\sigma)\}_\sigma$ constitutes a basis of another relative homology 
$H_n(\CC^n,(\CC^n)_{-T};\ZZ)$ for large $T$, 
where $(\CC^n)_{-T}:=\{z\in\CC^n\,|\,\Re\,f(z)\leq -T\}$ is the ``bad'' regions.   
Of course, one cannot use $\KK(z_\sigma)$ 
as an integration cycle for $\ee^{-\frac{1}{g}\Re\,f(z)}$\,, because 
$\Re\,f\to -\infty$ at the ends of $\KK(z_\sigma)$ (i.e., for $\tau\to -\infty$). 
Rather, the importance of $\KK(z_\sigma)$ lies in the fact that $n_\sigma$ 
appearing in \eqref{eq:23} can be computed as the number of (oriented) intersections 
between $\mathcal{D}$ and $\KK(z_\sigma)$.%
\footnote{In $2n$ dimensions ($\CC^n$), two general hypersurfaces with dimension $n$ 
intersect at isolated points. Consider two lines on a plane, for example. } 
While the calculation of $\{n_\sigma\}$ is tractable for low-dimensional toy models, 
it becomes a formidable task in the case of infinite-dimensional QFTs. In recent numerical 
Monte Carlo approaches to Lefschetz thimbles, only a single thimble associated with 
the perturbative vacuum was taken into account on the basis of universality \cite{Cristoforetti:2012su,Cristoforetti:2013wha,Fujii:2013sra,
Cristoforetti:2014gsa,Mukherjee:2014hsa}. 
Although numerical results so far look quite promising, 
the validity of this approach still remains to be clarified.    

In general applications of the Lefschetz thimble technique, we often face integrals 
that are not of a pure exponential form, 
but rather $\displaystyle \int_\mathcal{D}\!\! \dd x~h(x)\ee^{-\frac{1}{g}f(x)}$ 
with another function $h(x)$. 
When $h(x)$ is just an observable in field theories, it does not affect 
the saddle point analysis of the integral  
because $f(x)$ grows extensively in the thermodynamic limit while $h$ is 
an $\calO(1)$ quantity. In contrast, if $h$ grows in the thermodynamic limit 
this has to be taken into account in the saddle point analysis: examples of such $h$ are 
the quark determinant in QCD and the Vandermonde determinant 
in matrix integrals. In both cases $h$ has zeros and a simple rewriting 
$h\ee^{-\frac{1}{g}f} \to \ee^{\log h-\frac{1}{g}f}$ looks subtle. 
Moreover $h$ may also have poles. In the remainder of this section, 
we discuss how the Lefschetz-thimble approach works in these cases.  

Let us first consider the case when $h$ is holomorphic on $\CC^n$. 
Then $\log h$ is defined except at zeros of $h$.%
\footnote{Note that the set $\{z\in\CC^n \,|\,h(z)=0\}$ is typically 
of real dimension $2n-2$. The zeros of 
$h$ thus form a hypersurface in higher dimensions.} 
Although $\log h$ is ambiguous up to integer multiples of $2\pi i$, 
the gradient flow equations \eqref{eq:floweq} and \eqref{eq:flowup} are 
well-defined because the above ambiguity disappears by taking derivatives of $\log h$, and 
the thimbles can be defined in the same way as before, provided that all 
the critical points of $\log h-\frac{1}{g}f$ are non-degenerate. 
What is crucial here is that the ``good'' regions must be modified, 
by incorporating the vicinity of zeros of $h$. This can be explained most easily on examples.  

As a trivial case, consider $\int_{\RR}\dd x~x \ee^{-x^2/2}$. Of course this integral is identically 
zero, but this is irrelevant for the purpose of illustration. Casting this into the form 
$\int_{\RR}\dd x~\ee^{-f(x)}$ with $f(x)=-\log x + x^2/2$, we see that the critical points  
are located at $x=\pm 1$. The Lefschetz thimbles $\{\JJ(1)$, $\JJ(-1)\}$ and the upward flow lines 
$\{\KK(1), \KK(-1)\}$ can then be defined. 
%%%%%%%%%%%%%%%%%%%%%%%%%%%%%%%%%%%%%%%
   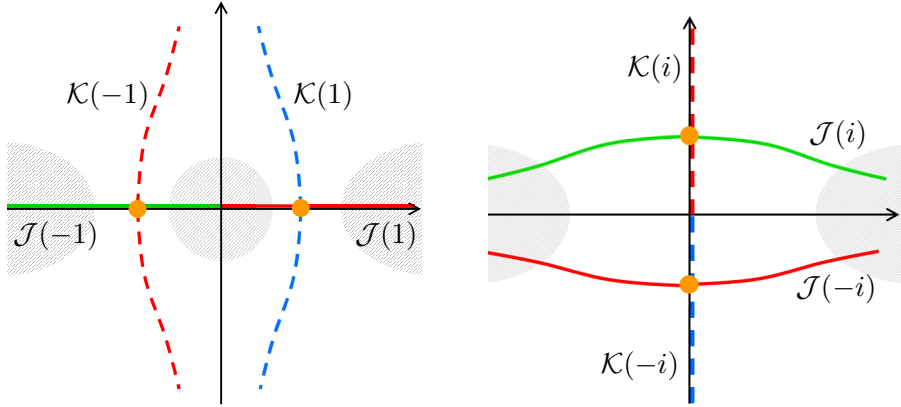
\begin{figure}[!t]
   \begin{center}
     \includegraphics[width=0.36\columnwidth]{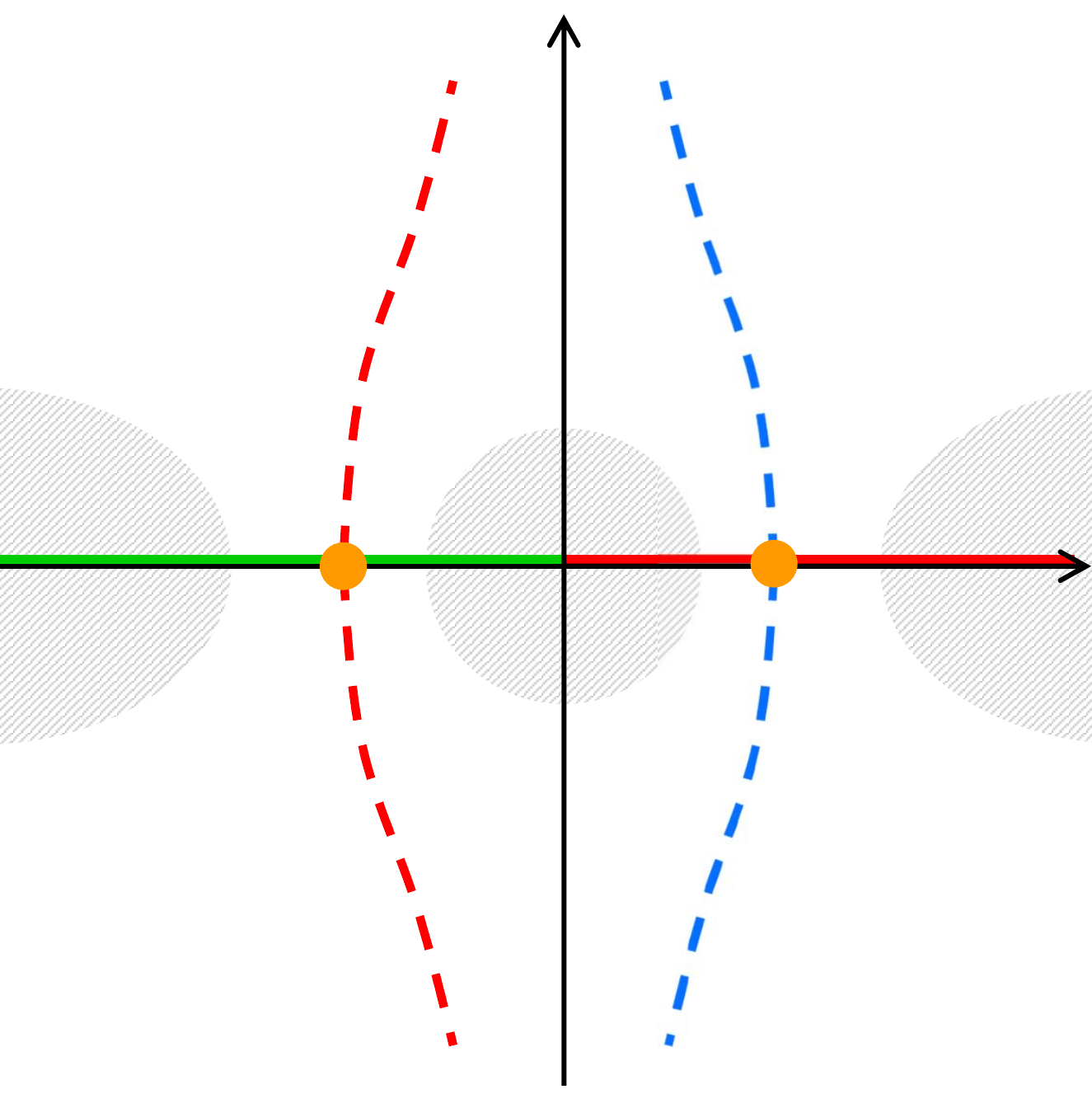}
     \qquad
     \includegraphics[width=0.36\columnwidth]{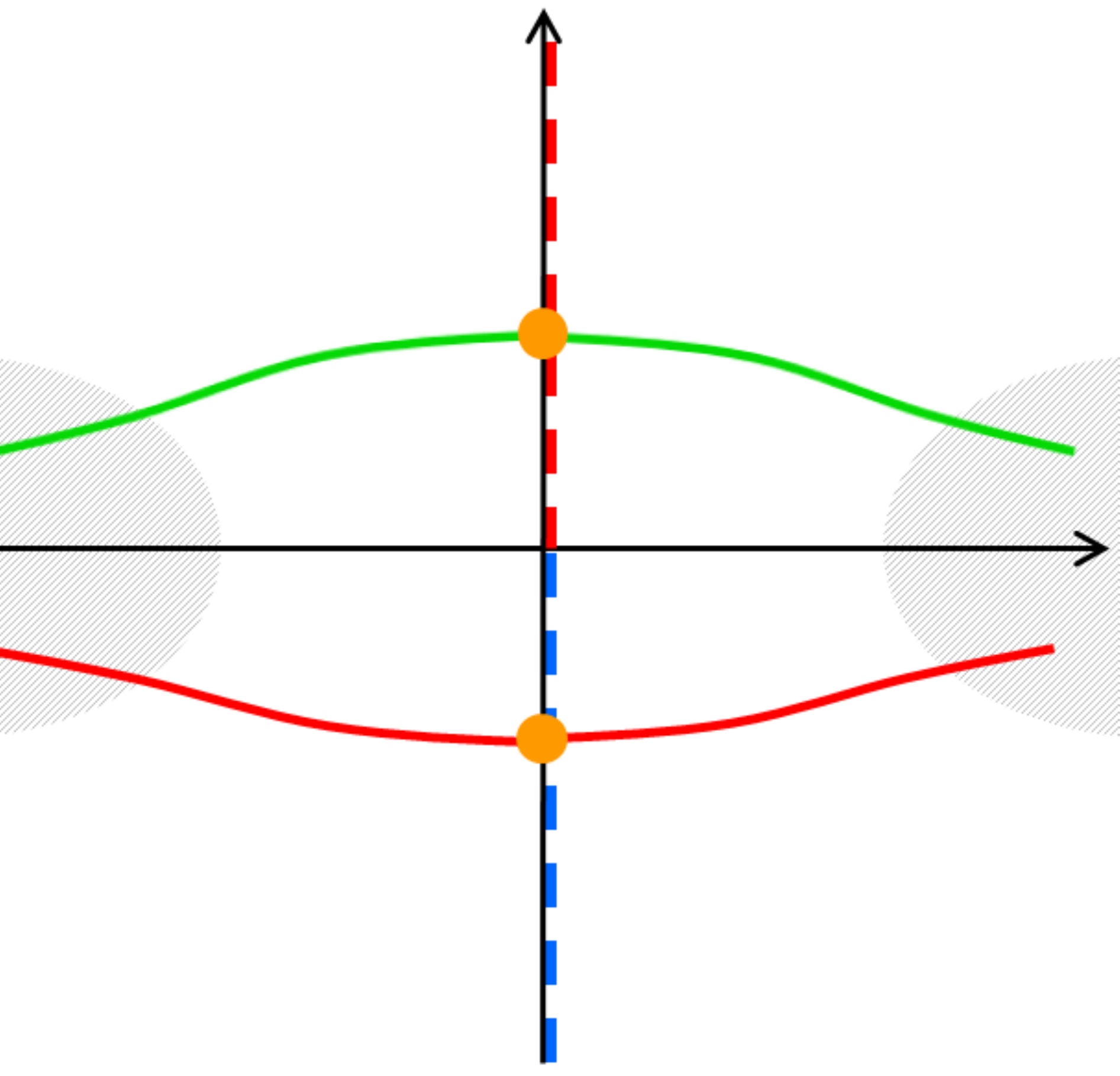}
     %%%
     \put(-205,62){$\JJ(1)$}
     \put(-333,62){$\JJ(-1)$}
     \put(-228,115){$\KK(1)$}
     \put(-313,115){$\KK(-1)$}
     %%%
     \put(-40,42){$\JJ(-i)$}
     \put(-35,100){$\JJ(i)$}
     \put(-103,125){$\KK(i)$}
     \put(-113,13){$\KK(-i)$}
   \end{center} 
   \vspace{-1.2\baselineskip}
   \caption{
     \label{fg:toy}
     Lefschetz thimbles $\JJ$ (solid lines) and their duals $\KK$ (dashed lines) 
     for simple one-dimensional integrals. 
     Orange blobs are critical points of the action. Hatched areas are ``good'' regions 
     where the integrand tends to zero. In both panels, the origin is a singularity of the flow. 
   } 
   \end{figure}
%%%%%%%%%%%%%%%%%%%%%%%%%%%%%%%%%%%%%%%
They are schematically shown in Figure \ref{fg:toy} (left panel). 
The three hatched areas are ``good'' regions, which includes 
the vicinity of $z=0$. The Lefschetz thimbles $\JJ(\pm 1)$ run from one good region to another, 
thus providing a basis of the relative homology 
$H_1(\CC,\CC^T;\ZZ)$ for very large $T$ in accordance with the general argument. By contrast, 
$\KK(\pm 1)$ run from one ``bad'' region $z\sim -i\infty$ to another $z\sim i\infty$, 
thus providing a basis of $H_1(\CC,\CC_{-T};\ZZ)$ for very large $T$.  Since $\KK(\pm 1)$ 
intersect with the real axis, the integral receives contributions from both $\JJ(1)$ and $\JJ(-1)$. 
This is also intuitively obvious, for the original contour $\RR$ is a union of $\JJ(1)$ and $\JJ(-1)$. 
As a whole, the general framework of Lefschetz-thimble approach does not seem to be obstructed 
by the presence of zeros of the integrand. 

There is one side remark here. We noted above that $\Im\,f$ is conserved along a flow. 
While this is generally true, it does \emph{not} imply that $\Im\,f$ is constant over an entire 
upward/downward flow cycle.  For illustration, let us note that $\KK(-1)$ is comprised of 
two distinct flow lines: one is stretching from $-1$ to $+i\infty$ and the other from $-1$ 
to $-i\infty$.  It is easily seen that $\Im\,f$ is $-\pi$ along the former and $\pi$ along 
the latter, owing to the fact that $z=-1$ sits right on the branch cut of logarithm.  
Thus we conclude that although $\Im\,f$ is locally conserved along a flow, it can jump 
by a multiple of $2\pi$ at a point where two flows meet. (See also \cite{Aarts:2014nxa}).    

Next, let us turn to the case when $h(z)$ is a meromorphic function with poles. An example of this is 
given by a bosonic functional determinant in QFTs.  
It is useful to once again employ a simple example to illustrate the general applicability of Lefschetz 
thimbles. Consider an integral 
$\displaystyle\int_{\RR+i\epsilon}\!\!\! \dd z~\frac{1}{z} \ee^{-z^2/2}$ where 
the contour is slightly uplifted from the real axis to avoid the pole at $z=0$. Now, writing this as 
$\int_{\RR+i\epsilon}\dd z~\ee^{-f(z)}$ with $f(z)=\log z + z^2/2$, we see that the critical points  
are located at $z=\pm i$. The Lefschetz thimbles $\{\JJ(i)$, $\JJ(-i)\}$ and the upward flow lines 
$\{\KK(i), \KK(-i)\}$ can then be defined with respect to the flow equations for $f(z)$. 
They are shown in Figure \ref{fg:toy} (right panel). 
Compared to the previous example, the geometrical structure of $\JJ$ and $\KK$ are exchanged. 
Interestingly, now, $\KK(i)$ and $\KK(-i)$ end at the origin because the area around $z=0$ 
was turned into a ``bad'' region by a pole. 
One can easily confirm that $\JJ$ and $\KK$ again constitute the bases of relative homology. 
(Note that $\JJ(i)$ and $\JJ(-i)$ are independent cycles, for they cannot be continuously moved 
to each other across the singularity at $z=0$.) Since $\KK(i)$ intersects with $\RR+i\epsilon$ 
while $\KK(-i)$ does not, the integral only receives contribution from $\JJ(i)$ and not from $\JJ(-i)$. 
In summary, the presence of a pole in the integrand does not undermine 
the applicability of the Picard-Lefschetz theory.  

We note that our discussion so far is concerned about mathematical aspects of the formalism and 
does not immediately suggest feasibility of numerical algorithms for Lefschetz thimbles. 
Actually the fermion determinant in QCD at finite density has dense zeros in the gauge configuration 
space and this could lead to a practical difficulty for the Lefschetz-thimble approach 
\cite{Cristoforetti:2012su}.

\section{Gross-Neveu-like model}
\label{sc:gross}

As a toy model for discrete chiral symmetry breaking, 
let us consider a zero-dimensional fermionic model 
similar to the Gross-Neveu (GN) model \cite{Gross:1974jv}.  
The partition function reads
\begin{align}
  Z_N(G,m) = \int \dd \bar\psi \dd\psi~\exp\mkakko{ 
  \sum_{a=1}^{N}\bar\psi_a (i\slashed{p}+m) \psi_a 
  + \frac{G}{4N}\Big(\sum_{a=1}^{N}\bar\psi_a \psi_a \Big)^2 }\,,
  \label{eq:Zgrossneveu}
\end{align}
where $\psi_a$ and $\bar{\psi}_a$ are $2$-component Grassmann variables 
with $N$ colors, $G>0$ is a coupling constant,  
$m$ is a bare mass, and $\slashed{p}\equiv \sum_{i=1}^{2}p_i \gamma_i$ is 
a constant $2\times 2$ matrix that mimics the effect of nonzero-momentum modes 
in higher dimensions. 
(As an explicit basis we use $\gamma_1=\sigma_1$, $\gamma_2=\sigma_2$ and 
$\gamma_5=\sigma_3$ in the following.)   
At $m=0$, the action in \eqref{eq:Zgrossneveu} is invariant under a $\ZZ_2$ chiral transformation 
$\psi\to\gamma_5\psi$. With a Hubbard-Stratonovich transformation,
\begin{align}
  Z_N(G,m) & = \sqrt{\frac{N}{\pi G}} \int \dd\bar\psi \dd\psi \dd\sigma~
  \exp\mkakko{ \sum_{a=1}^{N}
    \bar\psi_a (i\slashed{p}+m+\sigma) \psi_a - \frac{N}{G}\sigma^2 
  }
  \\
  & = \sqrt{\frac{N}{\pi G}}
  \int_\RR \dd\sigma~ {\det}^N(i\slashed{p}+m+\sigma) 
  \exp\mkakko{- \frac{N}{G}\sigma^2} 
  \equiv 
  \sqrt{\frac{N}{\pi G}}\int_\RR \dd\sigma\,\ee^{-NS (\sigma)}  \,, 
  \label{eq:ZGN}
\end{align}
where
\begin{align}
  S (\sigma) & \equiv \frac{\sigma^2}{G}-\log[p^2+(\sigma+m)^2] 
  \qquad \text{with}\quad p^2\equiv p_1^2+p_2^2 >0 \,.
\end{align}
The derivation of \eqref{eq:ZGN} is not only valid for $G>0$ but also 
for complex $G$ as long as $\Re\,G>0$. 
Now, let us discuss the cases with $m=0$ and $m>0$ separately.

\subsection{Massless case}
\label{sc:GNmassless}

\subsubsection{Lefschetz thimbles}

For $m=0$, the minimum of the action 
$S(\sigma)=\sigma^2/G-\log(p^2+\sigma^2)$ takes place at $\sigma=0$ 
for $0<G\leq p^2$ and at $\sigma\ne 0$ for $G>p^2$. A continuous ``chiral transition'' occurs 
at $G=p^2$.

In order to analyze this transition from the viewpoint of Lefschetz thimbles, we lift $\sigma\in\RR$ 
to a complex variable $z\in\CC\setminus\{\pm ip\}$. 
The critical points of the action are obtained as
\begin{align}
  0=\frac{\der S(z)}{\der z} = \frac{2z}{G}-\frac{2z}{p^2+z^2} ~~\Longrightarrow~~
  z=0,\ \pm\sqrt{G-p^2}\,.
\end{align}
Let us denote $z_\pm\equiv \pm\sqrt{G-p^2}$. Since the three critical points coalesce 
at $G=p^2$ and it makes the Lefschetz thimbles ill-defined, 
we will hereafter assume $G\ne p^2$. 

It can be easily checked that for $G>0$  
\begin{align}
  S(0)-S(z_\pm) & = -1 + \frac{p^2}{G}-\log \frac{p^2}{G} \geq 0\,, 
  \label{eq:Sinequality}
\end{align}
where the equality holds if and only if $G=p^2$. Thus the nontrivial critical points $z_\pm$ 
\emph{always} have a lower action than $z=0$ for $G>0$, regardless of $G\gtrless p^2$. 

An important fact is that $\Im\,S(z)=0$ at all these critical points. This means that 
we are right on the Stokes ray: there are flow lines that connect distinct 
critical points \cite{Berry1989,Witten:2010cx}. 
To avoid the Stokes ray and make Lefschetz thimbles well-defined, 
we endow $G$ with a phase factor 
$\ee^{i\theta}$ with $0<|\theta|\ll 1$ so that the degeneracy of $\Im\,S(z)$ 
among the critical points is lifted.%
\footnote{It is not appropriate to rotate the entire action as $S(z)\to \ee^{i\theta}S(z)$ 
because it allows the $2\pi$-ambiguity in the imaginary part of $S(z)$ to induce 
an ambiguity in the \emph{real part} of  $\ee^{i\theta}S(z)$, which renders the integral 
ill-defined. } A more comprehensive analysis of Stokes phenomena for complex $G$ 
will be presented in Section \ref{sc:monodromy}. 

%%%%%%%%%%%%%%%%%%%%%%%%%%%%%%%%%%%%%%%
   \begin{figure}[!t]
   \begin{center}
     \includegraphics[width=0.4\columnwidth]{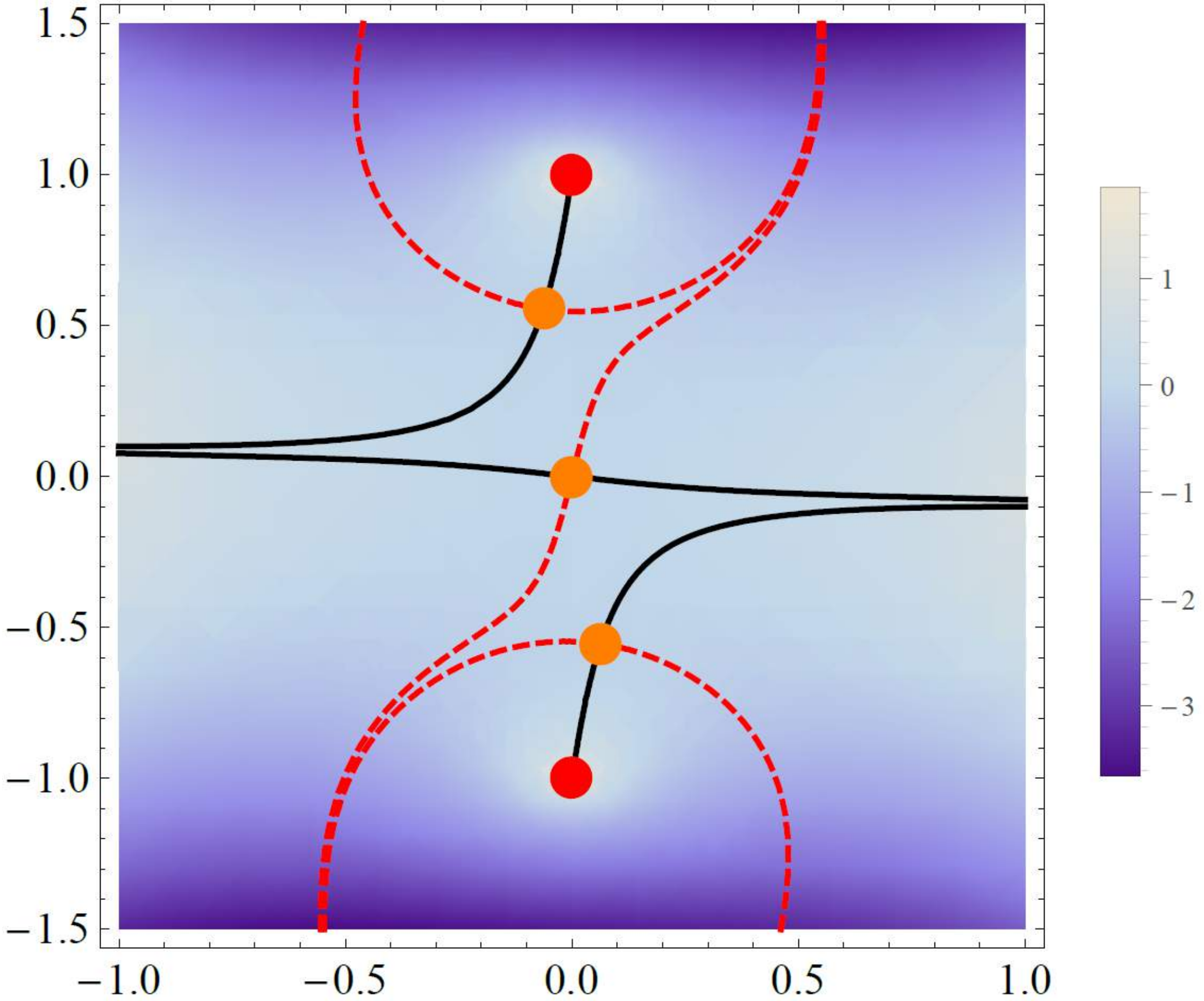}
     \quad
     \includegraphics[width=0.4\columnwidth]{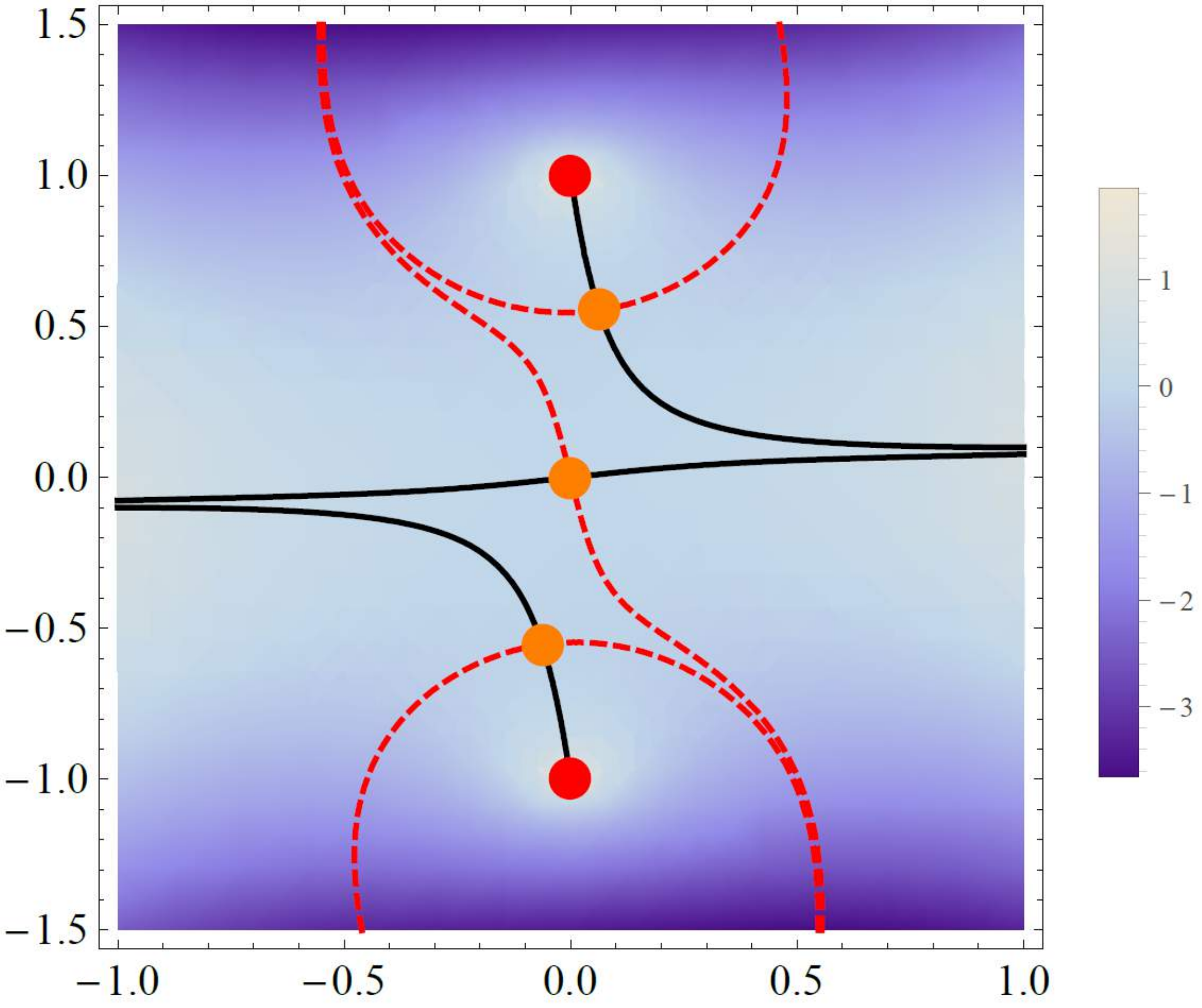}
     \put(-158,151){$G=0.7\ee^{+0.1i},~p=1,~m=0$}
     \put(-345,150){$G=0.7\ee^{-0.1i},~p=1,~m=0$}
   \end{center} 
   \vspace{-.8\baselineskip}
   \caption{
     \label{fg:GN1}
     Lefschetz thimbles (black lines) and upward flow lines (red dashed lines) 
     on the complex $z$-plane for the GN-like model in the chiral limit. 
     The three orange blobs at $z=0$ and $z_\pm=\pm\sqrt{G-1}\approx \pm 0.55i$ 
     are the critical points of $S(z)$, while the  
     two red blobs at $z=\pm i$ are the points where $S(z)$ diverges. 
     The background color scale describes $\Re\,S(z)$. 
   } 
   \end{figure}
%%%%%%%%%%%%%%%%%%%%%%%%%%%%%%%%%%%%%%%

In Figure \ref{fg:GN1} we show downward flow lines (i.e., Lefschetz thimbles) and 
their duals on the complex $z$ plane 
for $p=1$ and $G=0.7\ee^{i\theta}$ with $\theta=\pm 0.1$. 
(Since $G<p^2$, the system is in a chirally symmetric phase.) 
The displayed flow lines were obtained by first 
plotting the contours fulfilling $\Im\,S(z)=\Im\,S(0)$ and 
$\Im\,S(z)=\Im\,S(\pm \sqrt{G-p^2})$, and then discarding components that are 
not connected to the critical points. 

There are several remarks concerning Figure \ref{fg:GN1}.  
First of all, there are three Lefschetz thimbles  
$\JJ(0)$, $\JJ(z_+)$ and $\JJ(z_-)$ and three upward flow lines 
$\KK(0)$, $\KK(z_+)$ and $\KK(z_-)$,  passing through each critical point.  
Among the three $\KK$'s, only $\KK(0)$ intersects with the original integration 
cycle $\RR$, which means that $\JJ(0)$ is the only Lefschetz thimble contributing to $Z_N(G,0)$.%
\footnote{This is consistent with the general fact that Lefschetz thimbles associated with 
critical points that lie on the initial integration cycle contribute with a unit coefficient 
\cite{Witten:2010cx}.} 
This is so even though $S(z_\pm)$ is lower than $S(0)$. 

Secondly, as we can see, $\JJ(z_\pm)$ and $\KK(0)$ jump as $\theta$ crosses zero. 
This is a phenomenon called the Stokes jump. Since $\JJ(0)$ does not jump 
at $\theta=0$, the partition function $Z_N(G,0)$ itself is analytic at $\theta=0$.  

Finally, it is worth an attention that the Lefschetz thimbles indeed provide a homological 
basis of convergent integration cycles 
even in the presence of logarithmic singularities, 
in agreement with our argument in Section \ref{sc:generic}. Here what we call 
convergent integration cycles are contours that start from and end in ``good'' regions 
(bright regions in Figure \ref{fg:GN1}) 
which are the regions where $\Re\,S(z)$ grows to $+\infty$.   
In the present case, there are four ``good'' regions: 
(i) the vicinity of $z=z_+$, (ii) the vicinity of $z=z_-$, 
(iii) $\{z\in\CC\,|\, \Re\,z\gg 1,~\Im\,z\sim\calO(1)\}$, and  
(iv) $\{z\in\CC\,|\, \Re\,z\ll -1,~\Im\,z\sim\calO(1)\}$.  
As seen from Figure \ref{fg:GN1}, the three $\JJ$'s form a basis in the space of 
cycles that start from and end in those four ``good'' regions. This is the premise 
of the Lefschetz-thimble approach to complex integrals. 

%%%%%%%%%%%%%%%%%%%%%%%%%%%%%%%%%%%%%%%
   \begin{figure}[!t]
   \begin{center}
     \includegraphics[width=0.4\columnwidth]{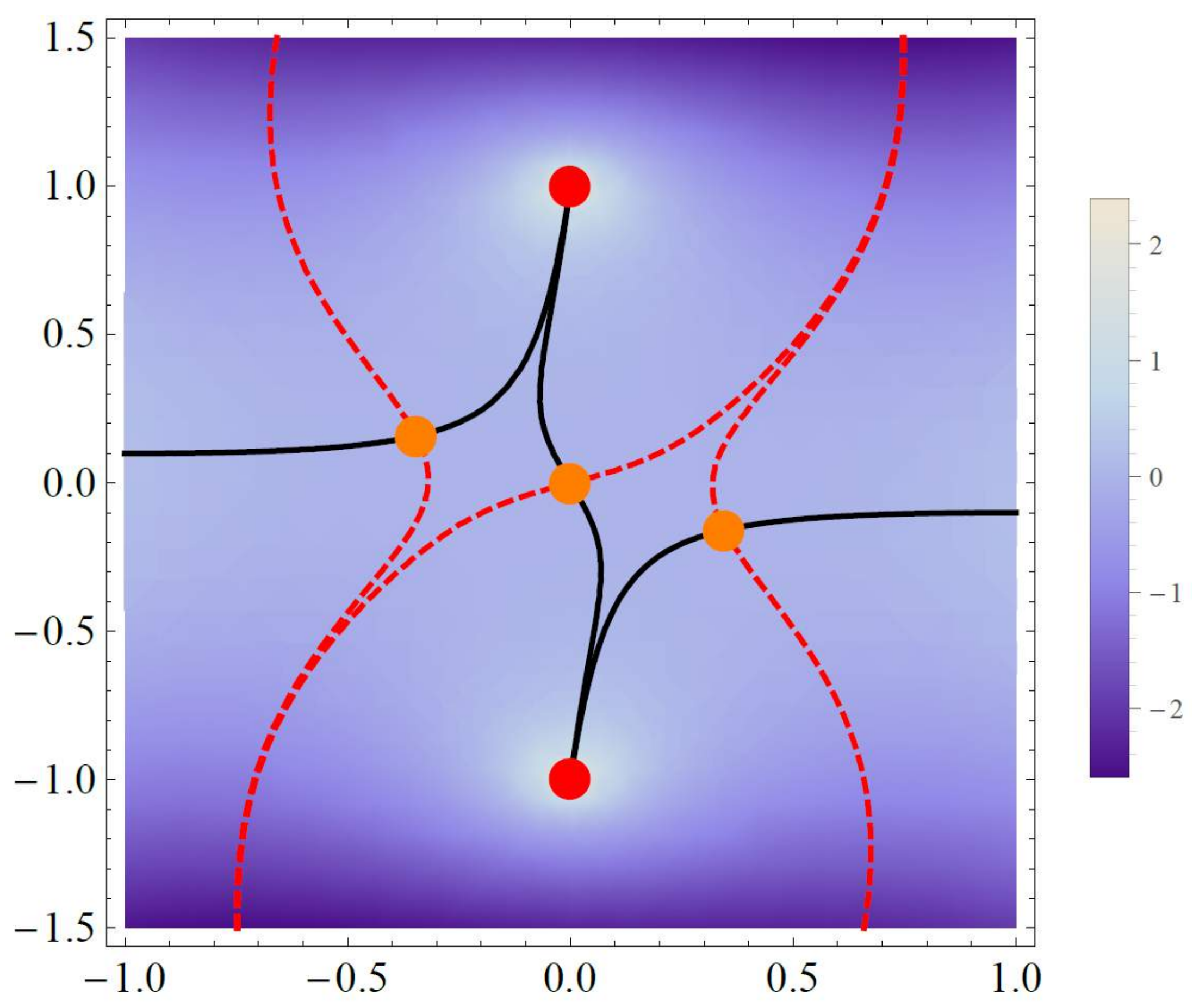}
     \quad
     \includegraphics[width=0.4\columnwidth]{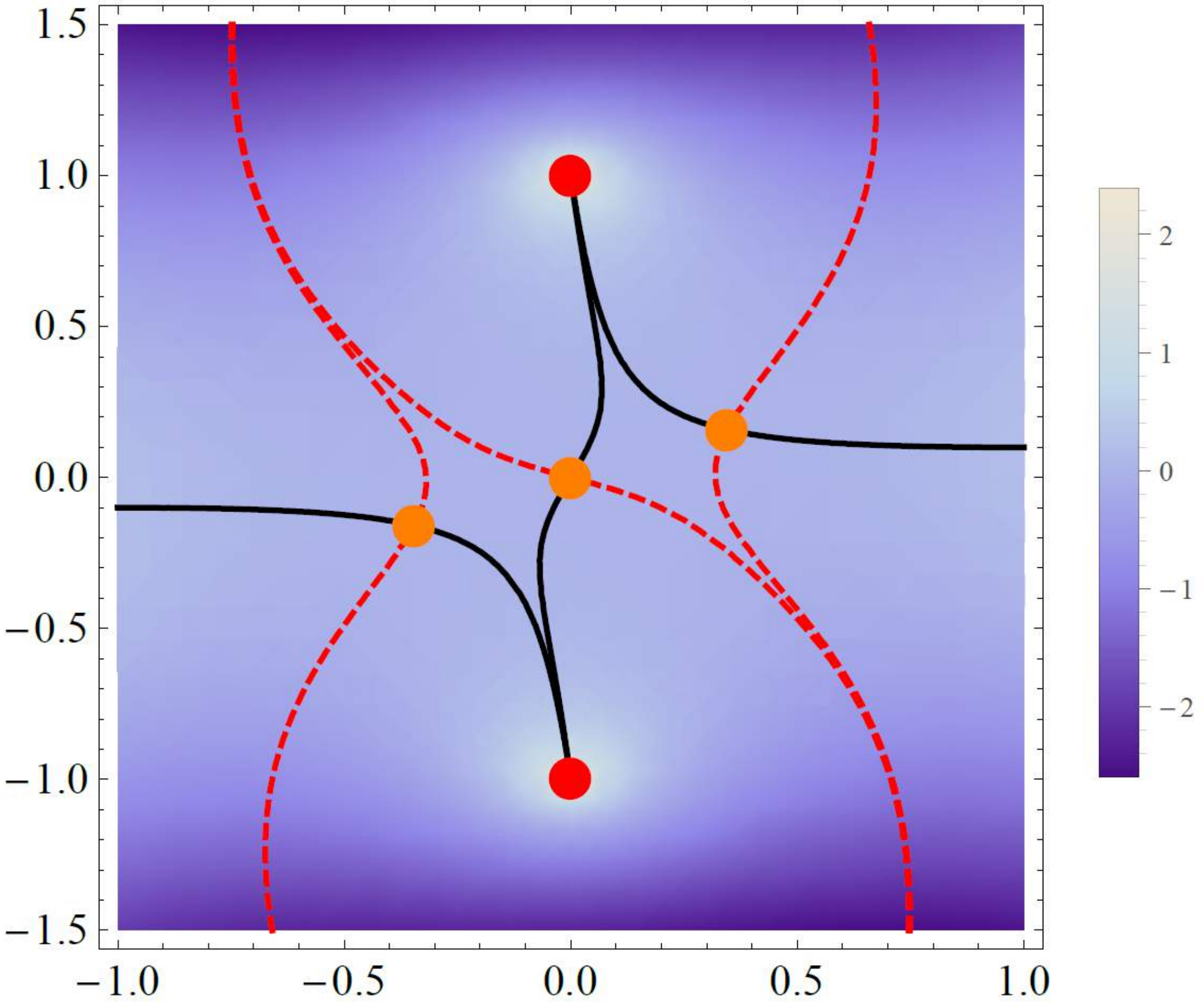}
     \put(-158,150){$G=1.1\ee^{+0.1i},~p=1,~m=0$}
     \put(-344,150){$G=1.1\ee^{-0.1i},~p=1,~m=0$}
   \end{center} 
   \vspace{-.8\baselineskip}
   \caption{
     \label{fg:GN2}
     Same as Figure \ref{fg:GN1} but with $|G|=1.1$. 
   } 
   \end{figure}
%%%%%%%%%%%%%%%%%%%%%%%%%%%%%%%%%%%%%%%

Next, we consider a chirally broken phase with $G=1.1>p^2=1$. 
To avoid the Stokes ray, we again attach a phase factor to $G$. 
The resulting Lefschetz thimbles are shown in Figure \ref{fg:GN2}. 
Notably, $\JJ(0)$ has rotated almost 90 degrees in comparison to Figure \ref{fg:GN1}, 
and now it connects the two singular points at $z=\pm i$. 

One can observe in Figure \ref{fg:GN2} that 
all the three $\KK$'s (red lines) intersect with the real axis, 
implying that $Z_N(G,0)$ now receives contributions from 
all of $\JJ(0)$, $\JJ(z_+)$ and $\JJ(z_-)$.  Indeed it is visually clear that the union of 
the three thimbles is homologically equivalent to $\RR$. 
Since $S(z_\pm)$ is lower than $S(0)$ (recall \ref{eq:Sinequality}), the nontrivial saddles 
will completely dominate the behavior of $Z_N(G,0)$ in the large-$N$ limit, and fermions 
acquire a dynamical mass that breaks discrete chiral symmetry. The second-order  
chiral transition in this model along the $G>0$ line thus occurs through a jump in the 
number of contributing thimbles at $G=p^2$.  
By contrast, we will see in Section \ref{sc:monodromy} that the chiral transition 
for $G\in\CC$ is generically first order and exhibits qualitatively new features.

We shall analyze the Stokes jump in Figure \ref{fg:GN2} in some details.  
Let us fix the orientation of $\JJ(z_\pm)$ as the direction of increasing $\Re\,z$ 
and of $\JJ(0)$ as the direction from $z=-ip$ to $z=+ip$. Then the homological 
jump of thimbles as $\theta$ is dialed from $0^-$ to $0^+$ can be summarized as
\begin{align}
  \begin{pmatrix}
    \JJ(z_+)\\\JJ(z_-)\\\JJ(0)
  \end{pmatrix} \to  
  \begin{pmatrix}
    ~1~&0&1 \\ 0&1&1 \\ 0&0&~1~
  \end{pmatrix}
  \begin{pmatrix}
    \JJ(z_+)\\\JJ(z_-)\\\JJ(0)
  \end{pmatrix}\,. 
  \label{eq:JJJ}
\end{align}
The meaning of this is that $\JJ(z_+)$ in the left panel of Figure \ref{fg:GN2} 
is equal to $\JJ(z_+)+\JJ(0)$ in the right panel of Figure \ref{fg:GN2}, and so on. 
This actually implies that the real cycle $\RR$ can be expressed in two 
different ways, according to how we approach the $\theta\to 0$ limit of $G\ee^{i\theta}$: 
\begin{align}
  \RR = \left\{\begin{array}{ll}
    \JJ(z_+) - \JJ(0) + \JJ(z_-) & \quad \text{for}~\theta= 0^- \,,
    \\ 
    \JJ(z_+) + \JJ(0) + \JJ(z_-) & \quad \text{for}~\theta= 0^+ \,.
  \end{array}\right. 
  \label{eq:RJJJ}
\end{align}
There are two remarks on \eqref{eq:JJJ} and \eqref{eq:RJJJ}. 
\begin{itemize}
  \item 
  The fact that $\JJ(0)$ does not jump across the Stokes ray 
  (as one can see in Figures \ref{fg:GN1} and \ref{fg:GN2}) has an intuitive explanation. 
  From the definition of a downward flow and \eqref{eq:Sinequality}, we have 
  $\Re\,S(z)\geq \Re\,S(0)>\Re\,S(z_\pm)$ for $^\forall z\in\JJ(0)$. This implies  
  that the flow along $\JJ(0)$ has no chance to touch $z_\pm$, so $\JJ(0)$ is 
  insensitive to the presence of $\JJ(z_\pm)$ and shows no Stokes jump at all.       
  By the same token, one can explain why $\KK(z_\pm)$ do not jump. 
  \item 
  The fact that the coefficients of $\JJ(z_\pm)$ in \eqref{eq:RJJJ} do not jump 
  across $\theta=0$ can be deduced in a simple way. 
  For $G>p^2$, the asymptotic behavior of $Z_N(G,0)$ at $N\gg 1$ is 
  dominated by the saddles $z=z_\pm$ since $S(z_\pm)<S(0)$.  
  Recalling that $Z_N(G,0)$ is a holomorphic function of $G$ at any finite $N$, it follows that 
  the contributions from $\JJ(z_\pm)$ cannot jump discontinuously. By contrast, 
  such an argument does not constrain the exponentially smaller contribution 
  from $\JJ(0)$, and indeed the coefficient of $\JJ(0)$ does jump in \eqref{eq:RJJJ}. 
  In short, the jump of coefficients can only occur for thimbles associated with subleading 
  saddle points. 
\end{itemize}
Both arguments have been presented by Witten \cite[Sect.~3]{Witten:2010cx} 
in the context of bosonic integrals, and here we have highlighted their usefulness 
in a fermionic model.

\subsubsection{Stokes lines and monodromy}
\label{sc:monodromy}

So far we have described the Stokes phenomenon in the massless GN-like model 
for $G>0$. It has been shown that the thimbles can be made well-defined if $G$ 
is given a small complex phase $\ee^{i\theta}$ with $0<|\theta|\ll 1$.  
In this subsection, we study the Stokes phenomenon and some technical issues 
for a generic coupling $G\in\CC$.%
\footnote{A complex four-fermion coupling appears in studies of the $\theta$ vacuum in QCD 
\cite{Boer:2008ct,Boomsma:2009eh}. } 

\begin{figure}[!t]
   \begin{center}
     \includegraphics[width=0.4\columnwidth]{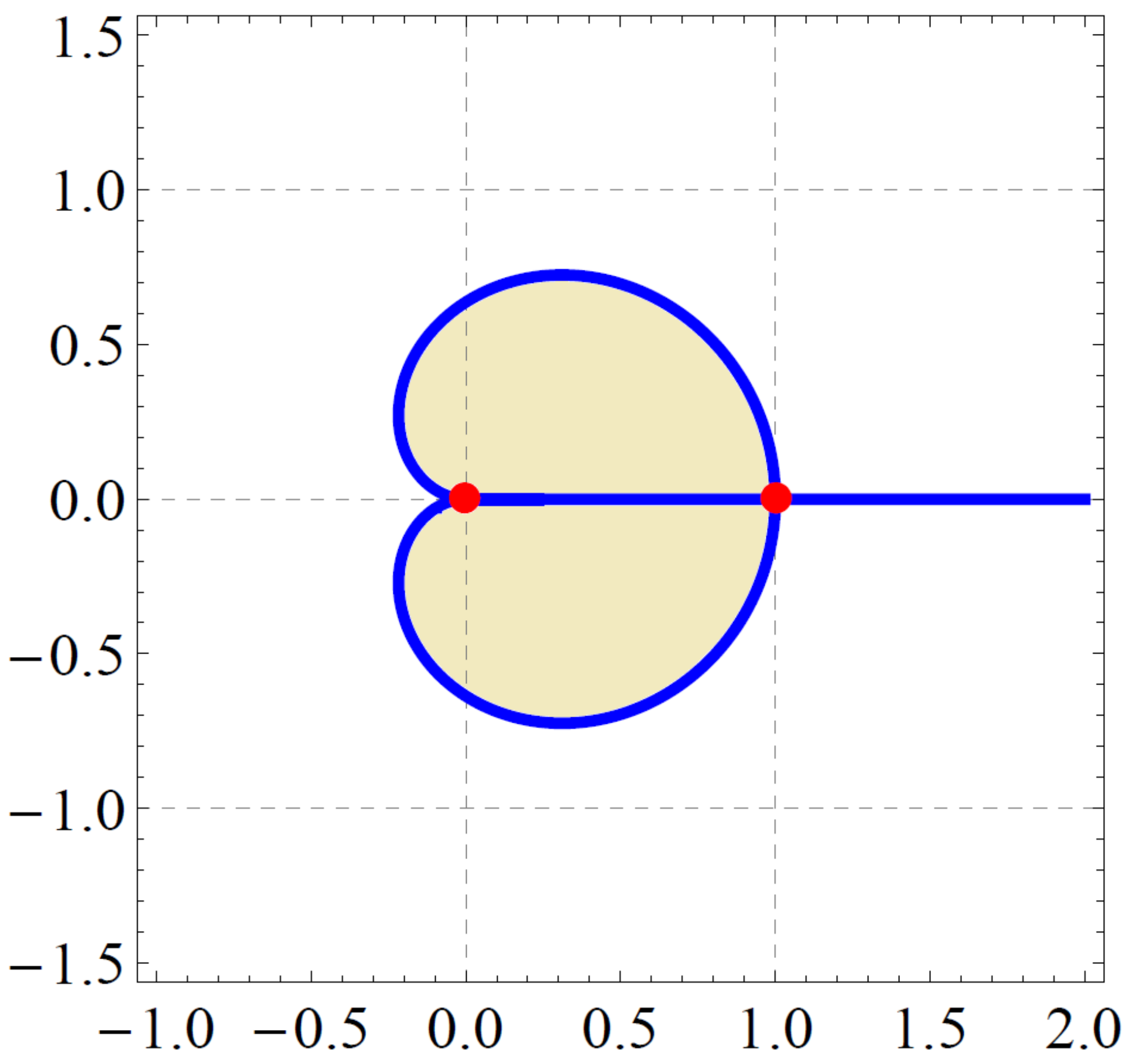}
     \put(-195,82){$\Im\,G$}
     \put(-90,-12){$\Re\,G$}
   \end{center} 
   \vspace{-.8\baselineskip}
   \caption{
     \label{fg:stokeslines}
     Stokes lines for the GN-like model with $p=1$ and $m=0$ (blue lines). 
     The global topology of Lefschetz thimbles and their duals changes  
     across the Stokes lines: $\JJ(z_\pm)$ and $\KK(0)$ jump across the horizontal line, while 
     $\JJ(0)$ and $\KK(z_\pm)$ jump across the round curve.  
     In the shaded area, only $\JJ(0)$ contributes to $Z_N(G,0)$. 
     Outside the shaded area, all thimbles contribute.   
     \newline 
     (The points $G=0$ and $1$ are excluded because the action $S(z)$ is singular at $G=0$ and 
     the critical points merge at $G=1$.)
   } 
   \end{figure}

The necessary condition for a Stokes jump to occur is that the imaginary part 
of the action is degenerate for multiple critical points. 
In the present model with $m=0$, this condition reads 
\begin{align}
  0\overset{!}{=}\Im\big[S(0)-S(z_\pm)\big] 
  & = \Im\Big[ \! -1 + \frac{p^2}{G} - \log \frac{p^2}{G} \Big] \,.  
\end{align}
This can be solved by $G\in\RR_{>0}$ and by 
$G\in\{\,p^2 r \ee^{i\phi}\,|\,r=\frac{\sin\phi}{\phi} ~\text{and}~-\pi<\phi\leq \pi\}$. 
The union of these sets is displayed in Figure \ref{fg:stokeslines} as a blue line. 
%%%%%%%%%%%%%%%%%%%%%%%%%%%%%%%%%%%%%%%
   \noindent
%%%%%%%%%%%%%%%%%%%%%%%%%%%%%%%%%%%%%%%
When these lines are crossed, some of the flow lines jump discontinuously. 
Which line jumps and which does not can be deduced along the line of arguments 
at the end of the last subsection. From \eqref{eq:Sinequality}, 
$S(z_\pm)<S(0)$ along the axis $G>0$, hence 
$\JJ(z_\pm)$ and $\KK(0)$ jump across this line while the others do not. 
It can also be checked numerically that $\Re\,S(0)$ is lower than $\Re\,S(z_\pm)$ 
along the kidney-shaped contour in Figure \ref{fg:stokeslines}. Therefore, this time, 
$\JJ(0)$ and $\KK(z_\pm)$ jump across this contour. 
We verified these expectations numerically. 

In the shaded area of Figure \ref{fg:stokeslines}, $\JJ(0)$ is the only thimble 
contributing to $Z_N(G,0)$. The structure of thimbles looks like  
Figure \ref{fg:GN1}. Outside the shaded area, $\JJ(0)$ and $\JJ(z_\pm)$ 
all contribute to $Z_N(G,0)$ and their structure resembles Figure \ref{fg:GN2}. 
The boundary of the shaded area is where a jump occurs in the number of 
contributing thimbles ($1\leftrightarrows 3$).

It has to be emphasized that the boundary curve in Figure \ref{fg:stokeslines} 
has nothing to do with the chiral phase transition. Roughly speaking, 
inside the curve $Z_N(G,0)\sim \ee^{-NS(0)}$, whereas outside the curve 
$Z_N(G,0)\sim \ee^{-NS(0)}+\ee^{-NS(z_+)}+\ee^{-NS(z_-)}$, at $N\gg 1$. 
In either case $Z_N(G,0)$ is entirely dominated by $\ee^{-NS(0)}$ since 
$\Re\,S(0)<\Re\,S(z_\pm)$ on the boundary curve; a phase transition 
does not occur. It is known that the emergence of such subdominant exponentials 
across a Stokes line occurs smoothly \cite{Berry1989,Berry1989b}, i.e.,  
$Z_N(G,0)$ is \emph{analytic} around the Stokes lines. The correct identification of the chiral 
transition line will be made in Section \ref{sc:LY}. 

So far we have tacitly assumed that $Z_N(G,0)$ originally defined for $G>0$ by 
the integral \eqref{eq:ZGN} can be extended to complex $G$, but this requires some care. 
While \eqref{eq:ZGN} is convergent for $\Re\,G>0$, it apparently 
\emph{diverges} for $\Re\,G<0$. Nevertheless, one can still expand 
\eqref{eq:Zgrossneveu} in Taylor series of $G$ and evaluate the partition function in a 
polynomial of order $N$ in $G$, which is of course analytic over the entire complex $G$-plane.  
The right procedure to fill this gap and define the integral \eqref{eq:ZGN} analytically 
for entire $G\in\CC$ is as follows%
\footnote{Pedagogical reviews of this procedure for the Airy integral 
can be found in \cite{Marino:2009dp,Witten:2010cx,Marino:2012zq}. }%
: as $G$ varies on the complex plane, the ``good'' regions on the 
complex $z$-plane also rotates simultaneously. In order for the integral to converge, 
the integration contour (initially $\RR$) must have ends in those good regions, hence 
the contour should be rotated hand-in-hand with the variation of $G$. 

To be more explicit, let us consider a phase rotation of $G$ 
by $2\pi$ on the complex plane, starting from some 
$G>1$ to avoid complications due to Stokes lines. 
Initially (with $G>0$), two of the good regions are located at $\Re\,z \gg1$ and 
$\Re\,z\ll -1$, where the weight $\exp(-z^2/G)$ goes to zero. 
When $G$ is dialed by $\pi/2$ and approaches the positive imaginary axis, 
these good regions are rotated by $\pi/4$. When $G$ is rotated by $\pi$, 
the good regions are rotated by $\pi/2$: they are now specified 
by $\Im\,z\gg 1$ and $\Im\,z\ll -1$. As $G$ returns to the positive real axis, 
the good regions return to their initial position. 

What is interesting here is that the two good regions are permutated by this rotation.  
In other words, they are rotated by $\pi$ when $G$ is rotated by $2\pi$, and 
so are the Lefschetz thimbles: they return to themselves only after $4\pi$ rotation of $G$. 
Therefore the monodromy of Lefschetz thimbles around $G=0$ is of order 2.  
Related to this, note that although the contour itself is homologically equivalent to $\RR$ 
after a $2\pi$-rotation of $G$, its orientation gets reversed. Thus the integral 
over the contour flips sign when $G$ is rotated by $2\pi$. 
At the same time, however, $\sqrt{G}$ in front of \eqref{eq:ZGN} changes sign 
(i.e., $\sqrt{G}\to \sqrt{G}\ee^{\pi i}$ when $G\to G\ee^{2\pi i}$) so that 
the integral \eqref{eq:ZGN} returns to its initial value. Thus the orientation of 
a thimble must be traced 
correctly to ensure the single-valuedness of the partition function for complex $G$.

\subsubsection{Anti-Stokes lines and Lee-Yang zeros}
\label{sc:LY}

\begin{figure}[!t]
   \begin{center}
     \includegraphics[width=0.42\columnwidth]{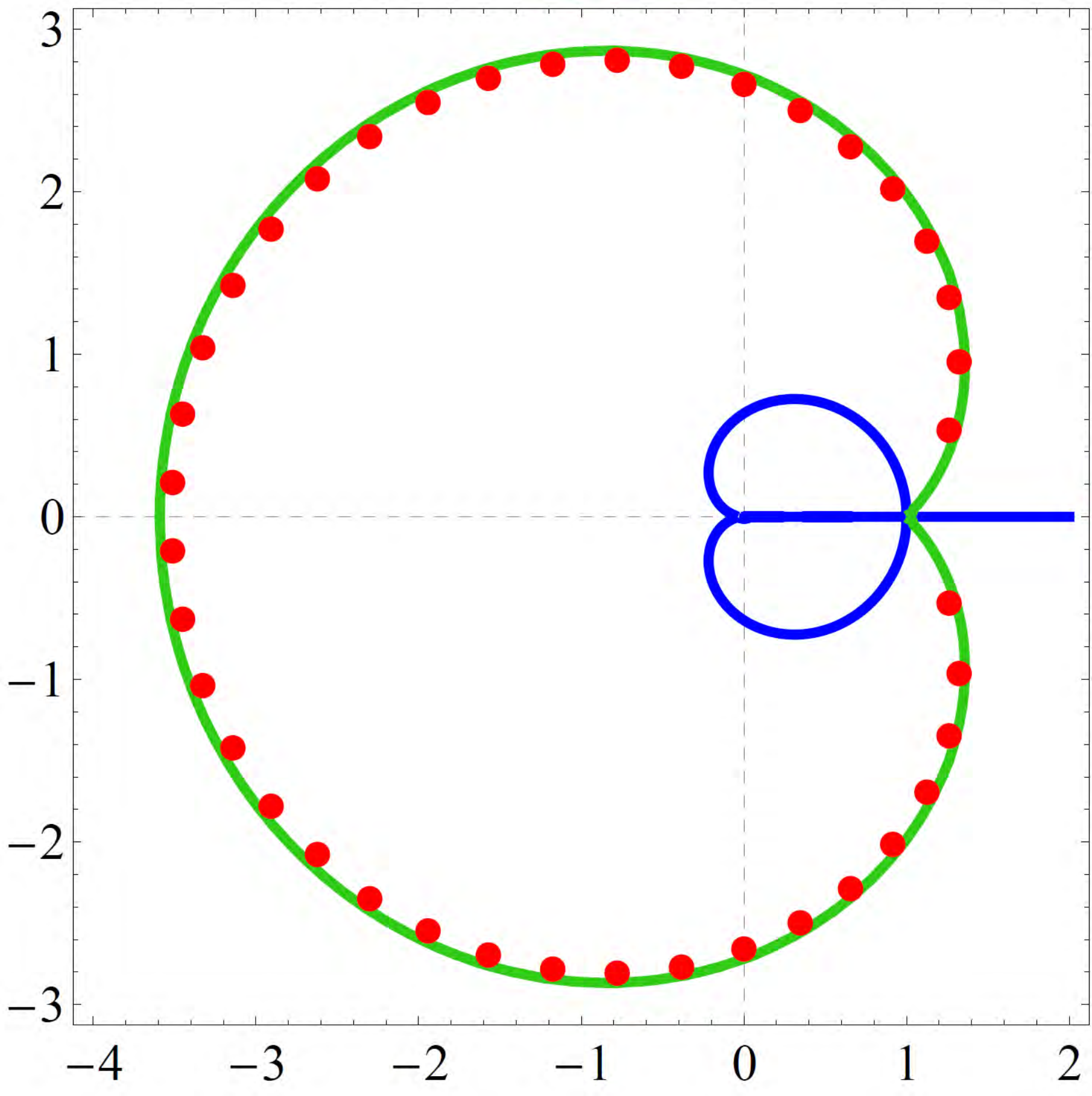}
     \put(-206,93){$\Im\,G$}
     \put(-96,-14){$\Re\,G$}
   \end{center} 
   \vspace{-.8\baselineskip}
   \caption{
     \label{fg:LYzero}
     Anti-Stokes line for the GN-like model with $p=1$ and $m=0$ (green curve),  
     overlaid with Lee-Yang zeros for $N=40$ (red bullets) and the Stokes line 
     in Figure \ref{fg:stokeslines} (blue curve).
   } 
   \end{figure}

Now that the partition function is well defined for complex coupling $G$, we can ask 
where is the boundary between a chirally broken phase and a chirally symmetric phase 
in the complex $G$-plane. Considering that $S(z)$ has three critical points $\{0,z_+,z_-\}$, 
we would get a nonzero condensate in the large-$N$ limit if the following 
two conditions are both met:
\begin{enumerate}
  \item $\JJ(z_\pm)$ contribute to $Z_N(G,0)$, and 
  \item $\Re\,S(z_\pm) < \Re\,S(0)$.  
\end{enumerate}
From the last subsection we know that the first condition is met for $G$ outside 
the shaded region in Figure \ref{fg:stokeslines}. The second condition is necessary 
for the symmetry-breaking saddles $z_\pm$ to dominate the partition function 
at large $N$. In general, a line on which exchange of dominance occurs between distinct 
saddles is called an \emph{anti-Stokes line}, which should not be confused with 
the Stokes lines. In the present case the anti-Stokes line is specified by
\begin{align}
  0 \overset{!}{=}\Re\,\big[S(0)-S(z_\pm)\big] 
  = \Re\,\Big[ \! -1 + \frac{p^2}{G} - \log \frac{p^2}{G} \Big] \,, 
  \label{eq:antistokes}
\end{align}
and is shown in Figure \ref{fg:LYzero} for $p=1$, 
together with the Stokes lines from Figure \ref{fg:stokeslines}.%
\footnote{The anti-Stokes line actually extends into the interior of the Stokes curve, but this part 
is not shown in Figure \ref{fg:LYzero} because 
$\JJ(z_+)$ and $\JJ(z_-)$ do not contribute to $Z_N(G,0)$ there.} 
%%%%%%%%%%%%%%%%%%%%%%%%%%%%%%%%%%%%%%%
   \noindent
%%%%%%%%%%%%%%%%%%%%%%%%%%%%%%%%%%%%%%%
Chiral symmetry is broken at large $N$ for $G$ outside the green anti-Stokes curve, 
and is restored for $G$ inside the curve.%
\footnote{It should be noted that a condensate for complex $G$ is a complex quantity and 
does not admit a physical interpretation as a dynamical mass of fermions.} Since multiple 
saddles exchange dominance, the phase transition 
along the anti-Stokes curve is generally \emph{first order}, with the only exception at $G=p^2$ 
where the transition is continuous. This point is quite special, as the Stokes curve 
and the anti-Stokes curve intersect there.  

Next, we would like to explore a connection between the anti-Stokes line and zeros of the 
partition function. Since the seminal work by Lee and Yang \cite{YL1952,LY1952}, it has been 
widely recognized that zeros of a finite-volume partition function in a complex parameter space, 
called \emph{Lee-Yang zeros}, 
provides rich information on the phase transition in the thermodynamic limit. 
See \cite{Bena2005} for a review and 
\cite{Fodor:2001pe,Fodor:2004nz,Ejiri:2005ts,Stephanov:2006dn,Denbleyker:2010sv} 
for applications to Yang-Mills theory and QCD. Connections between Stokes phenomenon 
and Lee-Yang zeros were investigated in \cite{Itzykson:1983gb,Pisani1993,Guralnik:2007rx}.  

In the present model, it is straightforward to evaluate \eqref{eq:Zgrossneveu} 
or \eqref{eq:ZGN} in the chiral limit to 
obtain a polynomial representation of the partition function:
\begin{align}
  Z_N(G,0) & = p^{2N} \sum_{k=0}^{N} 
  \begin{pmatrix}N \\ k\end{pmatrix}
  \begin{pmatrix}2k \\ k\end{pmatrix} k! \left(\frac{G}{4Np^2}\right)^k \,. 
\end{align}
We have numerically computed zeros of $Z_N(G,0)$ on the complex $G$-plane. 
The result for $N=40$ and $p=1$ is presented in Figure \ref{fg:LYzero}. 
Clearly all the zeros are distributed in the vicinity of the anti-Stokes curve.  For larger $N$,  
zeros are observed to align on the anti-Stokes curve more and more densely. It is expected that  
they will form a continuous cut in the large-$N$ limit, marking a boundary between a chirally 
symmetric phase and a chirally broken phase. The overall picture is quite 
consistent with the Lee-Yang picture of a phase transition.  

An important feature of the anti-Stokes curve in Figure \ref{fg:LYzero} is that 
it has a kink at $G=1$. It can be shown from \eqref{eq:antistokes} that 
the curve pinches the real axis at angle $\pm \pi/4$. According to a general 
theory of Lee-Yang zeros \cite{Bena2005}, a kink occurs when the transition 
for this point is of higher order, and the angle $\pi/4$ implies that it is 
a second-order phase transition with a mean-field critical exponent. This is exactly 
what happens in this model at $G=1$.

This completes our analysis of the GN-like model in the chiral limit.

\subsection{Massive case}
\label{sc:mGN}

When the fermion mass $m$ is nonzero, the $\ZZ_2$ chiral symmetry 
is explicitly broken and the ``condensate'' $\langle\sigma\rangle$ is nonzero 
for any $G>0$. Although a sharp phase transition is absent at $m\ne 0$, 
one can still find an interesting behavior of Lefschetz thimbles in the following.  
Without loss of generality, we assume $m>0$. 

To apply the Picard-Lefschetz theory we complexify $\sigma$ to $z\in\CC$. 
The critical points (i.e., saddles) of $S(z)$ are obtained as solutions to 
\begin{align}
  0 = \frac{\der S(z)}{\der z} & = 
  \frac{2z}{G} - \frac{2(z+m)}{p^2+(z+m)^2} \,. 
  \label{eq:mGNcrits}
\end{align}
This equation always has three roots, one of which is real and 
the other two are either both real or a complex-conjugate pair, 
depending on $G$, $p$ and $m$.  With a bit of algebra, we find that 
the jump in the number of real roots ($1\leftrightarrows 3$) occurs when 
\begin{align}
  2(\sqrt{\mathcal{D}}+2m)^2(\sqrt{\mathcal{D}}-m)-27Gm=0  
  \qquad \text{with}\quad 
  \mathcal{D} \equiv 3(G-p^2)+m^2\,, 
  \label{eq:31transition}
\end{align}
under the condition that $\mathcal{D}\geq 0$ and $m\geq 0$. 

%%%%%%%%%%%%%%%%%%%%%%%%%%%%%%%%%%%%%%%
   \begin{figure}[!t]
   \begin{center}
      \quad~ 
      \includegraphics[width=0.55\columnwidth]{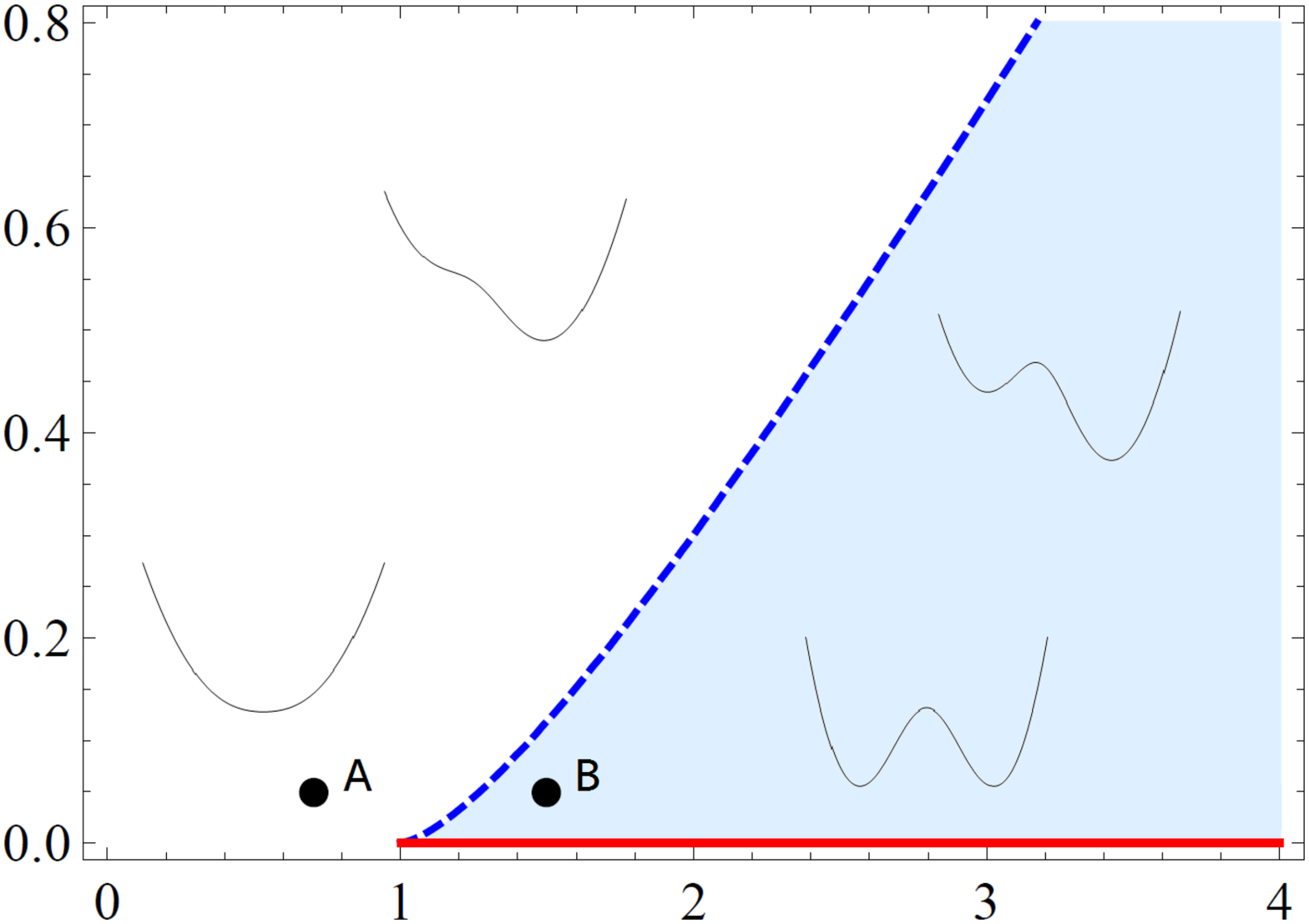}
      \ 
      \includegraphics[width=0.38\columnwidth]{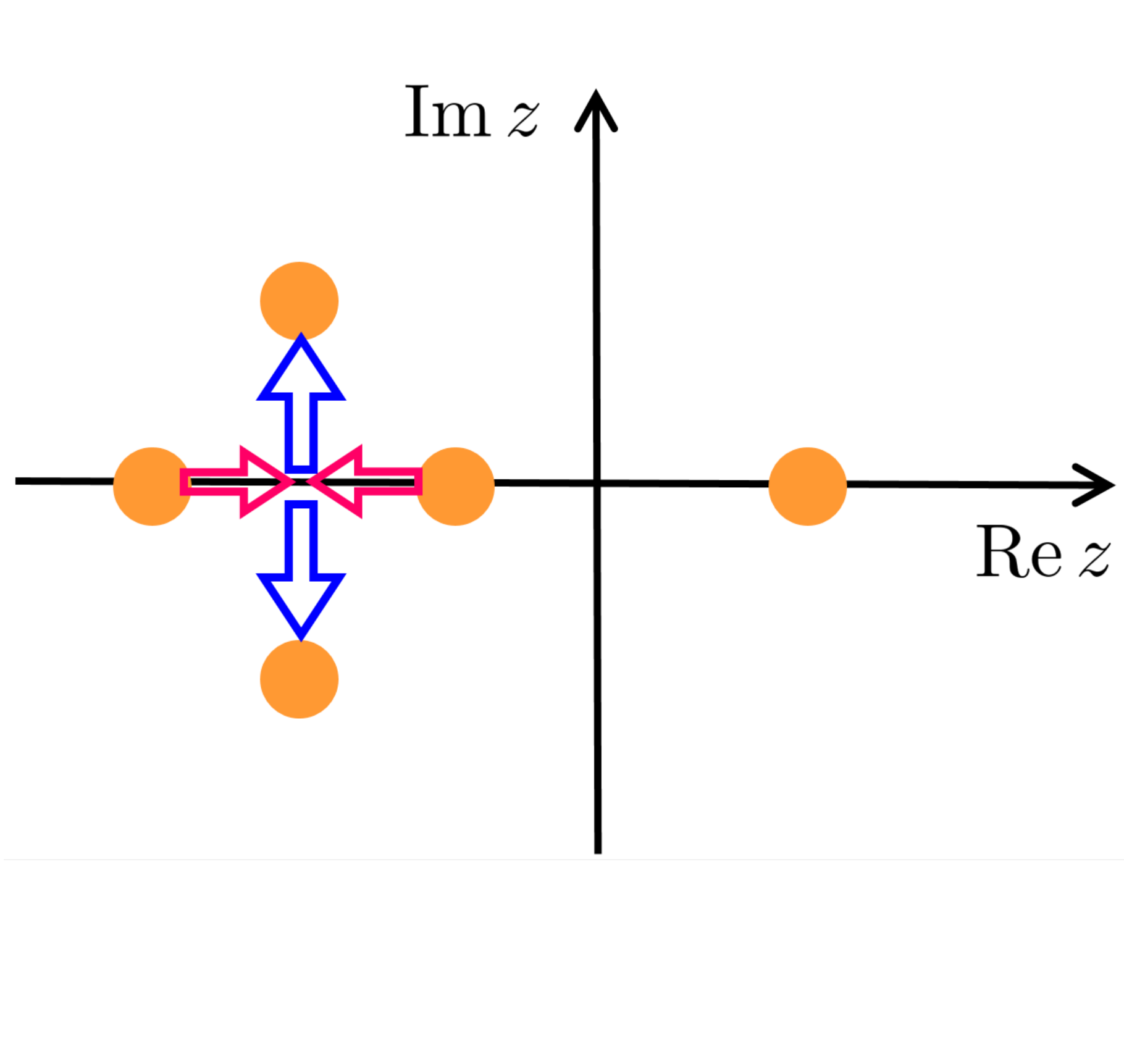}
      \put(-295,-15){\large $G/p^2$}
      \put(-423,88){\large $\displaystyle\frac{m}{p}$}
   \end{center} 
   \vspace{-.8\baselineskip}
   \caption{
     \label{fg:GNpd}
     \textbf{Left:} Phase diagram of the GN-like model. 
     (The $m<0$ part can be obtained by reflection about the horizontal axis.) 
     The red horizontal line represents a first-order phase transition line. 
     The blue dashed line is a limit of metastability (not a phase transition), i.e.,  
     below this line (in the blue region), there are two local minima for the action. 
     Above this line, the action has a single minimum. 
     Concerning the points {\bf A} and {\bf B}, see Figures \ref{fg:GN3} and 
     \ref{fg:GN4}. 
     \newline 
     \textbf{Right:} The behavior of saddle points of $S(z)$ when $m$ is increased 
     at fixed $G>p^2$. As we move out of the blue region in the left panel upward, 
     two of the three saddles on the real axis merge and then migrate into the complex plane.  
   } 
   \end{figure}
%%%%%%%%%%%%%%%%%%%%%%%%%%%%%%%%%%%%%%%

In Figure \ref{fg:GNpd} (left panel) 
we show the phase structure of the model, together with 
the typical shape of $S(\sigma)$ in each region. 
The domain having three (one) real saddles are painted blue (white), respectively. 
Across the blue dashed line given by \eqref{eq:31transition}, 
the number of saddles on $\RR$ jumps. This is not a phase transition, but 
corresponds to the disappearance (or emergence) of a metastable state.  
Figure \ref{fg:GNpd} (right panel) schematically shows the 
motion of critical points on the complex $z$-plane when 
the line of metastability is crossed from below.  

\begin{figure}[!t]
   \begin{center}
     \includegraphics[width=0.4\columnwidth]{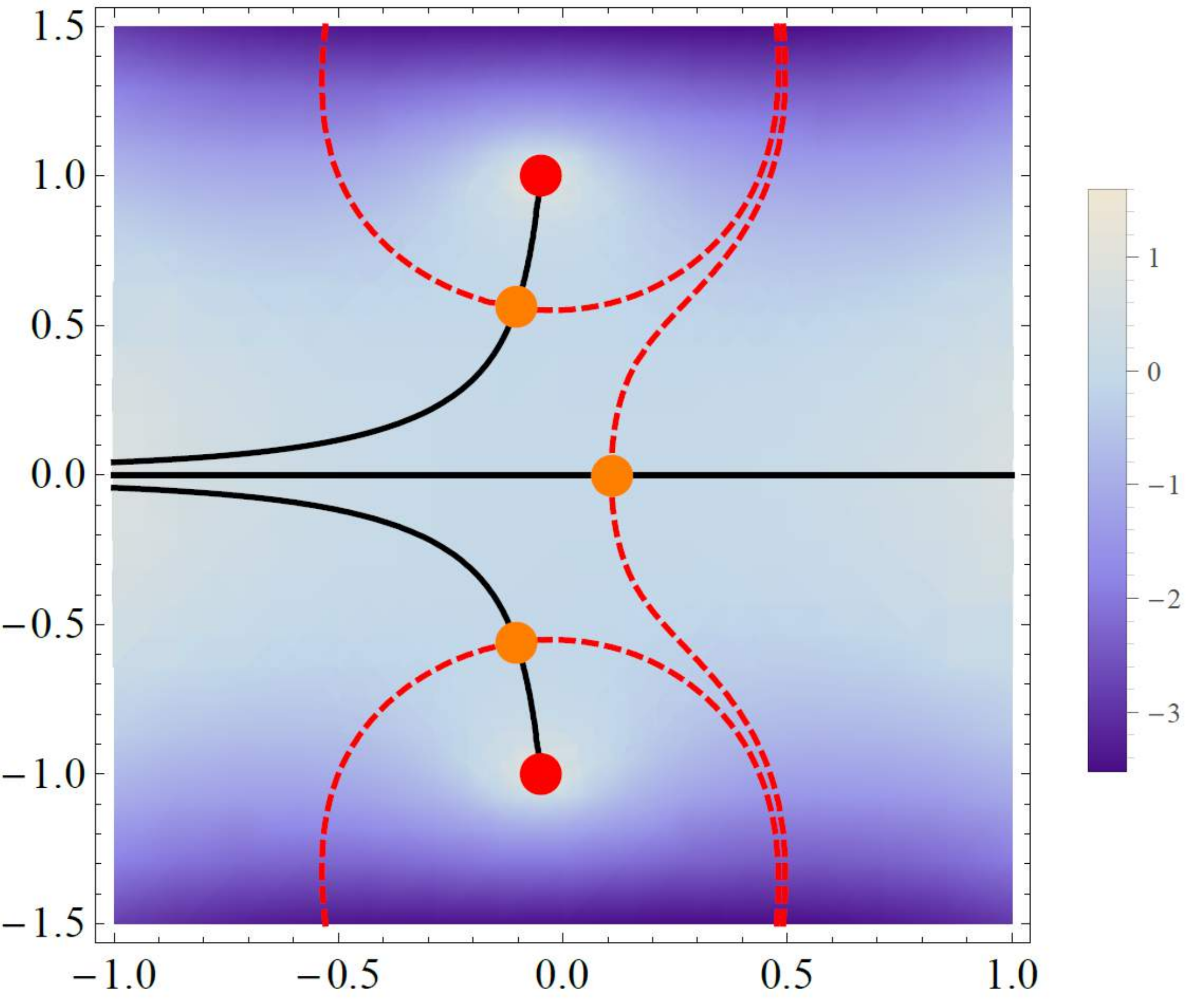}
     \put(-153,150){$G=0.7,~p=1,~m=0.05$}
   \end{center}
   \vspace{-.8\baselineskip}
   \caption{
     \label{fg:GN3}
     Same as Figure \ref{fg:GN1} but with $G=0.7$ and $m=0.05$, 
     corresponding to the point \textbf{A} in Figure \ref{fg:GNpd} (left panel). 
   }
   \vspace{.5\baselineskip}
   \begin{center}
     \includegraphics[width=0.4\columnwidth]{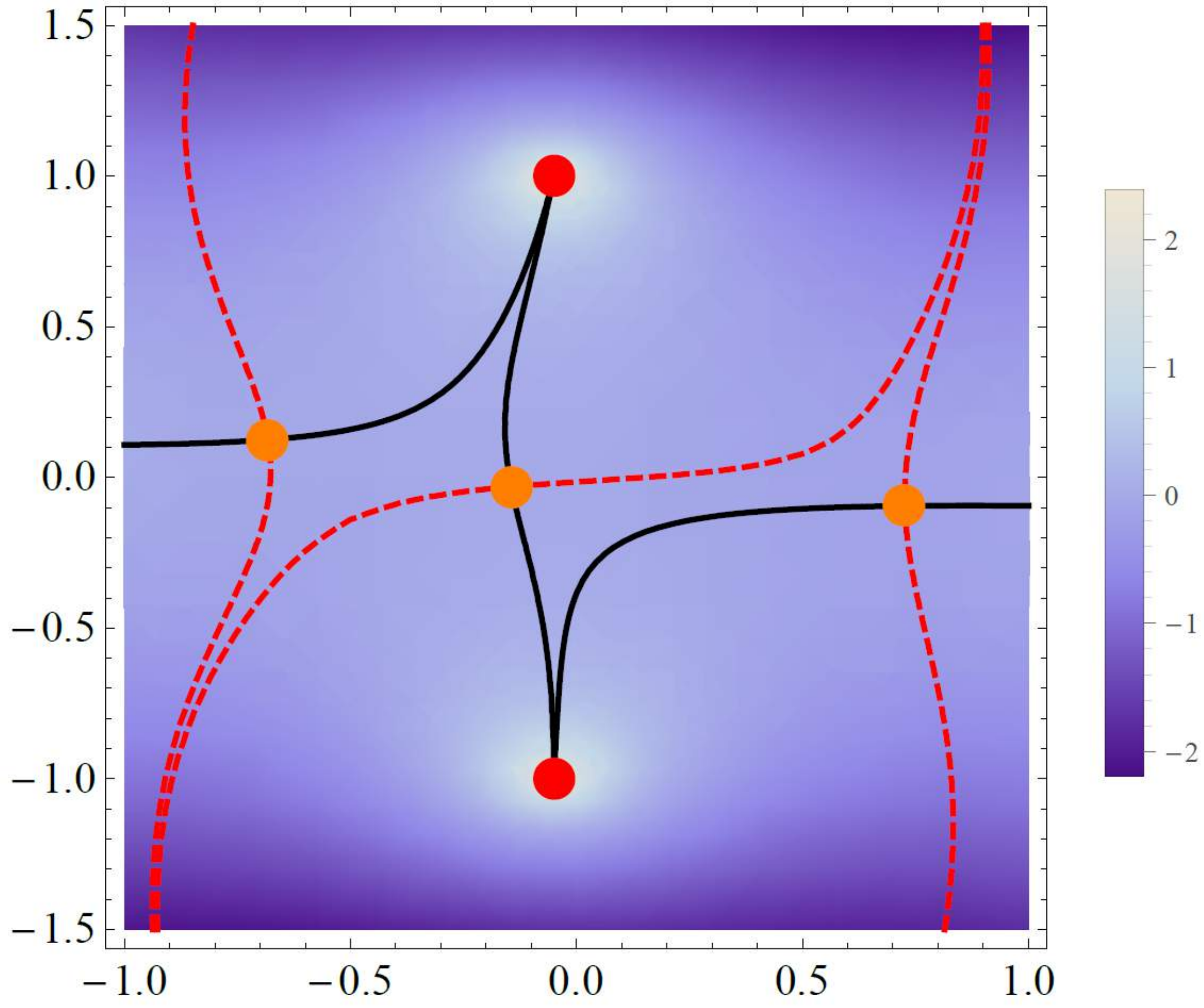}
     \quad
     \includegraphics[width=0.4\columnwidth]{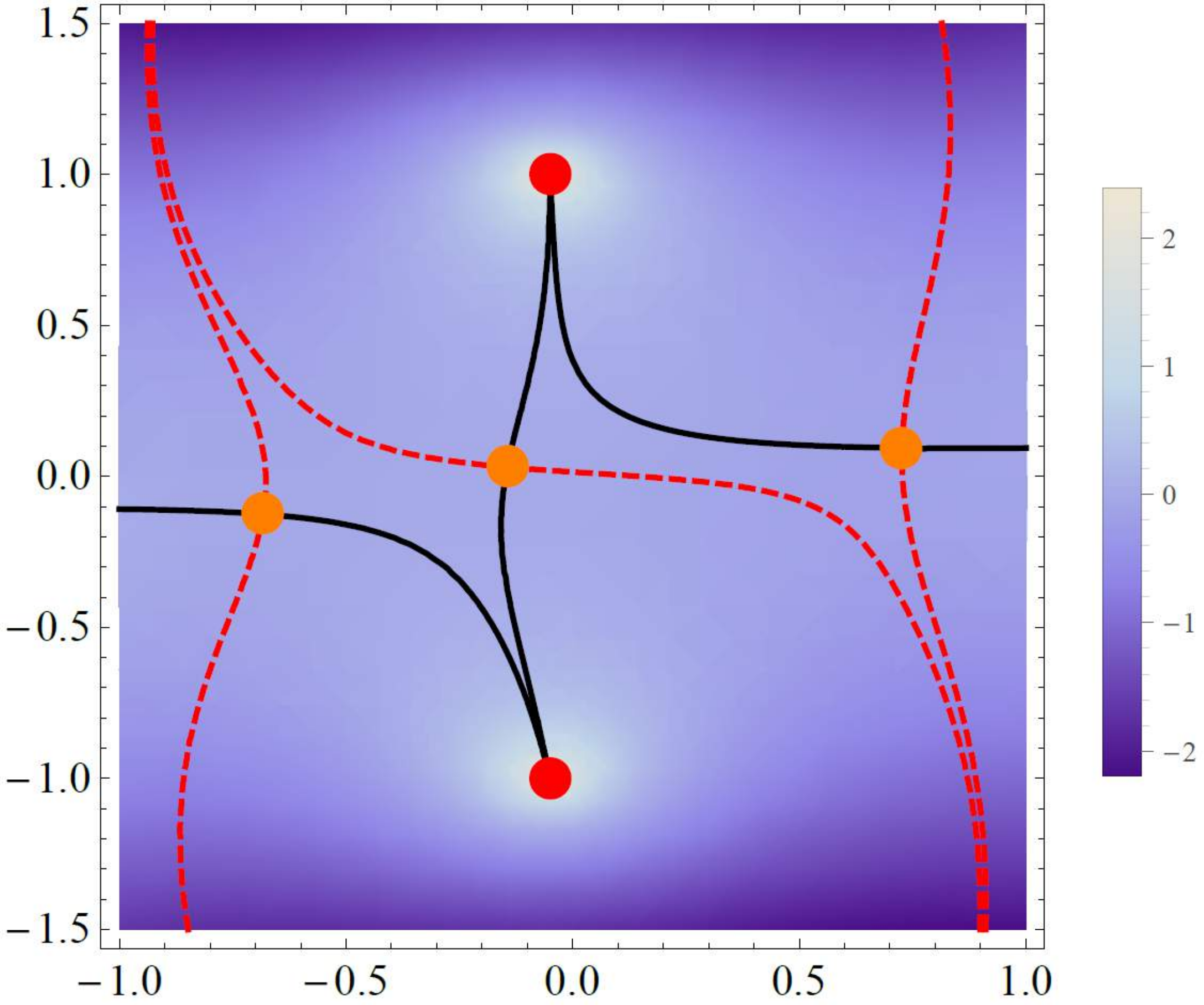}
     \put(-164,150){$G=1.5\ee^{+0.1i},~p=1,~m=0.05$}
     \put(-350,150){$G=1.5\ee^{-0.1i},~p=1,~m=0.05$}
   \end{center} 
   \vspace{-.8\baselineskip}
   \caption{
     \label{fg:GN4}
     Same as Figure \ref{fg:GN1} but with $|G|=1.5$ and $m=0.05$, 
     corresponding to the point \textbf{B} in Figure \ref{fg:GNpd} (left panel). 
   } 
   \end{figure}

Now we are in a position to reveal the behavior of Lefschetz thimbles. 
We take points {\bf A} and {\bf B} in Figure \ref{fg:GNpd} (left panel) as 
representatives of white and blue regions, respectively. 
Figure \ref{fg:GN3} shows the Lefschetz thimbles with $p=1$ at the point {\bf A}. 
Interestingly, at $m\ne 0$ no Stokes phenomenon occurs for $G\in\RR$ 
and one can safely take $G=0.7$ without a complex factor. 
For this $G$, the chirally condensate vanishes at $m=0$. 
We observe that, just as we saw in Figure \ref{fg:GN1}, 
there is only one thimble (the real axis, $\RR$) which contributes to the partition function. 
The saddle associated with this thimble gives rise to a condensate $\langle\sigma\rangle\ne 0$. 
The other two thimbles extend to the $z\to -\infty$ direction together, although 
they went in the opposite directions in Figure \ref{fg:GN1}. 
Note that Figure \ref{fg:GN3} will be horizontally reversed for $m<0$.

%%%%%%%%%%%%%%%%%%%%%%%%%%%%%%%%%%%%%%%
%%%%%%%%%%%%%%%%%%%%%%%%%%%%%%%%%%%%%%%

Figure \ref{fg:GN4} displays Lefschetz thimbles at the point \textbf{B} 
in Figure \ref{fg:GNpd} (left panel). This time all the three critical points lie on 
the real axis and have $\Im\,S=0$, so $G\in \RR$ is right on the Stokes ray.%
\footnote{More generally, a Stokes phenomenon occurs \emph{everywhere} in the shaded 
region of Figure \ref{fg:GNpd} (left panel), since all critical points are real there.} 
As before we rotated the phase of $G$ slightly to make the thimbles well-defined.
Figure \ref{fg:GN4} shows that now the three thimbles all contribute to the integral. 
The overall structure of the thimbles is the same as in Figure \ref{fg:GN2} at $m=0$. 
Among the three critical points, the right-most one has the lowest $\Re\,S$ and hence 
governs the partition function and condensate at $N\gg 1$.  

Between $G=0.7$ and $G=1.5$ there is a jump in the number of contributing thimbles. 
This occurs when one traverses the blue dashed boundary in Figure \ref{fg:GNpd} (left panel). 
This is not a phase transition, since the critical point that gives the dominant contribution 
always sits on the positive real axis and moves smoothly with $G$. Rather, there 
appears a new subleading contribution to the partition function, 
which is exponentially smaller than the leading one at $N\gg 1$. 

It is worth an emphasis that the three Lefschetz thimbles in Figures 
\ref{fg:GN3} and \ref{fg:GN4} indeed form a homological basis of cycles  
connecting ``good'' regions, in accordance with the general 
argument in Section \ref{sc:generic}. 

\begin{figure}[!t]
   \begin{center}
     \includegraphics[width=0.4\columnwidth]{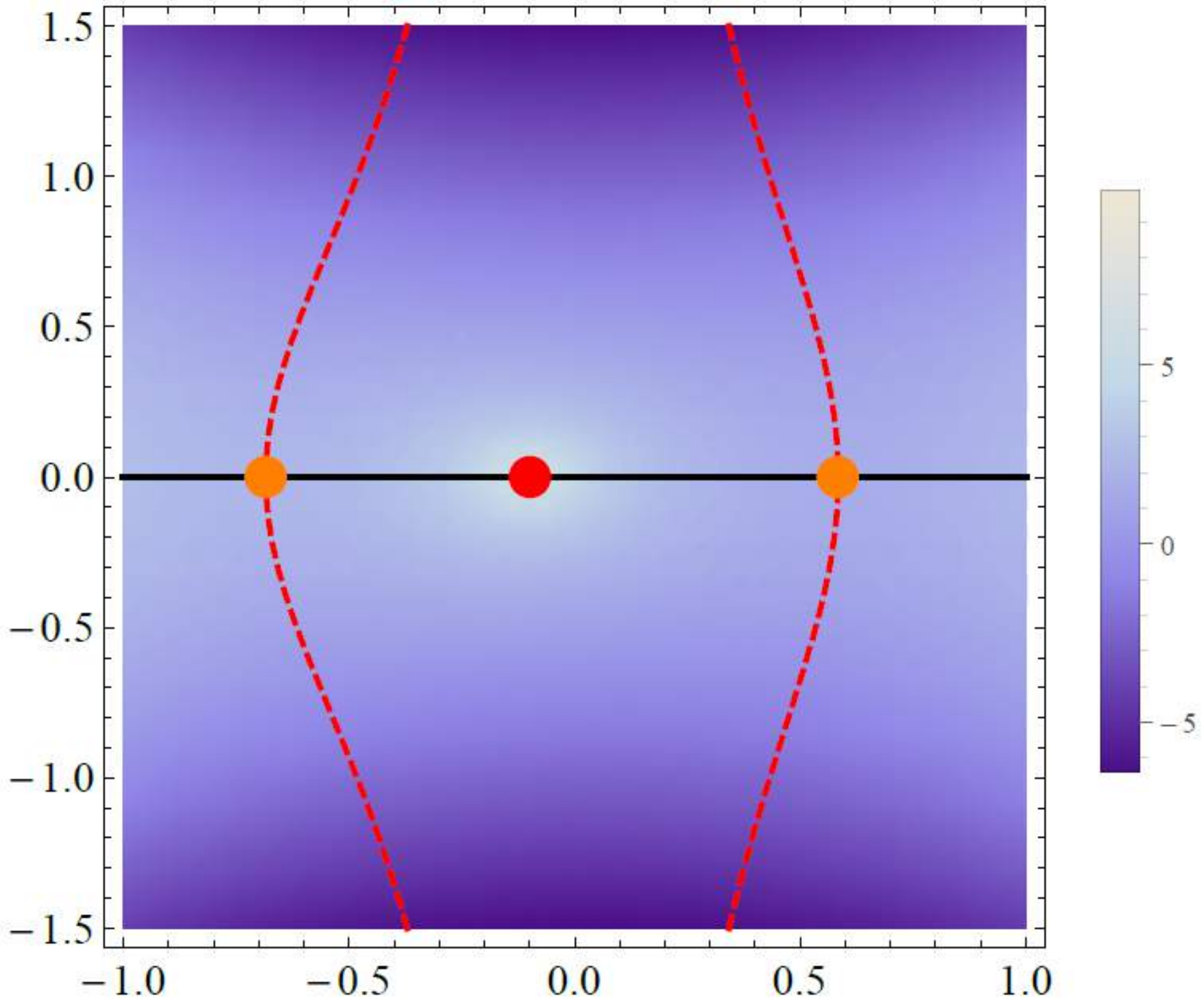}
     \put(-148,150){$G=0.4,~p=0,~m=0.1$}
   \end{center}
   \vspace{-.8\baselineskip}
   \caption{
     \label{fg:GN5}
     Lefschetz thimbles and their duals for the GN-like model at $p=0$.  
   }
   \end{figure}

Finally we consider a special limit $p\to 0$, in which the two singular points 
at $z=-m\pm ip$ merge into a single singularity at $z=-m$. The situation is 
simpler than for $p\ne 0$, because there are only two critical points 
at $z_\pm \equiv \frac{-m\pm\sqrt{m^2+4G}}{2}$. 
The corresponding thimbles are shown in Figure \ref{fg:GN5}. 
(Note that this is quite analogous to the example considered 
in Section \ref{sc:generic}.) The Lefschetz thimbles for $z_\pm$ are given by
\begin{align}
  \JJ(z_+)=\{ z\in\RR\,|\,z>-m \}\quad \text{and} \quad 
  \JJ(z_-)=\{ z\in\RR\,|\,z<-m \}\,,
  \label{eq:Jatp0}
\end{align}
They meet at the singular point $z=-m$ and together constitute the 
integration cycle $\RR$. In the limit $m\to 0$, 
$z_\pm$ move to $z=\pm \sqrt{G}$, hence chiral symmetry is 
spontaneously broken in the large-$N$ limit for any small $G>0$, 
with $\langle\sigma\rangle=\pm \sqrt{G}$. Such a non-analytic dependence on $G$ 
cannot occur at any finite order of expansion in $G$ and is a hallmark of nonperturbative physics. 

%%%%%%%%%%%%%%%%%%%%%%%%%%%%%%%%%%%%%%%
%%%%%%%%%%%%%%%%%%%%%%%%%%%%%%%%%%%%%%%

This completes our analysis of the Lefschetz thimbles in the 
zero-dimensional GN model.

\section{Nambu-Jona-Lasinio-like model}
\label{sc:NJLm}

\subsection{Model setup}

Next, we consider two zero-dimensional toy models having continuous chiral 
symmetries, in analogy to the Nambu-Jona-Lasinio(NJL) model 
\cite{Nambu:1961tp,Nambu:1961fr}. The first model is defined by the partition function 
\begin{align}
  Z_{\U(1)}=\int \dd\bar\psi \dd\psi~\exp \bigg(
  \sum_{a=1}^{N}\bar\psi_a (i\slashed{p}+m) \psi_a + 
  \frac{G}{4N}\bigg\{
    \Big(\sum_{a=1}^{N}\bar\psi_a \psi_a \Big)^2
    + \Big(\sum_{a=1}^{N}\bar\psi_a i \gamma_5\psi_a \Big)^2
  \bigg\} \bigg) \,.
  \label{eq:Z_njltype}
\end{align}
The definitions of symbols and variables are the same as in \eqref{eq:Zgrossneveu}. 
In the chiral limit $m=0$ 
the action is invariant under a $\U_{\rm A}(1)$ chiral rotation 
$\psi\to \ee^{i \theta\gamma_5}\psi$ and $\bar\psi\to \bar\psi\ee^{i \theta\gamma_5}$. 
By introducing auxiliary fields $\sigma$ and $\pi$ to bilinearize the action, we obtain
\begin{align}
  Z_{\U(1)} 
  & = \frac{N}{\pi G} \int_{\RR^2} \dd\sigma \dd\pi~{\det}^N
  (i\slashed{p} + m + \sigma+i\gamma_5\pi)
  \exp\mkakko{-\frac{N}{G}(\sigma^2+\pi^2)}
  \\
  & = \frac{N}{\pi G} \int_{\RR^2} \dd\sigma \dd\pi~\ee^{-N S(\sigma,\pi)} \,,
  \label{eq:Z_model1}
\end{align}
with
\begin{align}
  S(\sigma,\pi) & \equiv -\log 
  \kkakko{p^2+(\sigma+m)^2+\pi^2} + \frac{\sigma^2+\pi^2}{G} \,.
  \label{eq:Snjl}
\end{align}
With $m=0$ this action enjoys an $\O(2)$ symmetry that rotates $\sigma$ and $\pi$. 
If we set $\pi=0$ by $\O(2)$ rotation, then the present model reduces to the GN-like model 
in Section \ref{sc:gross}. This will become important in the analysis of Lefschetz thimbles later. 

The second model we consider is defined by the partition function
\begin{align}
  Z_{\SU(2)} = \int \dd\bar\psi \dd\psi~\exp\bigg(  
  \sum_{a=1}^{N}\bar\psi_a(i\slashed{p} + m)\psi_a + \frac{G}{4N}
  \bigg\{
    \Big(\sum_{a=1}^{N}\bar\psi_a \psi_a \Big)^2
    + \sum_{A=1}^{3} \Big(\sum_{a=1}^{N}\bar\psi_a i \gamma_5 \tau^A \psi_a \Big)^2
  \bigg\}  
  \bigg) \,, 
\end{align}
where $\psi$ and $\bar\psi$ are two-component Grassmann variables 
with two flavors and $N$ colors, $\{\tau^A\}$ are the Pauli matrices, and  
the summation over flavor indices is implicitly assumed. 
At $m=0$, this model has an exact $\SU_{\rm R}(2)\times \SU_{\rm L}(2)$ 
chiral symmetry, which is broken explicitly to $\SU_{\rm V}(2)$ by nonzero $m$. 
 
Proceeding as before, 
\begin{align}
  Z_{\SU(2)} 
  & = \mkakko{\frac{N}{\pi G}}^2
  \int_{\RR^4} \dd\sigma \dd\pi_A~
  {\det}^N(i\slashed{p}+m+\sigma+i\gamma_5\pi_A\tau^A)
  \exp\mkakko{-\frac{N}{G}(\sigma^2+\pi_A^2)}
  \\
  & = \mkakko{\frac{N}{\pi G}}^2
  \int_{\RR^4} \dd\sigma \dd\pi_A~\ckakko{ p^2+(m+\sigma)^2+\pi_A^2 }^{2N} 
  \exp\mkakko{-\frac{N}{G}(\sigma^2+\pi_A^2)} \,.
\end{align}
Under the chiral $\SU_{\rm R}(2)\times\SU_{\rm L}(2)\cong\O(4)$ symmetry,  
$(\sigma,\pi_1,\pi_2,\pi_3)$ rotates as a vector. 
One can rotate any such vector to $(\sigma,0,0,\pi_3)$ 
by means of an unbroken $\SU_{\rm V}(2)\cong\SO(3)$ rotation. 
The resulting integral over $\sigma$ and $\pi_3$ is essentially equivalent to 
the former model \eqref{eq:Z_model1} and does not entail a new feature. 
For this reason we focus on the first model in the following.

\subsection{Massless case}
\label{sc:NJLnomass}

For simplicity we begin with the chiral limit $m=0$ where the chiral symmetry is exact. 
The task is to identify the Lefschetz thimbles for $Z_{\U(1)}$ 
and to figure out how to decompose the original integration cycle 
$\RR^2$ of \eqref{eq:Z_model1} into a sum of Lefschetz thimbles in $\CC^2$.  
Upon a complexification of variables, the action becomes 
\begin{align}
  S(z,w)=-\log(p^2+z^2+w^2)+ \frac{z^2+w^2}{G}\,,
\end{align}
whose domain is $\{(z,w)\in\CC^2\,|\,z^2+w^2\ne -p^2\}$. 
The set of singularities of logarithm $\{(z,w)\in\CC^2\,|\,z^2+w^2 = -p^2\}$ forms 
a surface of real dimension 2 in $\CC^2$ rather than a set of isolated points.  
It is equal to the $\O(2,\CC)$-orbit of the singular points $(z,w)=(\pm ip,0)$ of 
the massless GN-like model, where 
$\O(2,\CC)\equiv\{g\in \mathrm{GL}(2,\CC)\,|\,g^{\rm T}g=\1\}$ 
is a complexification of the $\O(2)$ group. 

The downward flow equation reads
\begin{gather}
  \frac{\dd \bar{\scalebox{0.7}{$\mathstrut$} z}}{\dd\tau} 
  = \frac{2z}{p^2+z^2+w^2} - \frac{2z}{G} 
  \quad 
  \text{and}
  \quad 
  \frac{\dd \bar{\scalebox{0.7}{$\mathstrut$} w}}{\dd\tau } 
  = \frac{2w}{p^2+z^2+w^2} - \frac{2w}{G} \,.
  \label{eq:zwflows}
\end{gather}
An important property of this flow is that it is symmetric 
under an $\O(2)$ rotation of $(z,w)$ although 
it is \emph{not} under a general $\O(2,\CC)$ rotation. 
This will play a pivotal role in the construction of Lefschetz thimbles later. 

The critical points may be obtained by solving \eqref{eq:zwflows} with 
$\dd_\tau \bar{\scalebox{0.7}{$\mathstrut$} z}=
\dd_\tau\bar{\scalebox{0.7}{$\mathstrut$} w}=0$. 
To avoid accidental degeneracy of critical points, we assume $G\ne p^2$. 
The set of critical points then consists of two components: 
\begin{align}
  C_0:= \{(0,0)\} \quad ~\text{and} ~\quad 
  C_1:= \{(z,w)\in\CC^2\,|\,z^2+w^2=G-p^2\}\,. 
  \label{eq:criticalset}
\end{align}  
$C_1$ is equal to the $\O(2,\CC)$-orbit of the critical points $\big(\pm\sqrt{G-p^2},0\big)$ 
in the massless GN-like model.  It crosses the real plane $\RR^2$ if $G-p^2\in\RR_{>0}$ but 
has no crossing otherwise. 

We now discuss how to determine the Lefschetz thimbles and their duals associated with 
$C_0$ and $C_1$. Since we are considering a two-dimensional integral, the thimbles 
should be cycles of real dimension 2. 
As for $C_0$ we can apply the standard procedure 
since it has a Morse index 2, i.e., $\Re\,S(z,w)$ 
increases in two directions and decreases in the other two directions around $C_0$.  
Then the Lefschetz thimble $\JJ_0$ and its dual $\KK_0$ associated with $C_0$ can be 
defined as a union of downward and upward flows, respectively, flowing into $C_0$. 
Since the flow preserves $\O(2)$ symmetry, $\JJ_0$ may be simply obtained 
by rotating a Lefschetz thimble in the GN-like model by $\O(2)$ action:  
\begin{align}
  \JJ_0 & = \ckakko{
    \begin{pmatrix}z \\ w\end{pmatrix} = 
    \begin{pmatrix} 
      \cos\theta & -\sin\theta 
      \\ 
      \sin\theta & \cos\theta 
    \end{pmatrix} \!\! 
    \begin{pmatrix}z' \\ 0 \end{pmatrix}
    ~\Bigg|~-\pi<\theta\leq\pi~~\text{and}~~ z' \in \JJ(0)\big|_{\rm GN}
  }\,, 
  \label{eq:JJ0}
\end{align}
and the same goes for $\KK_0$. So much for $C_0$. 

The prescription for $C_1$ is a bit different. 
A general framework to handle a continuous manifold of critical points was 
developed in \cite[Sect.~3]{Witten:2010cx} and we shall outline how to 
apply this framework to the present fermionic model with $\O(2,\CC)$ symmetry. 
First of all, at a point of $C_1$, the Hessian matrix of 
$\mathrm{Re}\;S$ has one positive eigenvalue, one negative eigenvalue 
and two null eigenvalues. Therefore, the set of points that can be reached 
by a downward/upward flow from any given point in $C_1$ is of real dimension 1.%
\footnote{Note that no flow exists on $C_1$, since all points on $C_1$ have the same value of $\Re\,S$.} 
Therefore if we pick up a one-dimensional subset out of $C_1$, the set of points that 
can be reached by a downward flow from that subset of $C_1$ forms a two-dimensional cycle, 
which gives an element of $H_2(\CC^2,(\CC^2)^T;\ZZ)$ for very large $T$, 
with $(\CC^2)^T:=\{(z,w)\in\CC^2\,|\,\Re\,S(z,w)\geq T\}$. This cycle could be 
employed as the Lefschetz thimble for $C_1$, say, $\JJ_1$. 

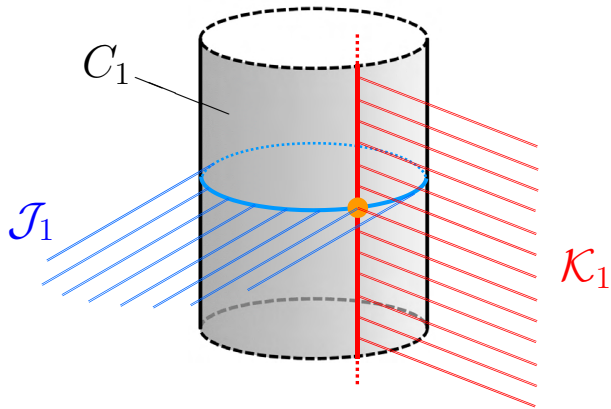
\begin{figure}[!t]
   \begin{center}
     \includegraphics[width=0.45\columnwidth]{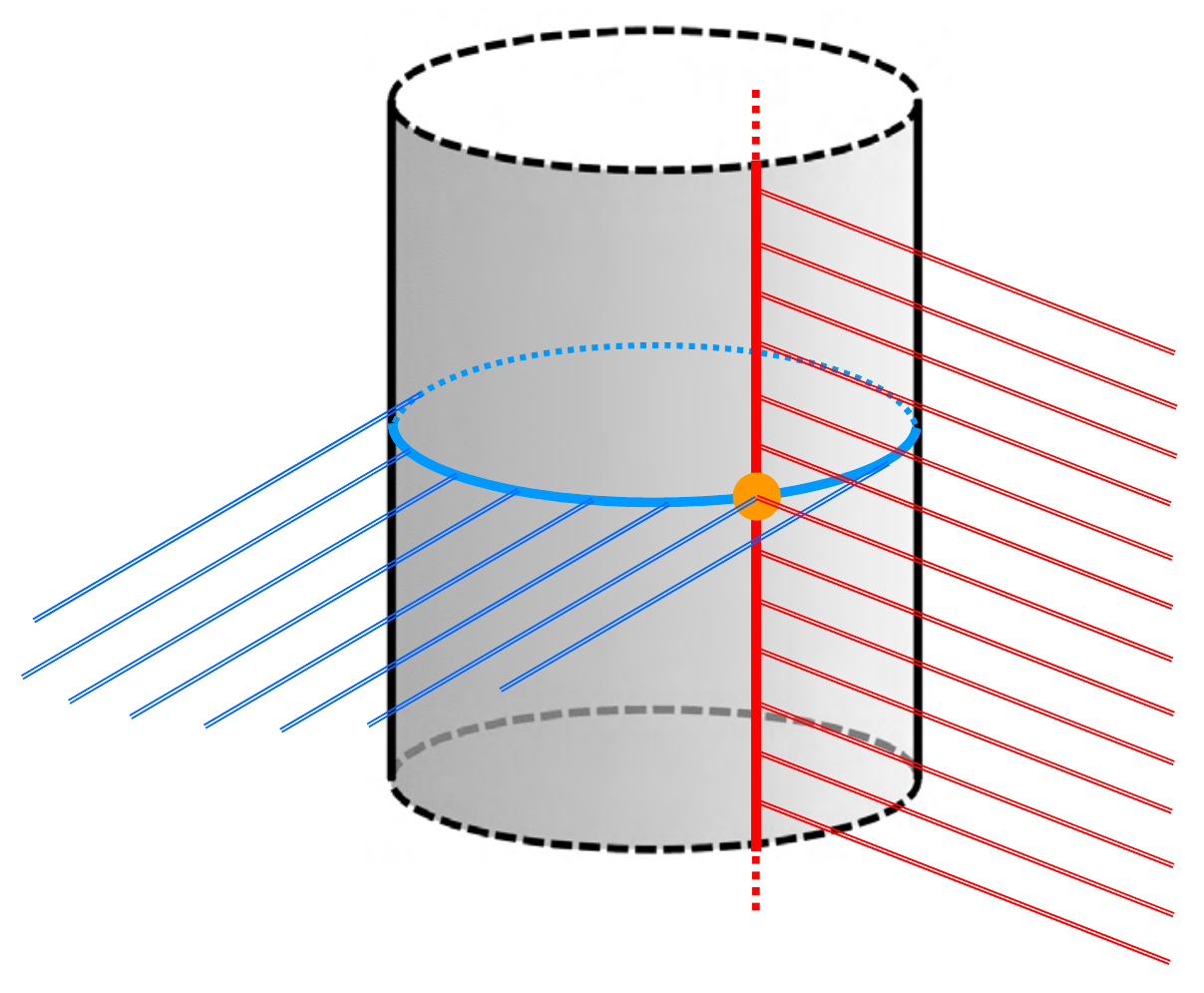}
     \put(-203,70){\LARGE ${\color{blue}\JJ_1}$}
     \put(4,52){\LARGE ${\color{red}\KK_1}$}
     \put(-120, 115){\line(-5,2){35}}
     \put(-175,129){\LARGE $C_1$}
   \end{center}
   \vspace{-\baselineskip}
   \caption{
     \label{fg:3dcycles}
     A schematic illustration of relations between the $\O(2,\CC)$-orbit $C_1$, 
     the Lefschetz thimble $\JJ_1$ and its dual $\KK_1$. The blue circle (red vertical line) 
     represents a compact orbit (non-compact orbit), respectively. The slant lines are 
     upward(red)/downward(blue) flow lines. 
   }
   \end{figure}

The next question is how to choose a one-dimensional subset in $C_1$ in the first place. 
It is known that if a critical orbit is ``semistable'', then it has a middle-dimensional homology 
of rank $1$ \cite{Witten:2010cx}. In the current setup, the condition of semistability for 
$C_1$ is that it should include a point where 
$\mu:= \bar{\scalebox{0.7}{$\mathstrut$}z}w - z\bar{\scalebox{0.7}{$\mathstrut$}w}$ vanishes. 
Since this condition is trivially met ($\mu=0$ at, say, $(\sqrt{G-p^2},0)\in C_1$),  
$C_1$ is semistable and consequently its one-dimensional homology is of rank $1$. 
This implies that the choice of a cycle in $C_1$ is essentially unique up to homologically 
equivalent ones. Adopting a canonical choice suggested in \cite{Witten:2010cx}, 
we shall take the set of points in $C_1$ where $\mu=0$. It is given by 
$\big\{(z,w)=\sqrt{G-p^2}(\cos\theta,\sin\theta)\,| -\pi<\theta\leq \pi \big\}$. As this cycle is 
the $\O(2)$-orbit of $(\sqrt{G-p^2},0)$, we shall call it a \emph{compact orbit}. It is 
pictorially shown as a blue circle in Figure \ref{fg:3dcycles}. 
%%%%%%%%%%%%%%%%%%%%%%%%%%%%%%%%%%%%%%%
%%%%%%%%%%%%%%%%%%%%%%%%%%%%%%%%%%%%%%%
The Lefschetz thimble $\JJ_1$ can now be defined as a union of downward flow lines 
emanating from the compact orbit. Recalling that the flow respects $\O(2)$ symmetry, 
$\JJ_1$ can be simply obtained as an $\O(2)$-revolution of the flow line emanating 
from $(\sqrt{G-p^2},0)$, which is nothing but $\JJ(z_+)$ considered in Section \ref{sc:gross}. 
We thus conclude that 
\begin{align}
  \JJ_1 & =  
  \ckakko{
    \begin{pmatrix} z \\ w \end{pmatrix} = 
    \begin{pmatrix} \cos\theta & - \sin\theta \\ \sin\theta & \cos\theta \end{pmatrix}\!\!
    \begin{pmatrix} z'' \\ 0 \end{pmatrix}
    ~\Bigg|~-\pi<\theta\leq \pi~~\text{and}~~ z'' \in \JJ(z_+)\big|_{\rm GN}
  }\,. 
  \label{eq:JJ1}
\end{align}
Since $\mu$ is conserved along a flow,%
\footnote{$\mu$ corresponds to the angular momentum if the Morse flow \eqref{eq:zwflows} 
is viewed as the Hamiltonian flow with Hamiltonian $\Im\,S$.}  
$\mu$ vanishes everywhere on $\JJ_1$. 

The dual cycle $\KK_1$ can also be constructed on the basis of \cite{Witten:2010cx}. 
Recall that we have used a rotation by $\O(2)\subset\O(2,\CC)$ 
to construct a compact orbit. Now, we shall use 
the complementary part of $\O(2,\CC)$ to build 
a \emph{non-compact orbit} of $(\sqrt{G-p^2},0)$%
\footnote{This could be replaced with any other point on the compact orbit.}: it 
is given by 
$\big\{(z,w)=\sqrt{G-p^2}(\cosh\phi,i\sinh\phi)\,|\,-\infty<\phi <\infty \big\}$ 
and is depicted in Figure \ref{fg:3dcycles} as a red vertical line. 
(Generally $\mu$ is nonzero and varies along the non-compact orbit.) 
Then $\KK_1$ can be defined as a union of upward flow lines emanating from 
the non-compact orbit; see Figure \ref{fg:3dcycles} for an illustration of this. 
By construction it is ensured that $\JJ_1$ and $\KK_1$ crosses exactly once, 
and thus intersection numbers $\{n_{\sigma}\}$ in (\ref{eq:23}) are well defined. 
This completes our determination of $\JJ_0,\JJ_1,\KK_0$ and $\KK_1$ 
in the massless NJL-like model. 
It is intriguing that the number of Lefschetz thimbles, $2$, is fewer 
than in the GN-like model, which can be 
attributed to the existence of $\O(2)$ chiral symmetry in the present model. 

Behaviors of the thimbles, \eqref{eq:JJ0} and \eqref{eq:JJ1}, for varying $G\in\CC$ can be 
learned from Section \ref{sc:gross} with no additional calculation. Here is a summary: 
\begin{itemize}
  \item 
  $\JJ_0$ and $\JJ_1$ jump on the Stokes lines in Figure \ref{fg:stokeslines}. 
  No modification of the figure is necessary. $\JJ_0$ jumps across the circular curve and 
  $\JJ_1$ jumps across the horizontal line, respectively.
  \item 
  For $G$ inside the shaded area of Figure \ref{fg:stokeslines}, 
  $\JJ_0$ is the only thimble that contributes to the partition function. 
  $\JJ_0$ has no boundary (it extends to infinity in $\CC^2$). 
  \item 
  For $G$ outside the shaded are of Figure \ref{fg:stokeslines}, both $\JJ_0$ and $\JJ_1$ 
  contribute to the partition function. Notably, $\JJ_0$ is now 
  a finite domain enclosed by a ring of logarithmic singularities, 
  $\{(z,w)=ip(\cos\theta,\sin\theta)\,| -\pi < \theta \leq \pi \}$.  
  This situation does not arise in a bosonic model.  
  \item 
  For $G$ inside the anti-Stokes line in Figure \ref{fg:LYzero}, $\JJ_0$ is the dominant thimble 
  at large $N$. As $G$ moves out across the anti-Stokes line, $\JJ_0$ is overtaken by $\JJ_1$ 
  and spontaneous breaking of the $\O(2)$ chiral symmetry sets in through a first-order transition 
  (except for the real axis, on which the transition is continuous). 
\end{itemize}
To conclude, we found that the Lefschetz thimbles for the NJL-like model \eqref{eq:Z_model1} 
are $\O(2)$-revolution of the thimbles for the GN-like model \eqref{eq:ZGN}. As a result, 
much details of chiral symmetry breaking are shared by these models. 
While what is presented above is formally similar to Witten's treatment of a bosonic 
$\SO(2)$-symmetric model \cite{Witten:2010cx}, we witnessed fairly richer phenomena 
in the fermionic models due to the presence of logarithm in the action.

\subsection{Massive case}
\label{sc:massnjl}

At $m\ne 0$ the $\O(2)$ chiral symmetry is explicitly broken, and the NJL-like model 
becomes qualitatively similar to the massive GN-like model: instead of a critical manifold 
\eqref{eq:criticalset}, there are only 3 critical points $\{(z,w)=(z_i,0)\,|\,i=1,2,3\}$ 
where $\{z_i\}_{i=1}^3$ are the three solutions to \eqref{eq:mGNcrits}. Associated with 
them are 3 Lefschetz thimbles and their duals --- no subtlety 
that arose in Section \ref{sc:NJLnomass} for the choice of integration cycle appears 
in the massive case. Interestingly, a comparison with the chiral limit shows that 
\emph{the number of Lefschetz thimbles jumps} from 2 $\to$ 3 when an arbitrarily small 
$m\ne0$ is turned on. This phenomenon also occurs in a bosonic model 
with symmetry \cite{Tanizaki:2014tua}. 

\begin{figure}[!t]
   \begin{center}
     \includegraphics[width=0.25\columnwidth]{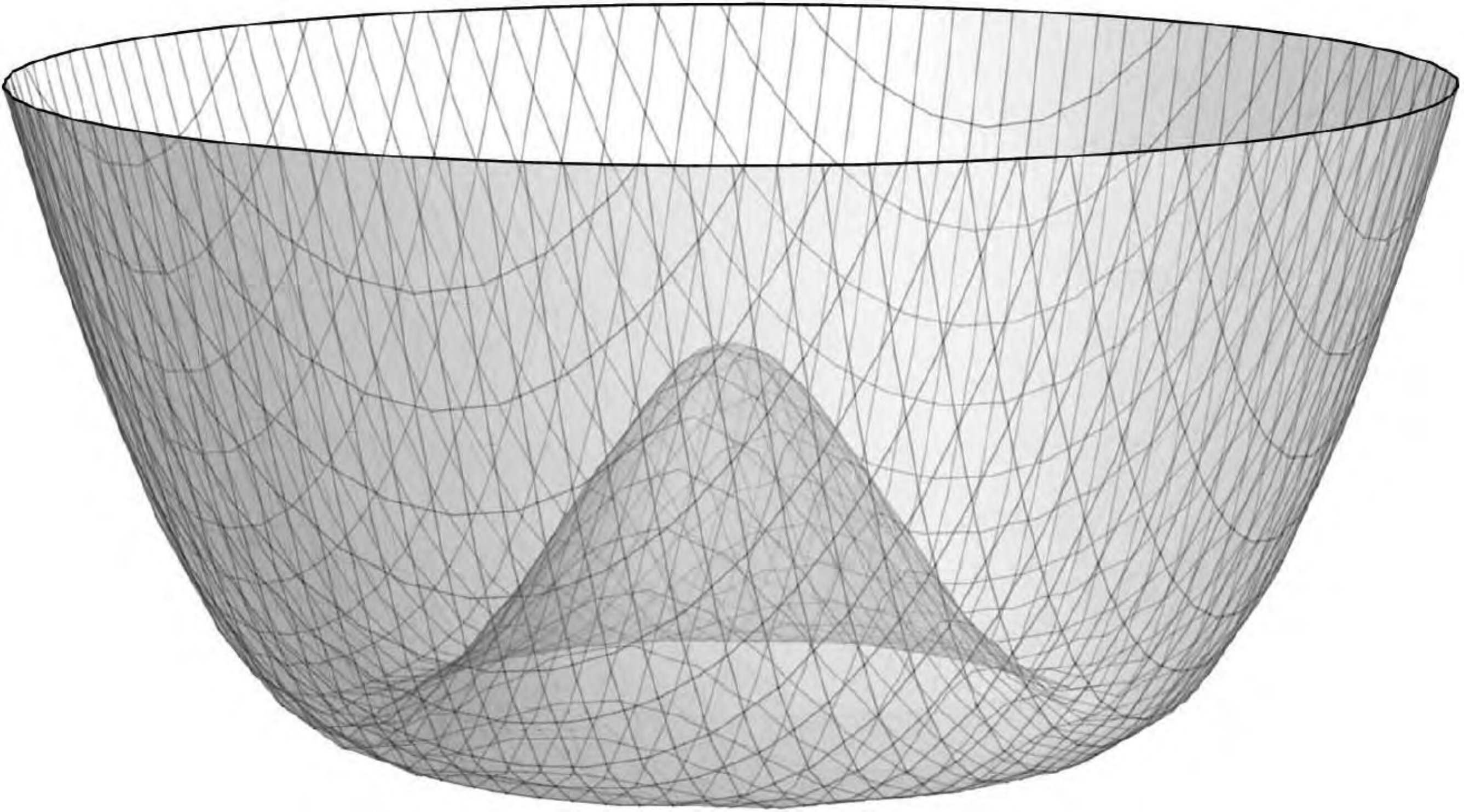}~~
     \includegraphics[width=0.25\columnwidth]{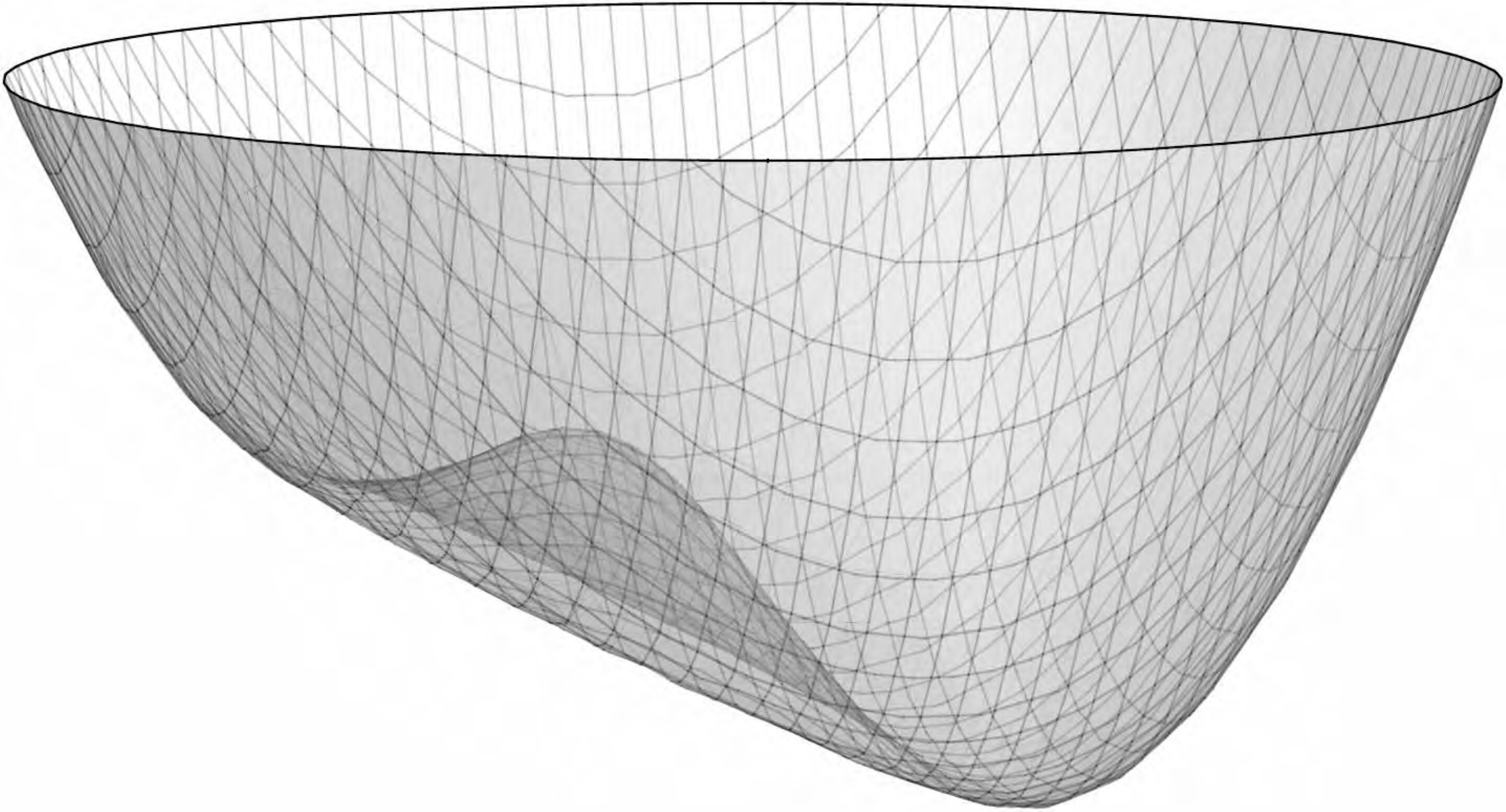}~~
     \includegraphics[width=0.25\columnwidth]{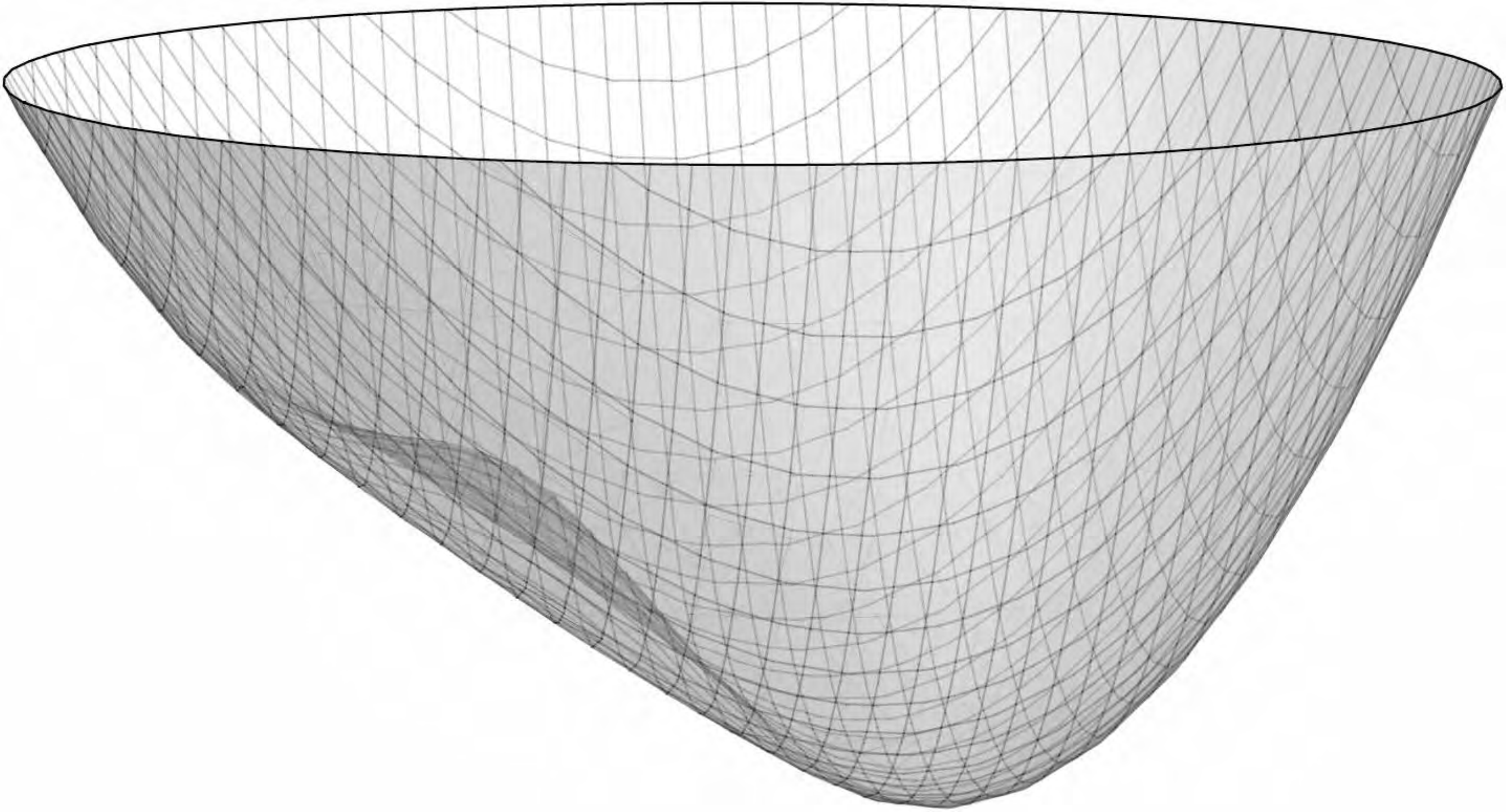}
     \put(-307,70){$m=0$}
     \put(-190,68){$m=0.4$}
     \put(-70,68){$m=1$}
   \end{center}
   \vspace{-.8\baselineskip}
   \caption{
     \label{fg:NJLpotential}
     Snapshots of the action $S(\sigma,\pi)$ in \eqref{eq:Snjl} for varying $m$ 
     at $G=3$ and $p=1$. 
     The number of saddle points jumps from $3$ to $1$ at $m\simeq 0.724$.
   }
   \end{figure}

The dependence of $\{z_i\}_{i=1}^3$ on $G$ and $m$ has already been laid out 
in Section \ref{sc:mGN} (especially in Figure \ref{fg:GNpd}). Its implication for the 
current NJL-like model is as follows. (To avoid complication we will assume $G\in\RR_{>0}$ 
throughout Section \ref{sc:massnjl}.) Roughly speaking, the number of critical points of 
$S(z,w)$ on $\RR^2$ jumps as $m$ is varied at fixed $G$ (see Figure \ref{fg:NJLpotential}), 
and this occurs in parallel with a jump in the number of thimbles contributing to the integral 
\eqref{eq:Z_model1}. Namely, at large $m$ just one thimble contributes, while at small $m$ 
or large $G$ all the three thimbles contribute. 
In the latter case the Stokes phenomenon occurs among the three thimbles, 
which is hard to visualize as it occurs in $\CC^2$.

%%%
%%%
%%%%%%%%%%%%%%%%%%%%%%%%%%%%%%%%%%%%%%%
%%%%%%%%%%%%%%%%%%%%%%%%%%%%%%%%%%%%%%%
%%%
%%%

So far we have stressed similarities between the NJL-like model at $m\ne 0$ and 
the GN-like model. However, the NJL-like model with sufficiently small $m$ 
has a unique feature that is absent in the GN-like model: quasi-fixed points of the flow. 
Namely, the flow along the set $\{(z,w)\in\CC^2\,|\,z^2+w^2=G-p^2\}$ is extremely slow at 
small $m$, owing to the approximate $\O(2,\CC)$ symmetry of the model. 
The nature of this quasi-stationary flow and the behavior of thimbles 
at small $m$ can be most easily revealed in the $p=0$ limit, in which a drastic simplification 
occurs as was shown for the GN-like model in Section \ref{sc:mGN}. Below is a summary of 
the main characteristics of the NJL-like model at $p=0$.
\begin{itemize}
  \item $m=0$:~~
  The thimble $\JJ_0$ in \eqref{eq:JJ0} disappears at $p=0$ as $\JJ(0)\big|_{\rm GN}$ 
  no longer exists; see Figure \ref{fg:GN5}. Therefore the $p=0$ limit of $\JJ_{1}$ 
  is the only thimble. From \eqref{eq:Jatp0} and \eqref{eq:JJ1}, we find 
  $\JJ_{1}=\{(z,w)\in\RR^2\,|\,(z,w)\ne (0,0)\}$, which is just a punctured plane. 
  %%%
  %%%
  \item $m>0$:~~ 
  The $\O(2,\CC)$-symmetric critical manifold is torn down to two critical points 
  at $\displaystyle (z,w)=\bigg(\frac{-m\pm \sqrt{m^2+4G}}{2},~0\bigg)=:(z_\pm,0)$, 
  with the associated thimbles that we call $\JJ_+$ and $\JJ_-$. 
  Since $(z_\pm,0)$ reside in the original integration cycle $\RR^2$, 
  both $\JJ_+$ and $\JJ_-$ contribute to the integral \eqref{eq:Z_model1} with unit coefficients. 
  As $\Re\,S(z_+,0)<\Re\,S(z_-,0)$, $\JJ_+$ gives a dominant contribution at $N\gg 1$.   
\end{itemize}
Thus an arbitrarily small $m\ne 0$ splits the thimble as $\JJ_1\to \JJ_+ \cup \JJ_-$.  
Our goal is to identify $\JJ_{\pm}$ and the quasi-stationary flows on them. 
Let us recall that $\JJ_+$ and $\JJ_-$ are unions of downward flows that 
run into $(z_+,0)$ and $(z_-,0)$ in the $\tau\to+\infty$ limit, respectively. 
The flow equations at $p=0$ read 
\begin{subequations}
  \begin{align}
    \frac{\dd\bar{\scalebox{0.7}{$\mathstrut$} z}}{\dd \tau} 
    & = \frac{2(z+m)}{(z+m)^2+w^2} - \frac{2z}{G}\,,
    \\
    \frac{\dd\bar{\scalebox{0.7}{$\mathstrut$} w}}{\dd \tau} 
    & = \frac{2w}{(z+m)^2+w^2} - \frac{2w}{G}\,. 
  \end{align}
  \label{eq:NJLflow}%
\end{subequations} 
It follows that if a flow starts from $(z,w)\in \RR^2$, 
then the flow stays in $\RR^2$ forever.  
Namely, the condition $\Im\,z=\Im\,w=0$ is conserved along the flow. 
%%
%%
%%%%%%%%%%%%%%%%%%%%%%%%%%%%%%%%%%%%%%%
   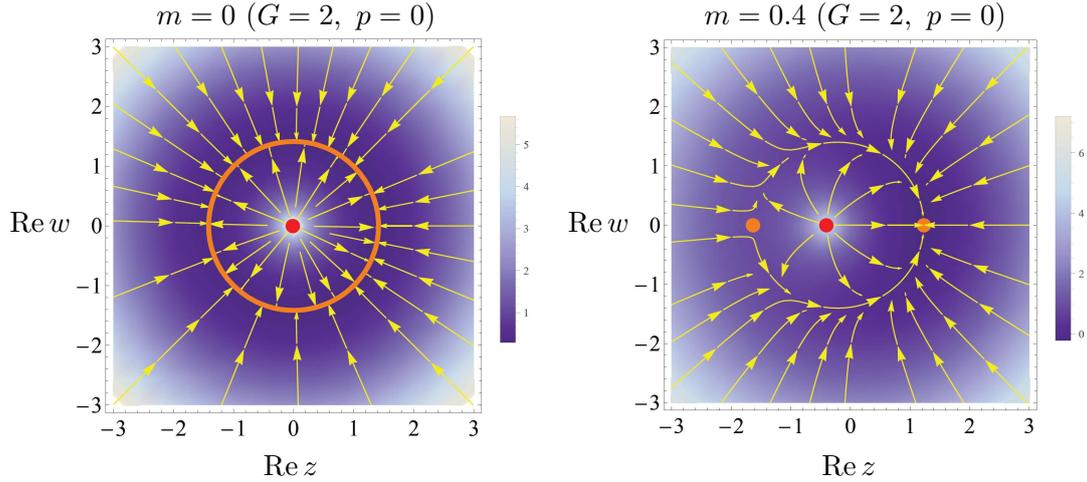
\begin{figure}[!t]
   \begin{center}
     \qquad~ 
     \includegraphics[width=0.4\columnwidth]{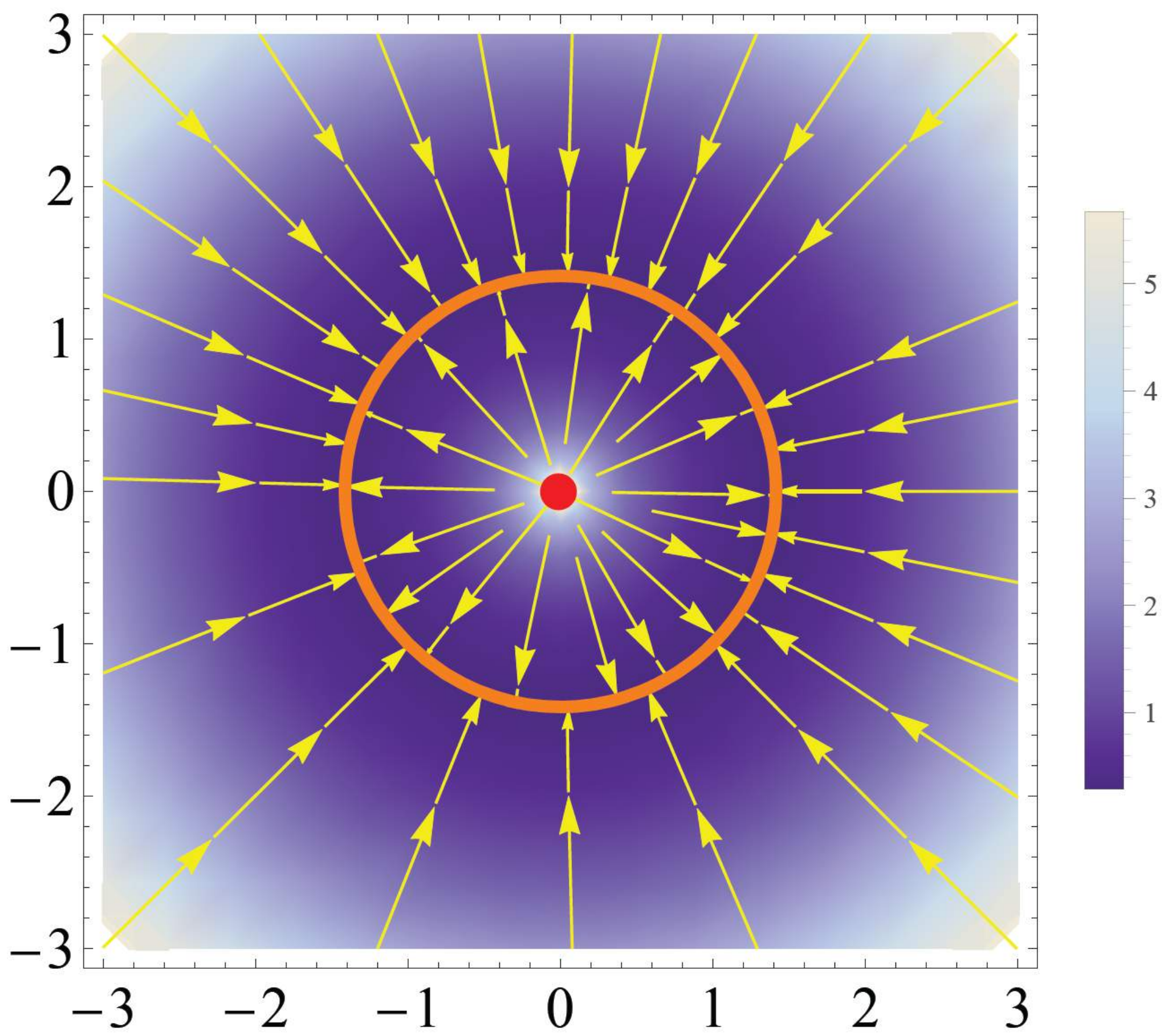}
     \qquad~~
     \includegraphics[width=0.4\columnwidth]{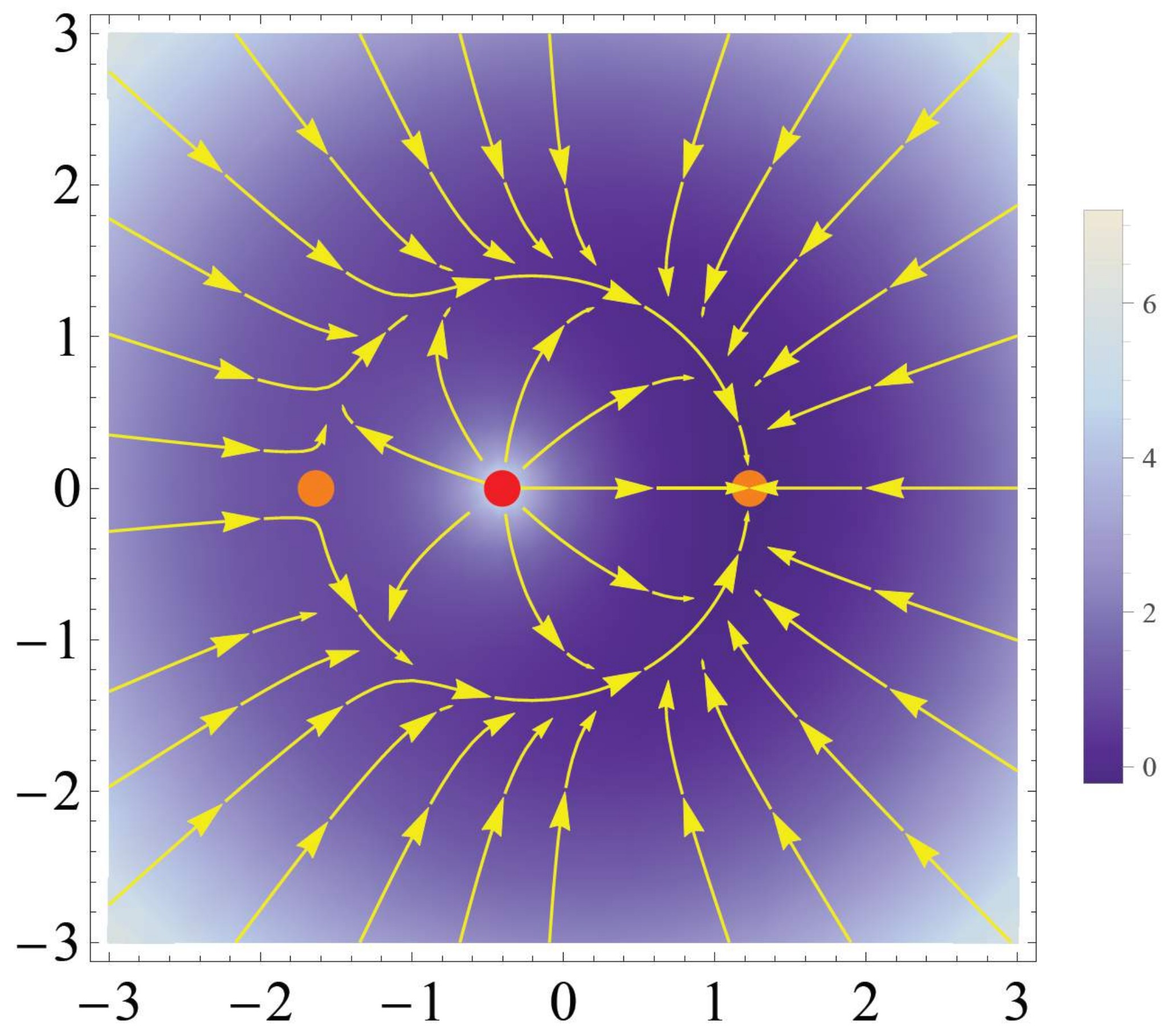}
     \put(-144,156){$m=0.4$ ($G=2,~p=0$)}
     \put(-349,156){$m=0$ ($G=2,~p=0$)}
     \put(-404,78){$\Re\,w$}
     \put(-195,78){$\Re\,w$}
     \put(-309,-14){$\Re\,z$}
     \put(-100,-14){$\Re\,z$}
   \end{center} 
   \vspace{-.6\baselineskip}
   \caption{
     \label{fg:NJL1}
     The downward flow \eqref{eq:NJLflow} of the NJL-like model 
     in a plane with $\Im\,z=\Im\,w\equiv 0$. 
     The background color scale represents $\Re\,S(z,w)$.  
     The red blob at $(-m,0)$ is a singularity of the flow. 
     The orange circle in the left panel and the orange blobs in the right panel 
     are fixed points of the flow. 
   } 
   \end{figure}
%%%%%%%%%%%%%%%%%%%%%%%%%%%%%%%%%%%%%%%
%%
%%
In Figure \ref{fg:NJL1} we show a sketch of the flow for $(z,w)\in\RR^2$ 
at $m=0$ (left panel) and $m>0$ (right panel). At $m=0$, there is a ring of 
fixed points $z^2+w^2=G$ which attracts all flows on the plane. This is exactly 
the compact orbit in Section \ref{sc:NJLnomass}. This $\RR^2$ plane 
(with the origin excised) represents $\JJ_{1}$. 

By contrast, for $m>0$ only two critical points exist, 
see the right panel of Figure \ref{fg:NJL1}. Now there is a slow but 
non-vanishing flow along the orbit $z^2+w^2\simeq G$. It is clear from the figure that 
$(z_+,0)$ attracts all the flows on this plane, except for the real axis to the left of 
$(-m,0)$ on which the flow is attracted to $(z_-,0)$.  Since $\JJ_+$ is by definition 
a union of all flows that sink in $(z_+,0)$, we conclude that the set 
\begin{align}
  \JJJ_A & : = \RR^2\setminus \{  (x,0)\,|\,x \leq -m  \} \subset \CC^2
\end{align} 
belongs to $\JJ_+$. 

\begin{figure}[!t]
   \begin{center}
     \qquad~ 
     \includegraphics[width=0.4\columnwidth]{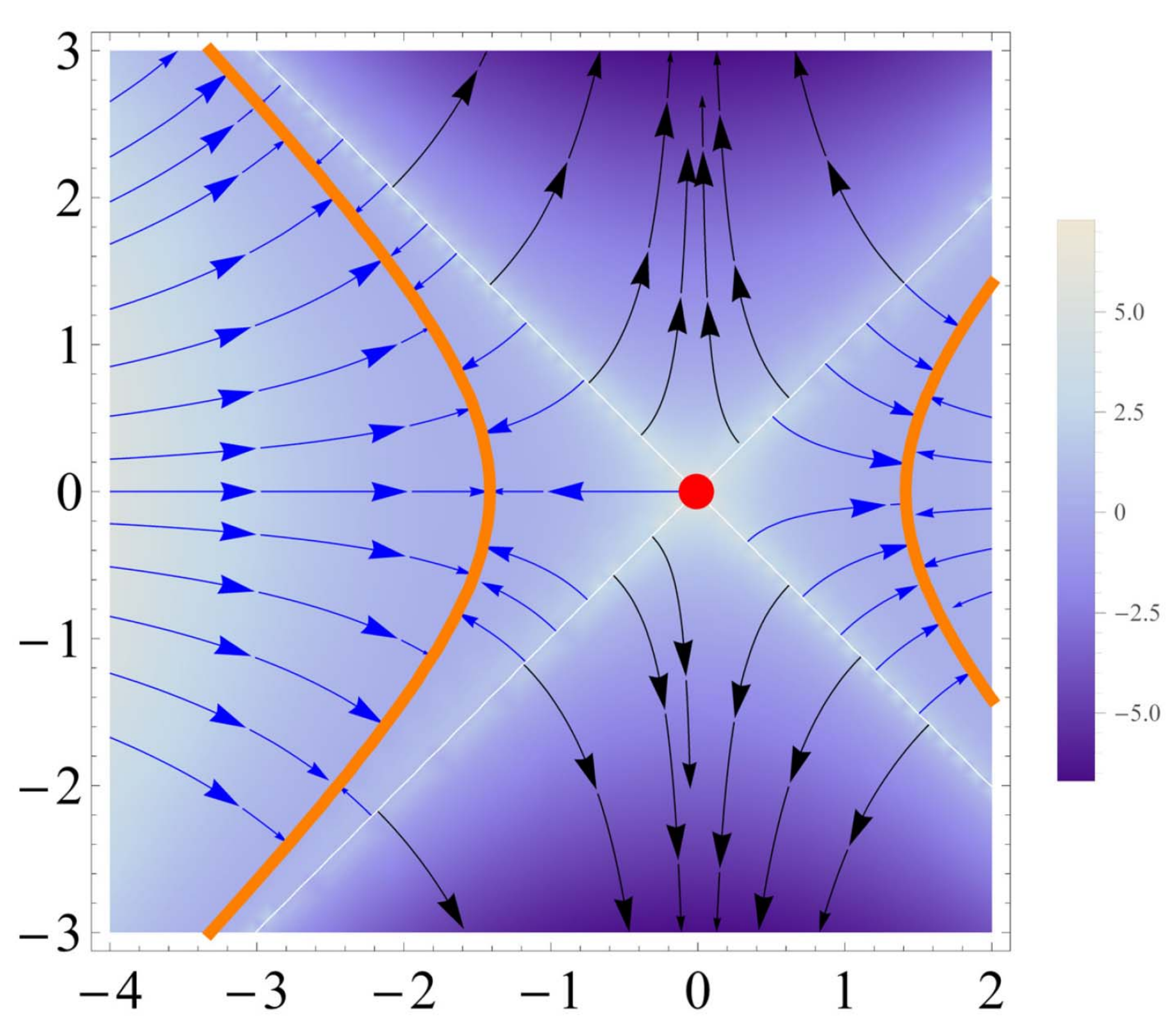}
     \qquad~~
     \includegraphics[width=0.4\columnwidth]{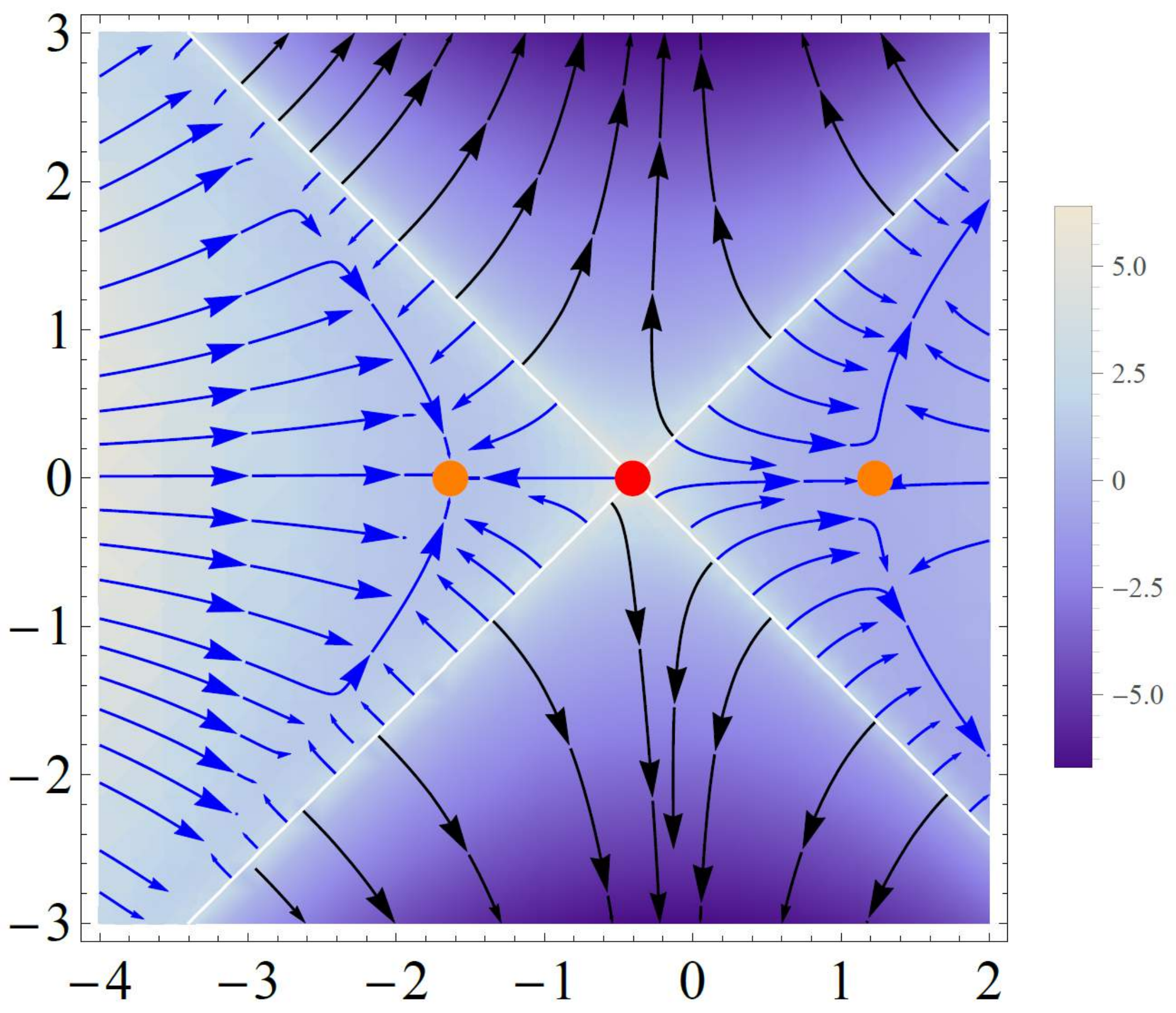}
     \put(-149,152){$m=0.4$ ($G=2,~p=0$)}
     \put(-352,151){$m=0$ ($G=2,~p=0$)}
     \put(-404,78){$\Im\,w$}
     \put(-195,78){$\Im\,w$}
     \put(-311,-14){$\Re\,z$}
     \put(-102,-13){$\Re\,z$}
   \end{center} 
   \vspace{-.6\baselineskip}
   \caption{
     \label{fg:NJL2}
     Same as Figure \ref{fg:NJL1} but with $\Im\,z=\Re\,w\equiv 0$. 
     The X-shaped white lines are points of logarithmic singularity of the action.  
   } 
   \end{figure}
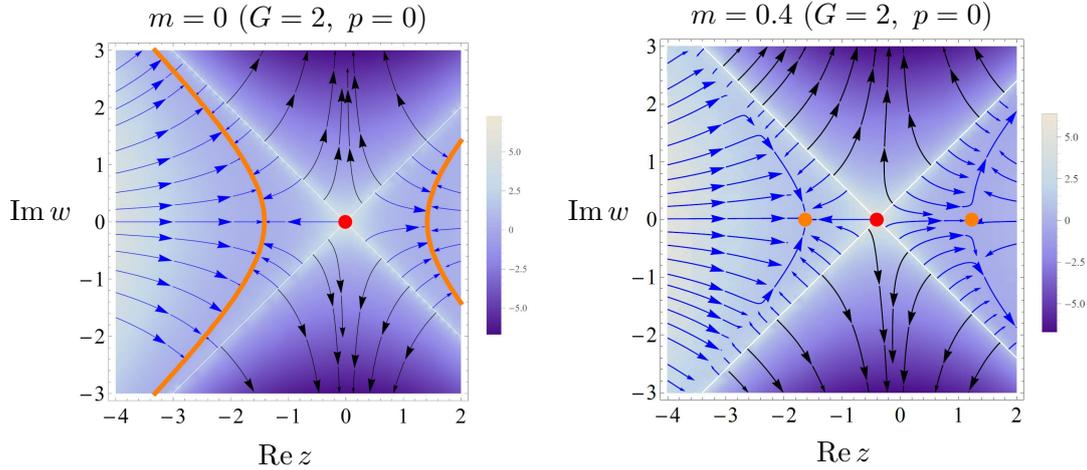

Next, returning to \eqref{eq:NJLflow}, we see that 
the condition $\Im\,z=\Re\,w=0$ is also conserved along the flow. 
Figure \ref{fg:NJL2} shows the flow pattern in the $(\Re\,z,\Im\,w)$ 
plane with $\Im\,z=\Re\,w=0$. 
%%
%%
%%%%%%%%%%%%%%%%%%%%%%%%%%%%%%%%%%%%%%%
%%%%%%%%%%%%%%%%%%%%%%%%%%%%%%%%%%%%%%%
%%
%%
At $m=0$ the fixed points of the flow form a hyperbola $(\Re\,z)^2-(\Im\,w)^2=G$, 
which is the non-compact orbit in Section \ref{sc:NJLnomass}. 
For large $|z|$ and $|w|$ the fixed-point lines asymptote to $\Re\,z = \pm \Im\,w$, 
on which the action diverges logarithmically. 

At $m>0$, a slow but non-vanishing flow appears along the hyperbola 
(see the right panel of Figure \ref{fg:NJL2}). The critical point $(z_-,0)$ 
is an attractive fixed point of all flows in the wedge-shaped region 
\begin{align}
  \JJJ_B:= \{(z,w)\in \CC^2\,|~ \Im\,z=0,~\Re\,w=0,~\Re\,z+m < \Im\,w < -\Re\,z-m \} \,,
\end{align}
which implies that $\JJJ_B$ is part of the Lefschetz thimble $\JJ_-$.%
\footnote{It may seem puzzling that the non-compact orbit, which used to be 
part of the dual cycle ($\KK_1$) at $m=0$, has suddenly become 
part of the Lefschetz thimble at $m>0$. 
This is not problematic because $m>0$ ensures that $\Re\,S\to+\infty$ 
in the far ends of the (vestige of) non-compact orbit passing through $(z_-,0)$. 
However, the increase of $\Re\,S$ is quite slow for small $m$, 
which leads to a subtlety in the $m\to 0$ limit --- see the discussion at the end of 
this section.} 

\begin{figure}[!t]
   \begin{center}
     \qquad~ 
     \includegraphics[width=0.7\columnwidth]{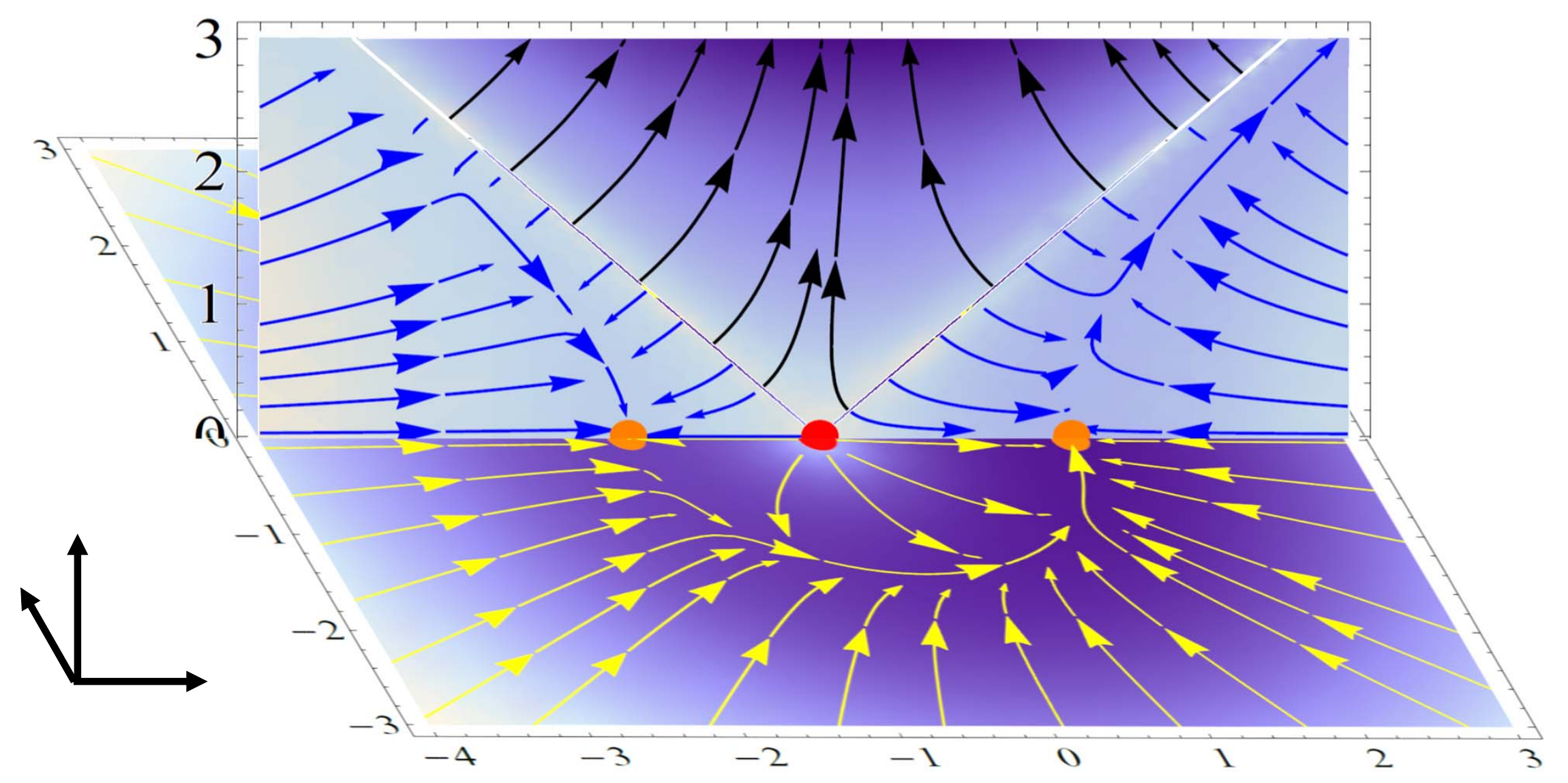}
     \put(-282,5){$\Re\,z$}
     \put(-322,22){$\Re\,w$}
     \put(-296,52){$\Im\,w$}
   \end{center} 
   \vspace{-.6\baselineskip}
   \caption{
     \label{fg:NJLintersect}
     The downward flow \eqref{eq:NJLflow} 
     in a 3-dimensional space with $\Im\,z\equiv 0$. The parameters are 
     $m=0.4$, $G=2$ and $p=0$ 
     (same as for the right panels in Figures \ref{fg:NJL1} and \ref{fg:NJL2}). 
   } 
   \end{figure}

In Figure \ref{fg:NJLintersect} we display a combination of flow plots  
in Figures \ref{fg:NJL1} and \ref{fg:NJL2} at $m=0.4$ within 
a three-dimensional subspace of $\CC^2$ specified by $\Im\,z =0$. 
As the figure shows, $\JJJ_A$ and $\JJJ_B$ intersect orthogonally.  
Such a singular behavior of thimbles is a typical signature of the Stokes 
jump --- $G=2$ is exactly on the Stokes ray. 
Although the singularity of thimbles can be 
smoothed out through complexification of $G$ as $G\ee^{i\theta}$, it makes 
the thimbles extend into the whole $\CC^2$ space and impedes our 
visual understanding of the Stokes jump.

%%
%%
%%%%%%%%%%%%%%%%%%%%%%%%%%%%%%%%%%%%%%%
%%%%%%%%%%%%%%%%%%%%%%%%%%%%%%%%%%%%%%%
%%
%%

To keep the discussion as simple as possible, we shall exploit 
a somewhat indirect argument to deduce the behavior of thimbles as 
$\theta\to 0^\pm$. First of all, since both $\JJ_+$ and $\JJ_-$ contribute 
to the partition function with unit coefficients,  the sum of  
$\JJ_+$ and $\JJ_-$ must be homologous to the original integration cycle, $\RR^2$.  
For this reason one cannot have $\JJ_+=\JJJ_A$ and $\JJ_-=\JJJ_B$, as 
$\JJJ_A+\JJJ_B\supsetneqq \RR^2$. Secondly, recall that 
a Lefschetz thimble associated with a least dominant critical point do not jump 
at $\theta=0$, while a Lefschetz thimble associated with the dominant critical point 
can jump with a multiple of subdominant thimbles (cf.~Section \ref{sc:GNmassless} 
and \eqref{eq:JJJ}). 
This suggests that $\JJ_+$ jumps with a multiple of $\JJ_-$ at $\theta=0^\pm$ 
whereas $\JJ_-$ does not jump at all. These considerations indicate that we have 
\begin{align}
  \begin{cases}\JJ_+=\JJJ_A \pm \JJJ_B \\ \JJ_-=\JJJ_B\end{cases} 
  \text{at}~~ \theta=0^- \quad \Rightarrow \quad 
  \begin{cases}\JJ_+=\JJJ_A \mp \JJJ_B \\ \JJ_-=\JJJ_B\end{cases} 
  \text{at}~~ \theta=0^+ \,, 
  \label{eq:NJLjump}
\end{align}
where the sign in front of $\JJJ_B$ depends on one's choice of orientation 
of the thimble. Taking the upper sign, one finds that the thimble decomposition 
of the integration cycle also changes discontinuously 
from $\RR^2=\JJ_+-\JJ_-$ at $\theta=0^-$ 
to $\RR^2=\JJ_++\JJ_-$ at $\theta=0^+$.  
It is important that the contribution of $\JJJ_B$ is cancelled exactly between 
$\JJ_+$ and $\JJ_-$, thus ensuring constancy of the integral at $\theta=0$. 
Moreover, this cancellation is necessary to keep the partition function real, 
because the integral over $\JJJ_B$ is purely imaginary (see below). 
This case highlights why it is imperative to sum up 
contributions of all thimbles in order to obtain a correct result.

It is intriguing to ask what happens near the chiral limit. 
As $m\to 0^+$, $\JJJ_A$ converges to 
$\JJ_{1}=\{(z,w)\in\RR^2\,|\,(z,w)\ne (0,0)\}$, 
which is the thimble at $m=0$. However, according to \eqref{eq:NJLjump}, 
neither $\JJ_+$ nor $\JJ_-$ converges to $\JJ_{1}$ 
in the chiral limit because of the presence of $\JJJ_B$. 
Therefore we conclude that the description of the integral 
based on Lefschetz thimbles is necessarily discontinuous between 
$m>0$ and $m=0$.  A similar observation is made in \cite{Tanizaki:2014tua} 
for an $\O(2)$-symmetric bosonic model.  A unique feature of 
the fermionic model in comparison to bosonic models is that 
one of the thimbles ($\JJ_-$) is framed by lines of logarithmic singularity.  

Near the chiral limit $(m\ll \sqrt{G})$, a quasi-stationary flow along the orbit 
$z^2+w^2\simeq G$ emerges (see Figures \ref{fg:NJL1} and \ref{fg:NJL2}), 
reflecting the $\O(2,\CC)$-symmetric degenerate minima of the action 
at $m=0$. The contribution from the vicinity of the compact orbit 
to the integral over $\JJJ_A$ is $\calO(1)$ and harmless in the chiral limit, 
since it is a compact domain. By contrast, 
the non-compact orbit on $\JJJ_B$ (Figure \ref{fg:NJL2}) extends to infinity and 
can give a large contribution near the chiral limit. 
More explicitly, we find for the contribution from $\JJJ_B$ 
\begin{align}
  Z\big|_{\JJJ_B} = ~ 
  & \frac{N}{\pi G}
  \int_{\JJJ_B}\dd z \dd w~\ckakko{(z+m)^2+w^2}^N \ee^{-\frac{N}{G}(z^2+w^2)}
  \\
  = ~ & i \frac{N}{\pi G} \int_0^\infty \dd R\,R^{2N+1} \int_{-\infty}^{\infty}\dd\Phi~
  \ee^{-\frac{N}{G}(m^2+2mR\cosh\Phi+R^2)}
  \\
  =~ & 2 i \frac{N}{\pi G}\ee^{-\frac{N}{G}m^2} \int_0^\infty \dd R\,R^{2N+1} 
  \ee^{-\frac{N}{G}R^2} K_0\Big(\frac{2NR}{G}m\Big)\,,
\end{align}
where we changed the variables as $z=-m-R\cosh\Phi$ and $w=-iR \sinh \Phi$. 
Using $\displaystyle K_0(x)=-\Big(\log \frac{x}{2}+\gamma\Big)I_0(x)+\calO(x^2)$ 
for $x\ll 1$ \cite{AbramowitzStegun}, we conclude that 
$Z\big|_{\JJJ_B}$ blows up as $\sim \log m$ in the chiral limit. 
As $\JJ_+$ and $\JJ_-$ include $\JJJ_B$ (cf.~\eqref{eq:NJLjump}), this 
infrared divergence shows up in the integral on each thimble, but it cancels out 
completely when the sum of both contributions is correctly taken. We suspect that 
this is a generic phenomenon that occurs when an explicitly broken continuous symmetry is handled 
with Picard-Lefschetz theory. 
It will be interesting to look for a similar behavior of thimbles for models in higher dimensions.

\section{Chern-Simons-like theory with fermions}
\label{sc:CSm}
\subsection{Model setup}

In Sections \ref{sc:gross} and \ref{sc:NJLm} we have studied Lefschetz thimbles 
in zero-dimensional models with chiral symmetry, in which the integration measure 
incurred a complex phase through a complex-valued interaction coupling. 
To gain more insights into complex path integral, 
it would be worthwhile to look into another model where a complex measure 
arises from an entirely different mechanism. 
In this section, we elaborate on Lefschetz thimbles in an exactly solvable $0+1$-dimensional 
Chern-Simons(CS) theory with fermions \cite{Dunne:1996yb,Das:1997gga}.%
\footnote{A bosonic version of this model was studied in \cite{Alexanian:2008kd}.} 
Despite its simplicity, the model captures essential aspects of three-dimensional CS gauge theory:   
it has a CS-like topological term in the action with an integer-quantized coefficient, which makes 
the integration measure complex and renders Monte Carlo techniques ineffective. Nonetheless 
one can find a viable integration contour on the complex plane by following the recipe of 
Picard-Lefschetz theory. 

The model we consider is an Abelian gauge theory on $S^1$, defined by
\begin{align}
  Z_N(k) & \equiv \int \dd A \dd\bar\psi \dd\psi \ 
  \exp\mkakko{
    \int_0^\beta \!\! \dd \tau \bigg\{
    \sum_{a=1}^{N}\bar\psi_a \big[\sigma_1(\der_\tau+iA)+m \big] \psi_a 
    + ik A \bigg\}
  } \,,
  \label{eq:Z_CS}
\end{align}
where $\psi_a(\tau)$ is a fermion field with $N$ flavors, 
$A(\tau)$ is a $\U(1)$ gauge field, $m>0$ is a mass term and $\sigma_1$ is a Pauli matrix. 
A periodic (anti-periodic) boundary condition is imposed on $A$ ($\psi_a$) along $S^1$, 
respectively. Under a $\U(1)$ gauge transformation 
($\psi_a\to \ee^{i\phi}\psi_a$ and $A\to A-\der_\tau \phi$), 
the fermion action is invariant, whereas the CS term transforms as
\begin{align}
  ik \int_0^\beta \!\! \dd\tau \,A \to ik \int_0^\beta \!\! \dd\tau 
  \mkakko{A-\der_\tau\phi} 
  = ik \int_0^\beta \!\! \dd\tau \,A + ik [\phi(0) - \phi(\beta)]\,.
\end{align} 
In general, $\phi(0)-\phi(\beta)$ has to be a multiple of $2\pi$ to ensure 
the single-valuedness 
of $\psi$. The measure in \eqref{eq:Z_CS} would then be invariant if $k$ is an integer. 
This is an analog of the quantization condition for the 
CS term in three-dimensional gauge theory.   

Now, fixing the gauge as $\der_\tau A=0$, $A$ becomes constant along $S^1$. 
By a suitable ``non-trivial'' gauge transformation of the form 
$\phi(\tau)=2\pi n\tau/\beta$ with $n\in\ZZ$,  one can bring any $A$ to 
the range $[0,2\pi/\beta]$, hence it is enough to integrate over $A$ in this range. 
After integrating out fermions, one gets
\begin{align}
  Z_N(k) & = \int_0^{2\pi/\beta} \!\!\!\!\! \dd A~
  {\det}^{N}\big[\sigma_1(\der_\tau+iA)+m \big] 
  \ee^{ik\beta A}
  \\
  & = \int_0^{2\pi/\beta} \!\!\!\!\! \dd A \prod_{n=-\infty}^{\infty}
  \ckakko{ (\omega_n+A)^2+m^2 }^N \ee^{ik\beta A} 
  \qquad [ \,\omega_n\equiv(2n+1)\pi/\beta\, ] \,. 
\end{align}
Note that while the integral measure at $k=0$ is positive definite, 
the CS term causes a complex phase problem at $k\ne0$. Now, we regularize 
the divergent infinite product by normalizing to a fermion determinant at $A=m=0$: 
\begin{align}
  Z_N(k) & = \int_0^{2\pi/\beta} \!\!\!\!\! \dd A \prod_{n=-\infty}^{\infty}
  \ckakko{ 
    \frac{ (\omega_n+A)^2+m^2 }{\omega_n^2} 
  }^N   \! 
  \ee^{ik\beta A} 
  \\ 
  & = \frac{1}{2^N} \int_0^{2\pi/\beta} \!\!\!\!\! \dd A 
  \mkakko{ \cos \beta A+\cosh \beta m }^N \!
  \ee^{ik\beta A} 
  \\
  & = \frac{1}{2^N} \int_0^{2\pi} \!\!\! \dd A~\ee^{-N S(\kk,A)} \,,
  \label{eq:Z_S_CS}
\end{align}
where in the last step we set $\beta=1$ for simplicity, and defined the ``effective action'' 
\begin{align}
  S(\kk, A) & \equiv - \log \mkakko{ \cos A+\cosh m } - i \kk A  \qquad 
  \text{with}\quad \kk\equiv \frac{k}{N}\,. 
  \label{eq:Saction}
\end{align}
In contrast to the GN-like and NJL-like models, the present model 
\eqref{eq:Z_S_CS} has a \emph{compact} domain of integration, 
which is reminiscent of a link variable in $\U(1)$ lattice gauge theory.  
At $N\gg 1$, the asymptotic behavior of the integral \eqref{eq:Z_S_CS} is dominated by 
critical points of $S(\kk, A)$, which are away from the real axis when $\kk\ne 0$ 
as we will see in Section \ref{sc:CSsaddles}. This is therefore a good testbed 
to see the importance of complex saddles. 

When $k\in\ZZ$, one can switch to another variable $z\equiv \ee^{iA}$ 
and rewrite \eqref{eq:Z_S_CS} as  
\begin{align}
  Z_N(k) & = -\frac{i}{2^{2N}} \oint_{|z|=1} \dd z~z^{k-1} \mkakko{z+\frac{1}{z}+2\cosh m}^N \,,
  \label{eq:Zvanish}
\end{align}
whose value is determined by the residue of $1/z$ in the integrand. It then follows  
that $Z_N(k)$ is identically zero for all $k\in\mathbb{Z}$ such that $|k|>N$. 
This peculiar feature suggests that the behavior 
of Lefschetz thimbles will be qualitatively different for $|\kk|>1$ and $|\kk|<1$, which 
will be confirmed in Sections \ref{sc:CSlefschetz1} and \ref{sc:CSlefschetz2}.

\subsection{Choice of the integration cycle}

The original integral domain $[0,2\pi]$ of $A$ is an interval with boundaries. 
In order to express the integration cycle as a sum of Lefschetz thimbles, 
we should replace $[0,2\pi]$ with 
an element of the relative homology $H_1(\CC, \CC^{T}; \ZZ)$ for very large $T$, with 
$\CC^{T}:=\{ A\in\CC\,|\,\Re\,S(\kk,A)\geq T \}$. 
In short, we demand that $\Re\,S\to +\infty$ at the ends of a legitimate integration cycle. 

\begin{figure}[!t]
   \begin{center}
     \includegraphics[width=0.45\columnwidth]{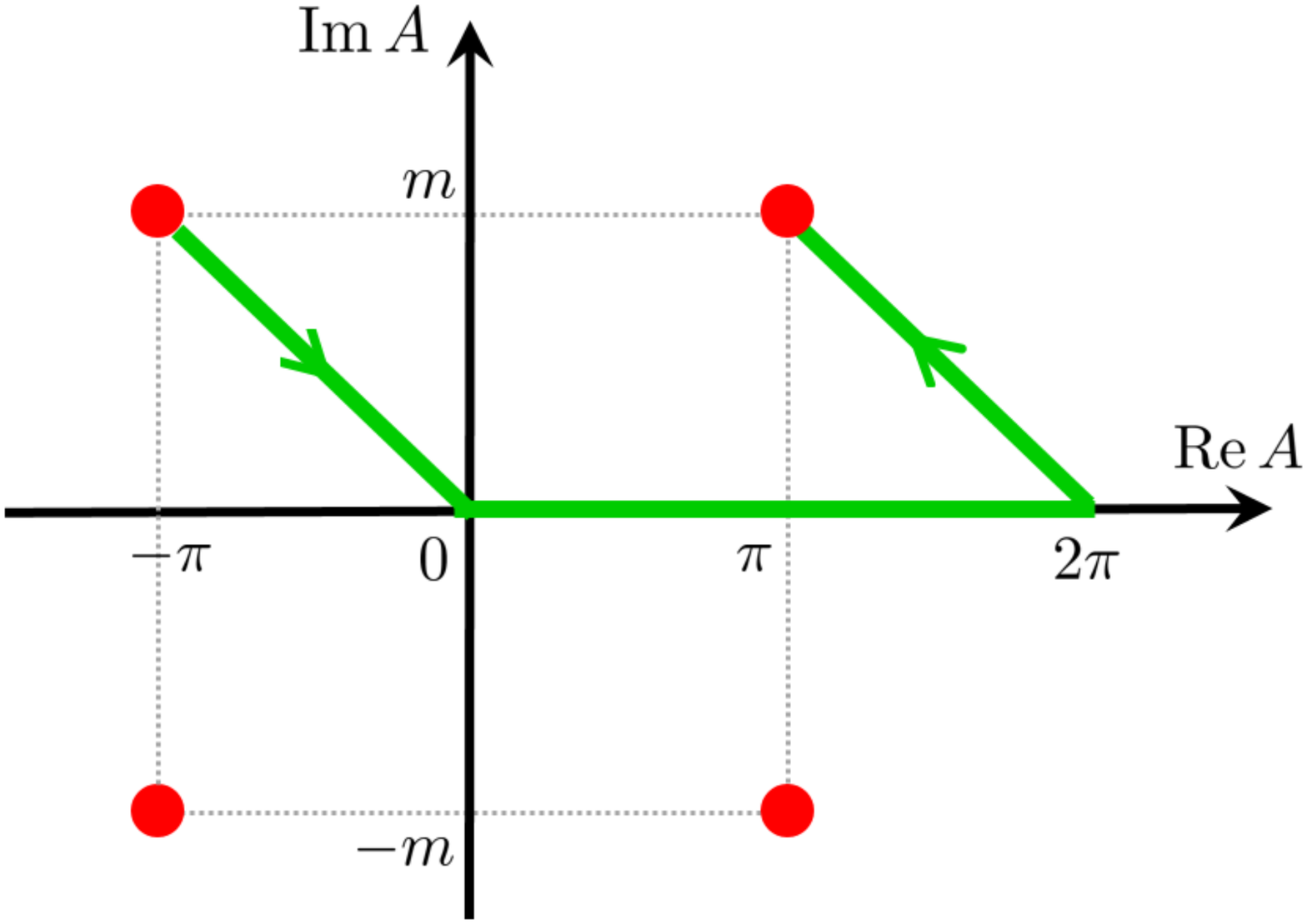}
     \put(-96,69){{\Large $\mathcal{C}$}}
   \end{center} 
   \vspace{-1.3\baselineskip}
   \caption{
     \label{fg:newcontour}
     A new integration cycle $\mathcal{C}$ for $Z_N(k)$. 
     Red blobs are zeros of the fermion determinant. 
   } 
   \end{figure}

For this we need to know where $\Re\,S\to +\infty$ on the complex $A$-plane. 
From \eqref{eq:Saction}, $S(\kk,A)$ clearly diverges if $\cos A+\cosh m=0$.  
This condition can be solved explicitly and yields an infinite number of singularities on the 
complex $A$-plane:
\begin{align}
  A = \pm i m + (2n+1)\pi \,, \qquad n\in \ZZ\,. 
  \label{eq:A_singu}
\end{align}
Then any contour emanating from and ending at these points 
gives an element of $H_1(\CC, \CC^{T}; \ZZ)$. 
To form such a contour we add two edges to the interval $[0,2\pi]$ 
as shown in Figure \ref{fg:newcontour} and define 
the integration contour $\mathcal{C}$ of $Z_N(k)$  
as the union of all three edges. (Note that this definition does not depend on $k$.) 
%%%%%%%%%%%%%%%%%%%%%%%%%%%%%%%%%%%%%%%
%%%%%%%%%%%%%%%%%%%%%%%%%%%%%%%%%%%%%%%
Namely 
\begin{align}
  Z_N(k) & = \frac{1}{2^N}\int_{\mathcal{C}}\dd A~\ee^{-NS(\kk,A)} \,. 
  \label{eq:ZNanalytic}
\end{align}
It can be easily checked that contributions from the appended edges precisely cancel 
if $k\in\ZZ$, ensuring that $Z_N(k)$ thus defined coincides with the original one 
\eqref{eq:Z_S_CS} when $k\in\ZZ$. Since $\mathcal{C}$ is an element of 
$H_1(\CC, \CC^{T}; \ZZ)$ one can express it as a sum of Lefschetz thimbles, 
as will be described later. 

The reader may feel that the above choice of $\mathcal{C}$ is rather \emph{ad hoc}: 
indeed Figure \ref{fg:newcontour} depicts just one of infinitely many ways 
to make the interval $[0,2\pi]$ into a closed cycle in $H_1(\CC, \CC^{T}; \ZZ)$. 
While different cycles will have different expressions as a sum of Lefschetz thimbles, 
the partition function itself is identical as long as the contributions from appended edges 
cancel exactly for all $k\in\ZZ$. Hence we will not consider contours other than 
the one in Figure \ref{fg:newcontour} in the following. 

As a side remark, we note that \eqref{eq:ZNanalytic} is actually giving an analytic 
continuation of $Z_N(k)$ from $k\in\ZZ$ to $k\in \CC$.  When $k\not\in\ZZ$, 
the integrals over added edges no longer cancel, implying that different choices 
of $\mathcal{C}$ provide different analytic continuations of $Z_N(k)$.  Such non-uniqueness 
of analytic continuation from $\ZZ$ to $\RR$ or $\CC$ has been well known 
in the literature of replica trick in random matrix theory 
\cite{Verbaarschot:1985qx,Zirnbauer1999}.

\subsection{Critical points}
\label{sc:CSsaddles}

In understanding the asymptotic behavior of $Z_N(k)$ for $N\gg 1$, it is crucial 
to know which critical point of $S(\kk,A)$ 
contributes to the integral \eqref{eq:ZNanalytic} and which does not. 
For simplicity, we will henceforth assume that $\kk$ is real. 
The saddle point equation of $S(\kk,A)$ reads as
\begin{align}
  0=\frac{\der S(\kk,A)}{\der A} = \frac{\sin A}{\cos A+\cosh m} - i \kk\,. 
  \label{eq:saddleCS}
\end{align}
Solutions are given by
\begin{align}
  A_{\pm} & = 
  - i \log \frac{-\kk \cosh m \pm \sqrt{1+\kk^2\sinh^2 m}}{1+\kk} + 2n\pi 
  \qquad \text{with}~~n\in \ZZ\,. 
\end{align}
The choice of the branch of the logarithm is immaterial when $k\in \ZZ$. 
Their limiting behaviors are as follows. 
\begin{itemize}
  \item $\kk\to \infty$:~ $A_+\to im+(2n+1)\pi$ and $A_-\to -im+(2n+1)\pi$. 
  Thus they merge with the singular points \eqref{eq:A_singu}. 
  \item $\kk\to 1$:~ $A_+\to +i\infty$ with $A_-\sim \calO(1)$.
  \item $\kk=0$:~ $A_+=2n\pi$ and $A_-=(2n+1)\pi$. 
  \item $\kk\to -1$:~ $A_+\to -i\infty$ with $A_-\sim \calO(1)$.
  \item $\kk\to -\infty$:~ $A_+\to -im+(2n+1)\pi$ and $A_-\to im+(2n+1)\pi$. 
  Again, they merge with the singular points \eqref{eq:A_singu}. 
\end{itemize}
It readily follows from \eqref{eq:saddleCS} that the critical points at $\kk<0$ 
and $\kk>0$ are related by complex conjugation.

\subsection[Lefschetz thimbles for $|\kk|<1$]{\boldmath Lefschetz thimbles for $|\kk|<1$}
\label{sc:CSlefschetz1}

Let us begin with the case $\kk=0$. The critical points are 
given by $A=n\pi$ with $n\in\ZZ$ as noted above. 
For each critical point $n\pi$ there is an upward flow line $\KK(n\pi)$ 
and a downward flow line $\JJ(n\pi)$. 
However, $S(0,n\pi)=-\log\mkakko{(-1)^n+\cosh m}$ 
is real for all $n\in\ZZ$, implying that a Stokes phenomenon occurs at $\kk=0$. 
To make the thimbles well-defined, $\kk$ must be slightly shifted away from $0$.  
Although critical points at small $\kk\ne 0$ slightly deviate from $n\pi$, we continue 
calling the flow lines as $\JJ(n\pi)$ and $\KK(n\pi)$ for simplicity of exposition. 

%%%%%%%%%%%%%%%%%%%%%%%%%%%%%%%%%%%%%%%
   \begin{figure}[!t]
   \begin{center}
     \includegraphics[width=0.48\columnwidth]{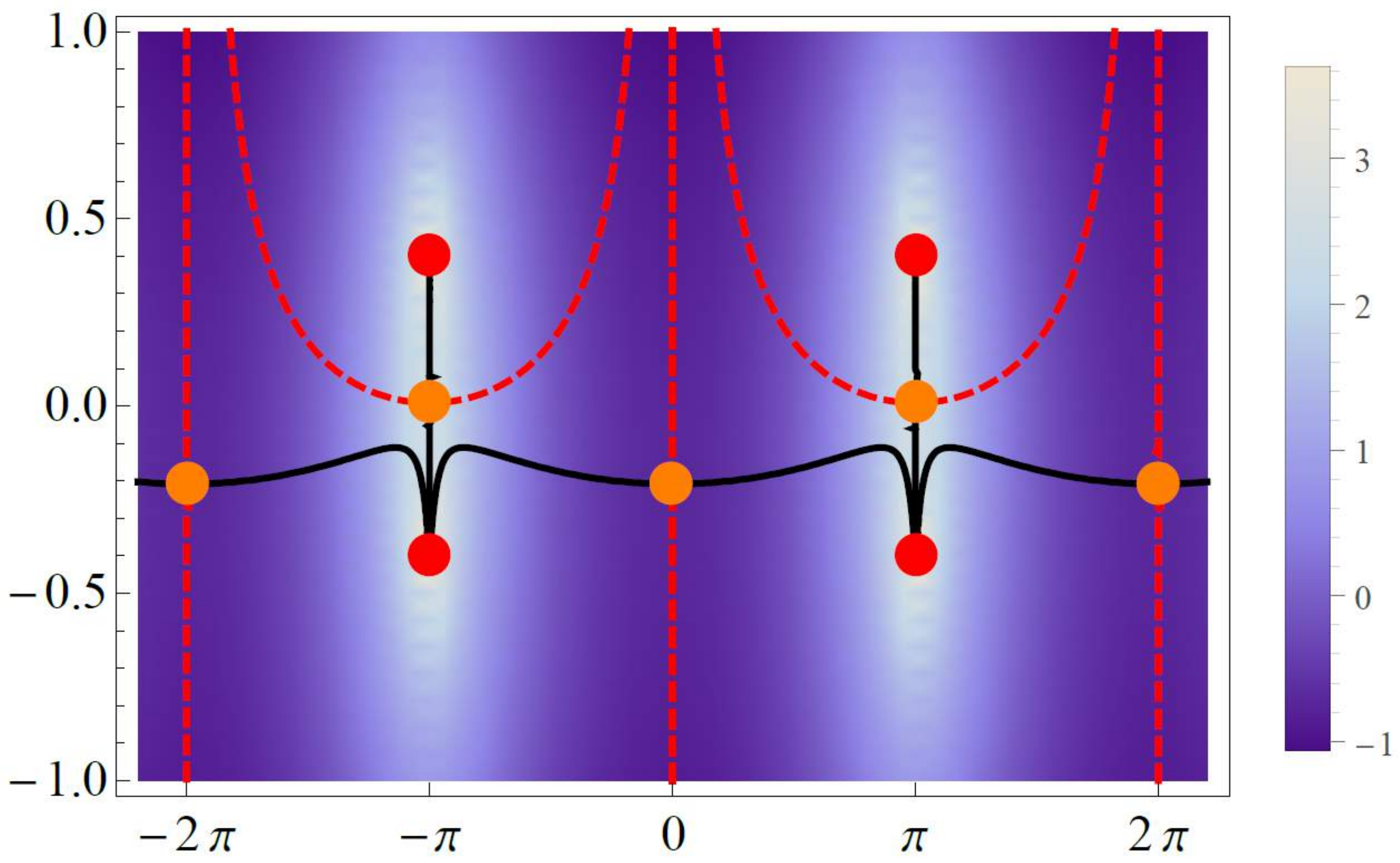}
    ~
     \includegraphics[width=0.48\columnwidth]{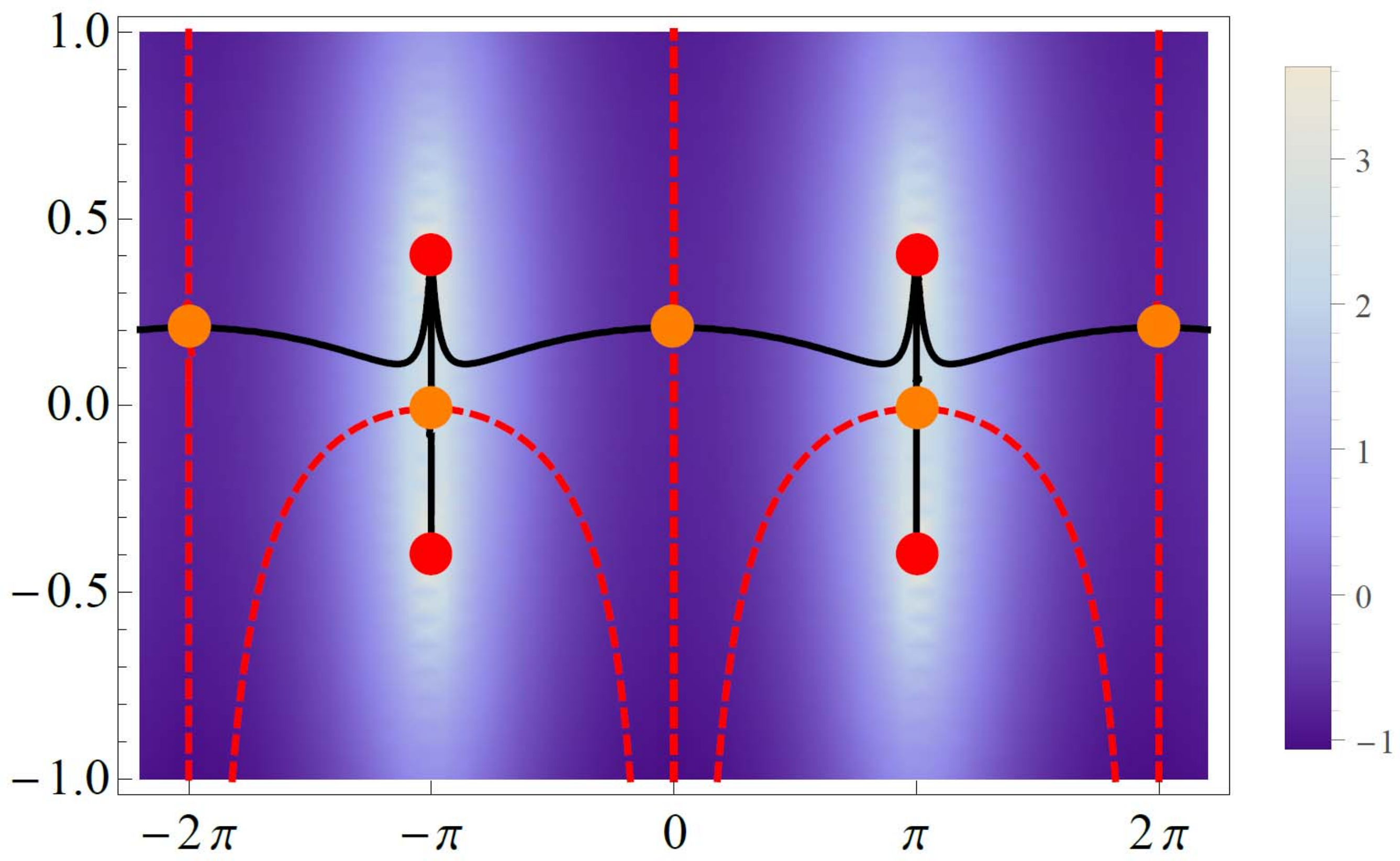}
     \put(-148,130){$\kk=0.1,~m=0.4$}
     \put(-369,130){$\kk=-0.1,~m=0.4$}
   \end{center} 
   \vspace{-.8\baselineskip}
   \caption{
     \label{fg:CS_smallk}
     Lefschetz thimbles (black lines) and their duals (red dashed lines) 
     with $\kk=\pm 0.1$ in the complex $A$-plane. 
     The background color scale represents $\Re\,S(\kk,A)$. 
   } 
   \end{figure}
%%%%%%%%%%%%%%%%%%%%%%%%%%%%%%%%%%%%%%%
In Figure \ref{fg:CS_smallk} we show flow lines for $\kk=-0.1$ (left panel) 
and $\kk=+0.1$ (right panel). As anticipated, a Stokes jump across $\kk=0$ is observed for  
$\JJ(2n\pi)$ and $\KK((2n+1)\pi)$, but not for $\JJ((2n+1)\pi)$ and $\KK(2n\pi)$, 
which may be attributed to $S(0,2n\pi) < S(0,(2n+1)\pi)$ (cf.~Section \ref{sc:GNmassless}). 
Notice also that $\KK(n\pi)$ can run to infinity whereas $\JJ(n\pi)$ cannot, due to the fact 
that the region $|\Im\,A|\gg 1$ is a  ``bad'' region with $\Re\,S(\kk,A)\ll -1$. 
The fact that the flow lines for $\kk>0$ and $\kk<0$ are simply related by complex 
conjugation can be understood from the flow equation%
\footnote{One should not confuse $\tau$ in \eqref{eq:CSfloweq} with the coordinate 
of $S^1$ in \eqref{eq:Z_CS}.}
\begin{align}
  \frac{\dd \bar{A}}{\dd \tau} = \frac{\sin A}{\cos A+\cosh m}-i\kk\,,
  \label{eq:CSfloweq}
\end{align}
in which the complex conjugation $A\leftrightarrow \bar{A}$ is equivalent to 
$\kk\leftrightarrow -\kk$. 
By multiplying $(-1)$ to the both sides with the complex conjugation, 
we can also observe invariance of the flow equation under $A\leftrightarrow -\bar{A}$, 
and it explains the reflection symmetry of Figure \ref{fg:CS_smallk} 
about the vertical axis for real $\kk$. 

For later use, we fix the orientation of thimbles as the increasing direction of $\Im\,A$ 
for $\JJ((2n+1)\pi)$ and as the increasing direction of $\Re\,A$ for $\JJ(2n\pi)$. 

Now we turn to the question of how the cycle $\mathcal{C}$ 
in Figure \ref{fg:newcontour} can be expressed as a sum 
of Lefschetz thimbles. This can be done by counting the intersection of $\mathcal{C}$ 
with upward flow lines from each critical point in Figure \ref{fg:CS_smallk}. 
The result is $\mathcal{C}=-\JJ(-\pi)+\JJ(0)+\JJ(\pi)$ for $\kk=-0.1$ and 
$\mathcal{C}=\JJ(0)$ for $\kk=+0.1$. In the first case there are three contributing 
thimbles; however, $\JJ(-\pi)$ and $\JJ(\pi)$ are $2\pi$ apart and their contributions 
cancel exactly when $k\in\ZZ$. Hence the partition function is solely given by 
the integral over $\JJ(0)$, both for $\kk=-0.1$ and for $\kk=+0.1$.  
With $\Im\,S$ constant along $\JJ(0)$, the complex phase problem has gone away.   

When $k\not \in \ZZ$, the integrals over $\JJ(-\pi)$ and $\JJ(\pi)$ 
no longer cancel exactly. To leading order at large $N$, the sum of 
their contributions is given by 
\begin{align}
  Z_N(k)\Big|_{-\JJ(-\pi)+\JJ(\pi)} & \sim \frac{1}{2^N}
  (\ee^{2\pi i k}-1) \ee^{-N S(\kk,A_\star)} \,, 
\end{align} 
where $A_\star\equiv-\pi-i\log 
\frac{\kk \cosh m + \sqrt{1+\kk^2\sinh^2 m}}{1+\kk}$ 
is the critical point located near $-\pi$. 

The description of Lefschetz thimbles for $\kk=\pm 0.1$ in this subsection needs  
no qualitative modification for $-1<\kk<1$. However the situation drastically changes 
for $\kk$ outside this domain.

\subsection[Lefschetz thimbles for $|\kk|>1$]{\boldmath Lefschetz thimbles for $|\kk|>1$}
\label{sc:CSlefschetz2}

As one approaches $\kk=1$ from below, nothing dramatic happens for the critical points 
near $(2n+1)\pi$, whereas the critical points with real parts equal to $2n\pi$ 
run away to $+i\infty$. As $\kk$ increases past $1$, the latter suddenly change their 
real parts to $(2n+1)\pi$ and then descend from $+i\infty$ gradually along the lines  
$\Re\,A=(2n+1)\pi$.  Similarly, as $\kk$ decreases past $-1$, critical points with real parts 
equal to $2n\pi$ first run away to $-i\infty$ and then crawl up the lines $\Re\,A=(2n+1)\pi$. 

The singular nature of $\kk=\pm 1$ can be viewed in another way. 
At $\kk=\pm 1$ the structure of ``good'' regions on the complex $A$-plane changes 
globally. For $\kk>1$, $\Re\,S\to +\infty$ as $\Im\,A\to +\infty$, hence the area with 
$\Im\,A\gg 1$ is a good region. Similarly, for $\kk<-1$, the area with 
$\Im\,A\ll -1$ is a good region. The emergence of these good regions implies that 
now the Lefschetz thimbles ($\JJ$) can extend to infinity. 

%%%%%%%%%%%%%%%%%%%%%%%%%%%%%%%%%%%%%%%
   \begin{figure}[!t]
   \begin{center}
     \includegraphics[width=0.47\columnwidth]{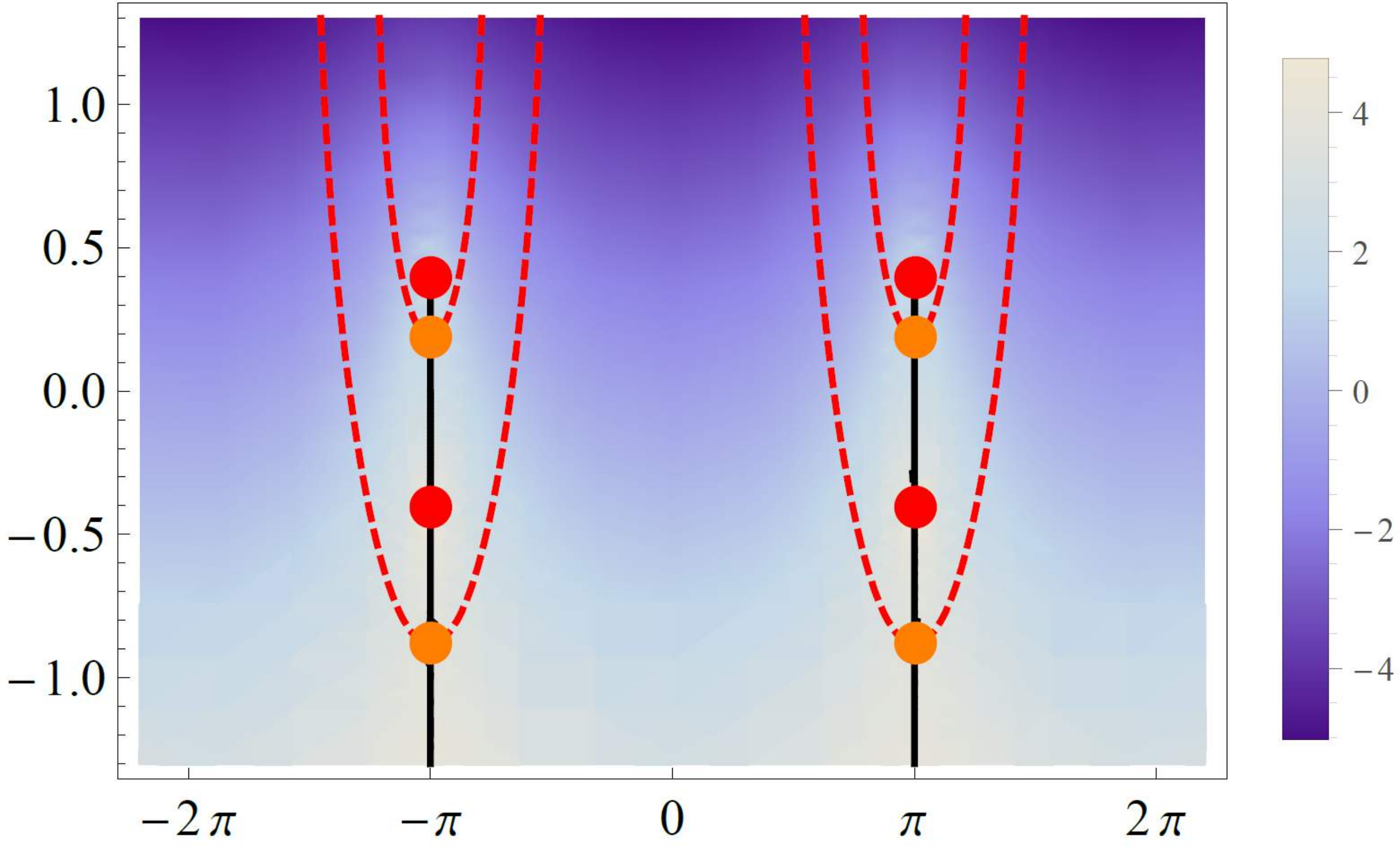}
     \quad 
     \includegraphics[width=0.47\columnwidth]{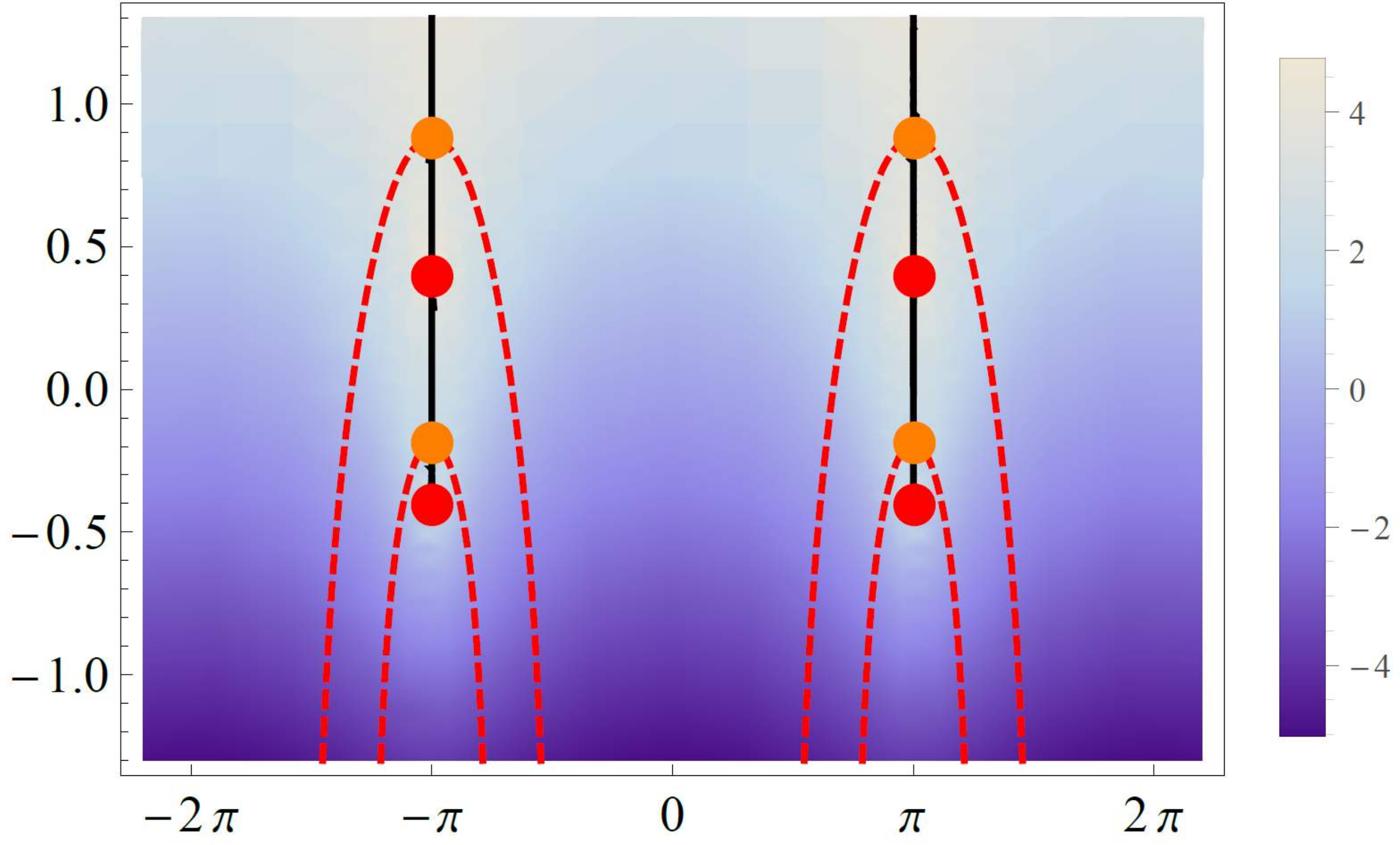}
     \put(-146,128){$\kk=3,~m=0.4$}
     \put(-367,128){$\kk=-3,~m=0.4$}
   \end{center} 
   \vspace{-.8\baselineskip}
   \caption{
     \label{fg:CS_largek}
     Same as Figure \ref{fg:CS_smallk} but with $\kk=\pm 3$. 
   } 
   \end{figure}
%%%%%%%%%%%%%%%%%%%%%%%%%%%%%%%%%%%%%%%

In Figure \ref{fg:CS_largek} we show $\JJ$ and $\KK$ for $\kk=\pm 3$. 
In each panel, there are four Lefschetz thimbles, consisting of 
two short ones that connect two adjacent logarithmic singularities, 
and two infinitely long ones. By carefully counting 
the intersection of $\mathcal{C}$ with the upward flow lines, we conclude that 
for $\kk=-3$ all the four thimbles contribute to $Z_N(k)$, while for $\kk=3$ 
only the two long ones contribute. In both cases, the total contribution sums up 
to zero for $k\in\ZZ$, because each thimble can be paired with a  
``partner thimble'' which is separated by $2\pi$ 
and has an opposite orientation. This cancellation is indeed required from 
the vanishing of the partition function as envisaged after \eqref{eq:Zvanish}. 
The case with $|\kk|>1$ thus offers a nice illustration of the fact that 
summing up contributions from multiple thimbles is mandatory for obtaining 
a correct result.

\section{Conclusion and perspective}
\label{sc:conclusion}

In this work, we have reported applications of the Picard-Lefschetz theory to 
a variety of fermionic models. While fermions can be integrated out from path integral, 
resulting effective actions are not entirely holomorphic and give rise to 
nontrivial behaviors of Lefschetz thimbles that are unseen in bosonic models. 
Taking the Gross-Neveu-like model with a complex four-fermion coupling as a test bed, 
we demonstrated how the partition function could be defined as a complex contour integral 
and how multiple complex saddles may be taken into account consistently using the 
Lefschetz-thimble techniques. The jumps of thimbles across the Stokes line were delineated, and 
the discrete chiral symmetry breaking was explained as an exchange of dominance between complex 
saddles. We also worked out the link between symmetry breaking, anti-Stokes lines and Lee-Yang zeros. 

Next, we outlined the determination of Lefschetz thimbles and their duals 
in the Nambu-Jona-Lasinio model with continuous chiral symmetry, showing that  
the method of \cite{Witten:2010cx} applies to fermionic systems with no extra difficulty. 
We then added a symmetry-breaking perturbation and explained that the structure of thimbles 
may be largely inferred from our former analysis of the Gross-Neveu-like model. 
It was uncovered that the number of thimbles changes discontinuously 
when a nonzero fermion mass is turned on, and that even for an infinitesimal mass, a smooth 
recovery of the symmetric limit is impeded by a highly intricate cancellation of large contributions between 
different thimbles.  

Finally we studied a $0+1$-dimensional model that mimics the Chern-Simons theory coupled to fermions. 
This model exhibits a ``quantum phase transition'' when the coefficient of the topological 
term is dialed. We examined this transition in detail and found that 
the global structure of Lefschetz thimbles undergoes a drastic change across the transition.

We believe that the methods and findings in this paper will have important implications for 
future application of the Picard-Lefschetz theory to path integral for QFTs. First and foremost, 
in QCD the two light quark flavors have small masses and there is an approximate 
$\SU(2)\times\SU(2)$ chiral symmetry. 
Also in lattice QCD, chiral symmetry of lattice fermions is usually broken explicitly by lattice artifacts. 
Our results for slightly broken chiral symmetry in Section \ref{sc:massnjl} would serve 
as a useful guide in these situations. For example, it is strongly suggested from 
our experience in toy models that summing up contributions of all thimbles is 
mandatory when there is an exact or approximate 
symmetry. This should be properly reflected in numerical implementation 
of the Lefschetz-thimble approach to QCD. 
It is worth noting that our treatment on 
continuously degenerate saddles in Section 
\ref{sc:NJLm} can be extended to other kinds of 
saddles in field theories, such as instantons, monopoles, vortices and kinks 
which all have various flat directions in field space.

Yet another direction of research to which our work is of potential relevance, is 
semiclassical analysis of QCD with adjoint quarks on $\RR^3\times S^1$. It was 
argued in \cite{Argyres:2012vv,Argyres:2012ka} 
that, at sufficiently small $S^1$, semiclassical expansion in this theory is well-defined and 
can be continued infinitely through an order-by-order cancellation of 
nonperturbative ambiguities. While it is tempting to conjecture that this offers a conceptual 
solution to the long-standing renormalon problem in QCD on $\RR^4$ \cite{Argyres:2012vv,Argyres:2012ka}, 
we have to recall that there is a chiral 
phase transition that separates a small-$S^1$ phase from a large-$S^1$ phase. So far not much work 
has been done as to the fate of semiclassical expansion across the chiral transition. It should be 
interesting to look into this issue using more sophisticated versions of our toy models. 

Finally we touch on the fundamental limitations of this work. Since the focus was on finite-dimensional 
integrals, we cannot tell exactly how our findings carry on to infinite-dimensional 
path integrals in field theories;  
in the first place it is still an open question whether a path integral in continuum QFTs 
can be decomposed into Lefschetz thimbles or not.%
\footnote{Of course a lattice regularization reduces a path integral to a finite-dimensional 
integral, to which the Lefschetz-thimble decomposition can be applied. 
However the next question is whether a sensible continuum and thermodynamic limit  
exists or not. This is by no means easier to answer.}  
Even if this proves possible, one has to figure out how to compute the coefficients 
$\{n_\sigma\}$ in \eqref{eq:23}. 
In general this would be a formidable task 
in the presence of infinitely many saddles in field theories.%
\footnote{Infinitely many contributing thimbles appear even in a finite-dimensional 
integral; see \cite{Berry1991b,Pasquetti:2009jg,Harlow:2011ny} for a nice illustration 
of this in the case of Gamma function.} 
We stress that this difficulty is not specific to fermionic theories but is common to any QFTs.  
In this paper we determined these numbers by plotting flow lines and counting their intersections, 
but this is of course impossible in higher dimensions and a novel approach is needed. 
After all, the Lefschetz-thimble approach to path integrals is still in its infancy and 
we should be optimistic.

%%%%%%%%%%   ACKNOWLEDGMENTS   %%%%%%%%%%

\acknowledgments

T.K. is supported by the RIKEN iTHES project and JSPS KAKENHI Grants Number 25887014. 
Y.T. is supported by Grants-in-Aid for JSPS fellows (No.25-6615) and 
the Program for Leading Graduate Schools, MEXT, Japan.

\bibliographystyle{JBJHEP}
\bibliography{draft_thimble.bbl}
%\bibliography{./Lef_thimble}
\end{document}